\documentclass[amssymb,aps,floats,preprint,prb]{revtex4}

\usepackage{bm}
\usepackage{graphics}
\usepackage{subfigure}
\usepackage{overpic}
\usepackage{graphicx}
\usepackage{bm}
\usepackage{color}
\usepackage{dsfont}
\usepackage{microtype}
\usepackage{exscale}

\newlength{\FigSpace}
\def\kv{{\bf k}}
\def\kp{k_\varphi}
\def\kz{k_z}
\def\qv{{\bf q}}
\def\da{\downarrow}
\def\ua{\uparrow}

\begin{document}

\title{Flux periodicities in loops and junctions with 
{\mbox{\boldmath$d$}}-wave superconductors}
\author{F. Loder, A. P. Kampf, and T. Kopp}
\affiliation{Center for Electronic Correlations and Magnetism, Institute of 
Physics, University of Augsburg, 86135 Augsburg, Germany}
\begin{abstract}
The magnetic flux periodicity in superconducting loops is reviewed. Whereas 
quantization of the magnetic flux with $hc/2e$ prevails in sufficiently thick 
loops with current free interior, the supercurrent in narrow loops is either 
$hc/2e$ or $hc/e$ periodic with the external magnetic flux. The periodicity 
depends on the properties of the condensate state, in particular on the Doppler
shift of the energy spectrum. For an $s$-wave superconductor in a loop with 
diameter larger than the coherence length $\xi_0$, the Doppler shift is small 
with respect to the energy gap, and the $hc/2e$ periodic behavior of its flux 
dependent thermodynamic properties is maintained. However, for smaller $s$-wave
loops and, more prominently, narrow $d$-wave loops of any diameter $R$, the 
Doppler shift has a strong effect on the supercurrent carrying state; as a 
consequence, the fundamental flux periodicity is in fact $hc/e$. It is shown 
analytically and numerically that the $hc/e$ periodic component in the 
supercurrent decays only algebraically as $1/R$ for large $d$-wave loops. For 
nodal superconductors the discrete nature of the eigenergies close to the Fermi
energy has to be respected in the evaluation of the Doppler shift. Furthermore,
we investigate, whether the Doppler shift modifies the supercurrent through 
Josephson junctions with $d$-wave superconductors. For transparent junctions, 
the Josephson current behaves similar to the persistent supercurrent in a loop.
These distinct physical phenomena can be compared, if the magnetic flux 
$\Phi=\phi\cdot hc/e$ is identified with the phase variation of the order 
parameter $\delta\varphi$ through $2\pi\phi=\delta\varphi/2$. Correspondingly, 
the Josephson current can display a $4\pi$ periodicity in $\delta\varphi$, if 
the Doppler shift is sufficiently strong which is true for transparent 
junctions of $d$-wave superconductors. Moreover, a $4\pi$ periodicity is also 
valid for the current-flux relation of field-threaded junctions. In the 
tunneling regime the microscopic theory reproduces the results of the 
Ginzburg-Landau description for sufficiently wide Josephson junctions. 
\end{abstract} 
\maketitle
\section{Introduction}
\label{secI2}
The quantum mechanical wave function  $\psi$ of particles moving in a multiply 
connected geometry has to be a unique function of the spatial coordinate. This
condition leads to a discrete energy spectrum, because the phase difference of 
the wave function accumulated on a closed path has to be $2\pi k$, where the 
integer $k$ serves as a quantum number of the wave function. For a circular 
geometry, this phase winding number $k$ represents the angular momentum $\hbar 
k$ of the particles. 

In the presence of a magnetic field ${\bf B}({\bf r})=\nabla\times{\bf A}({\bf 
r})$, an additional term adds to the phase of the wave function: 
$\psi'= \psi \exp(-{\rm i}\,2\pi (e/h c)\int_{{\bf r}_0}^{\bf r}{\rm d}{\bf r}'
\cdot{\bf A}({\bf r}'))$, where ${\bf A}({\bf r})$ is the vector 
potential, $e$ the charge of the electron, $c$ the velocity of light, $h$ is 
Planck's constant, and ${\bf r}_0$ an arbitrary space point within the system. 
The gauge transformed wave function $\psi'$ satisfies the Schr\"odinger 
equation with the vector potential ${\bf A}$ eliminated from the kinetic energy
term. The new condition is that $\psi'$ acquires the phase factor 
$\exp(-i2\pi (e/h c) \Phi)$ for a path $C$ enclosing the magnetic flux $\Phi=
\int_{C}{\rm d}{\bf r}\cdot{\bf A}({\bf r})$. This leads to a total phase 
difference of $2\pi(k-e\Phi/hc)$ on the closed path $C$. Because physical 
quantities are obtained by a thermal average over all possible $k$, they are 
periodic in $\Phi$ with the fundamental period
\begin{equation}
\Phi_0=hc/e ,
\end{equation}
which is the flux quantum in the normal state. In particular, the persistent 
current $J(\Phi)$ induced by the magnetic flux vanishes whenever $\Phi/\Phi_0$ 
is an integer.

The effect described above is present in any system with sufficient phase 
coherence, and best known from the periodic resistance modulations of a 
microscopic metallic loop, predicted first by Ehrenberg and Siday in 
1948~\cite{ehrenberg} and in 1959 by Aharonov and Bohm~\cite{AB}. Already ten 
years earlier, London predicted the manifestation of a similar effect in 
superconducting loops, where the phase coherence is naturally 
macroscopic~\cite{London}: the magnetic flux threading the loop is quantized in
multiples of $\Phi_0$, because the interior of a superconductor has to be 
current free. London did not know about the existence of $\Phi_0/2$ flux quanta
in superconductors, but he already speculated that the supercurrent might be 
carried by pairs of electrons with charge $2e$ and that the superconducting 
flux quantum and hence the flux periodicity of the supercurrent is rather 
$\Phi_0/2$. This point of view became generally accepted after the 
\textquoteleft Theory of Superconductivity' by Bardeen, Cooper, and Schrieffer 
(BCS) was published in 1957~\cite{bcs}. Direct measurements of magnetic flux 
quanta $\Phi_0/2$ trapped in superconducting rings followed in 1961 by Doll and
N\"abauer~\cite{Doll} and by Deaver and Fairbank~\cite{Deaver}, corroborated 
later by the detection of $\Phi_0/2$ flux lines in the vortex phase of type 
II superconductors~\cite{Abrikosov,Essmann}.

For thin superconducting loops with walls thinner than the penetration depth 
$\lambda$, finite currents are flowing throughout the entire superconductor. 
The magnetic flux is consequently not quantized, but London introduced instead
the quantity $\Phi'=\Phi+\Lambda/c\oint{\rm d}{\bf r}\cdot{\bf J}({\bf r})$, 
the quantized ``fluxoid''. The flux $\Phi$ is the total flux threading the 
loop, which already includes the current induced flux. $\Lambda$ is a 
phenomenological constant parametrizing the strength of the current response of
the superconductor to the applied magnetic field; $\Lambda$ is related to the 
penetration depth via $\Lambda=4\pi\lambda^2/c^2$ through the London 
equation~\cite{London}. Thin superconducting loops therefore react periodically
to the continuous variable $\Phi$.

\begin{figure}[tb]
\center{\includegraphics[width=4.5cm]{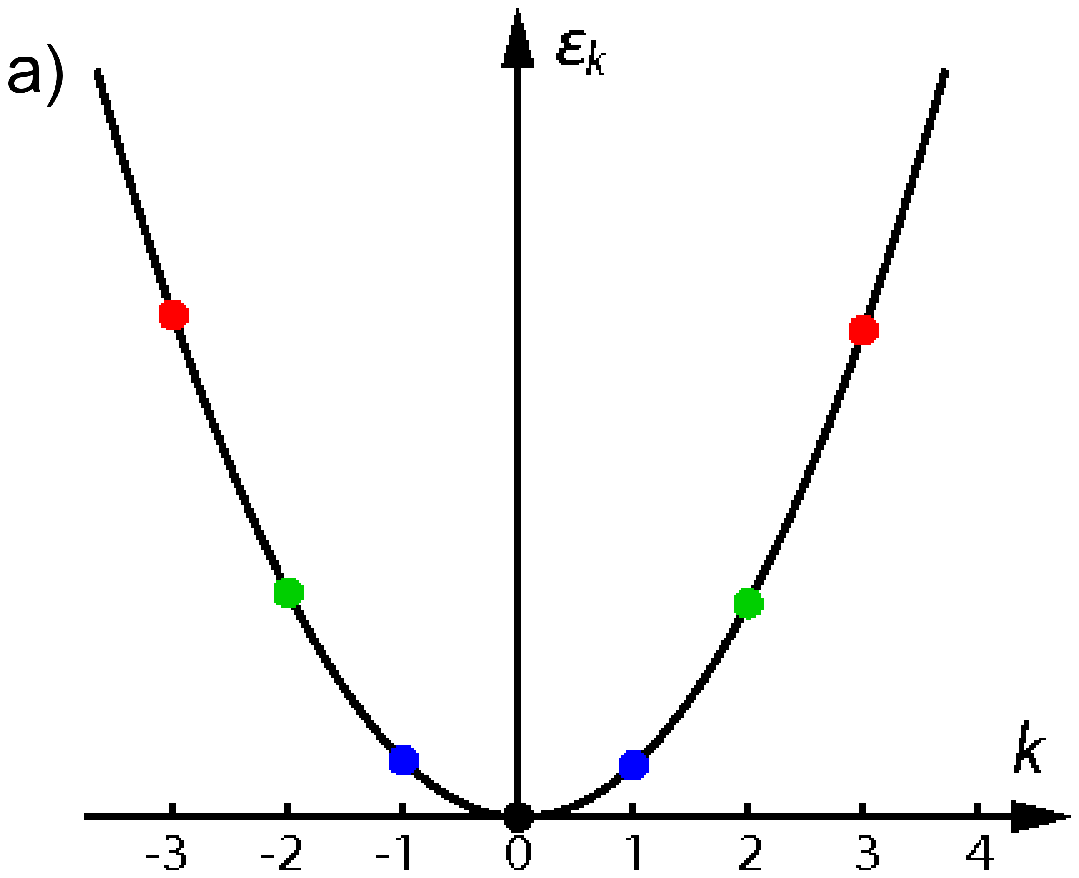}\hspace{1.8cm}\includegraphics[width=4.5cm]{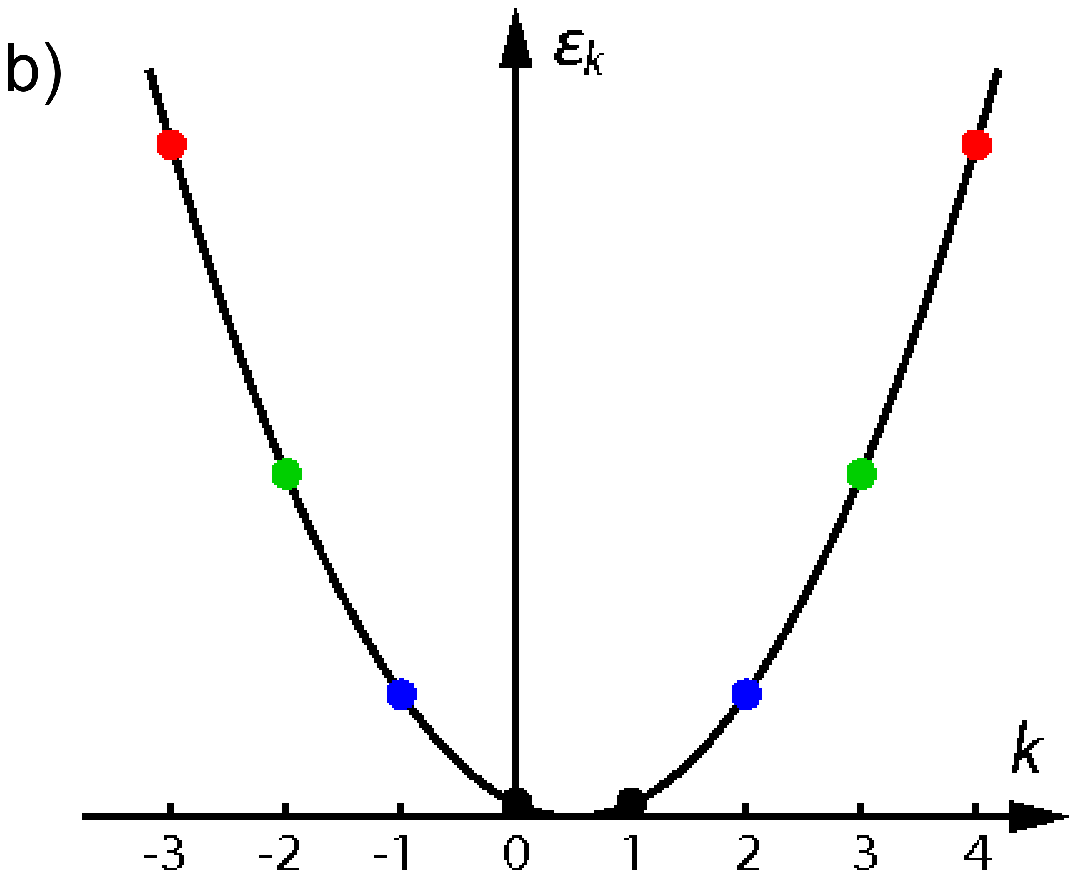}}
\caption{Scheme of the pairing of angular-momentum eigenstates in a one 
dimensional metal loop for (a) $\Phi=0$ and (b) $\Phi=\Phi_0/2$, as used by
Schrieffer in~\cite{schrieffer} to illustrate the origin of the $\Phi_0/2$ 
periodicity in superconductors. Paired are states with equal energy, which 
leads to pairs with a center-of-mass angular momentum $q=0$ in (a) and $q=1$ in
(b) in units of $\hbar$.}
\label{FigI1}
\end{figure}

It is tempting to relate the $\Phi_0/2$ flux periodicity of superconducting 
loops to the charge $2e$ of the Cooper pairs~\cite{onsager:61} which carry the 
supercurrent, but the pairing of electrons alone is not sufficient to explain 
the half-integer flux periodicity. A theoretical description of its true origin
was found independently in 1961 by Byers and Yang~\cite{Byers} and by 
Brenig~\cite{brenig:61} on the basis of the BCS theory by realizing that there 
are two distinct classes of superconducting wave functions that are not related
by a gauge transformation. An intuitive picture illustrating these two types of
states is contained in Schrieffer's book on 
superconductivity~\cite{schrieffer}, using the energy spectrum of a 
one-dimensional metallic ring: The first class of superconducting wave 
functions is related to pairing of electrons with angular momenta $k$ and $-k$
and equal energies without an applied magnetic field, as schematically shown in
Fig.~\ref{FigI1}~(a). The Cooper pairs in this state have a center-of-mass 
angular momentum $q=0$. The wave functions of the superconducting state for all
flux values $\Phi$, which are integer multiples of $\Phi_0$ and correspond to 
even pair momenta $q$, are related to the wave function for $\Phi=0$ by a gauge
transformation. For a flux value $\Phi_0/2$, pairing occurs between degenerate 
electrons with angular momenta $k$ and $-k+1$ [Fig.~\ref{FigI1}~(b)], and leads
to a pair momentum $q=1$. The corresponding wave function is again related by a
gauge transformation to the states for flux values $\Phi$ which are 
half-integer multiples of $\Phi_0$ and correspond to odd pair momenta.

\begin{figure}[tb]
\center{\includegraphics[width=6.0cm]{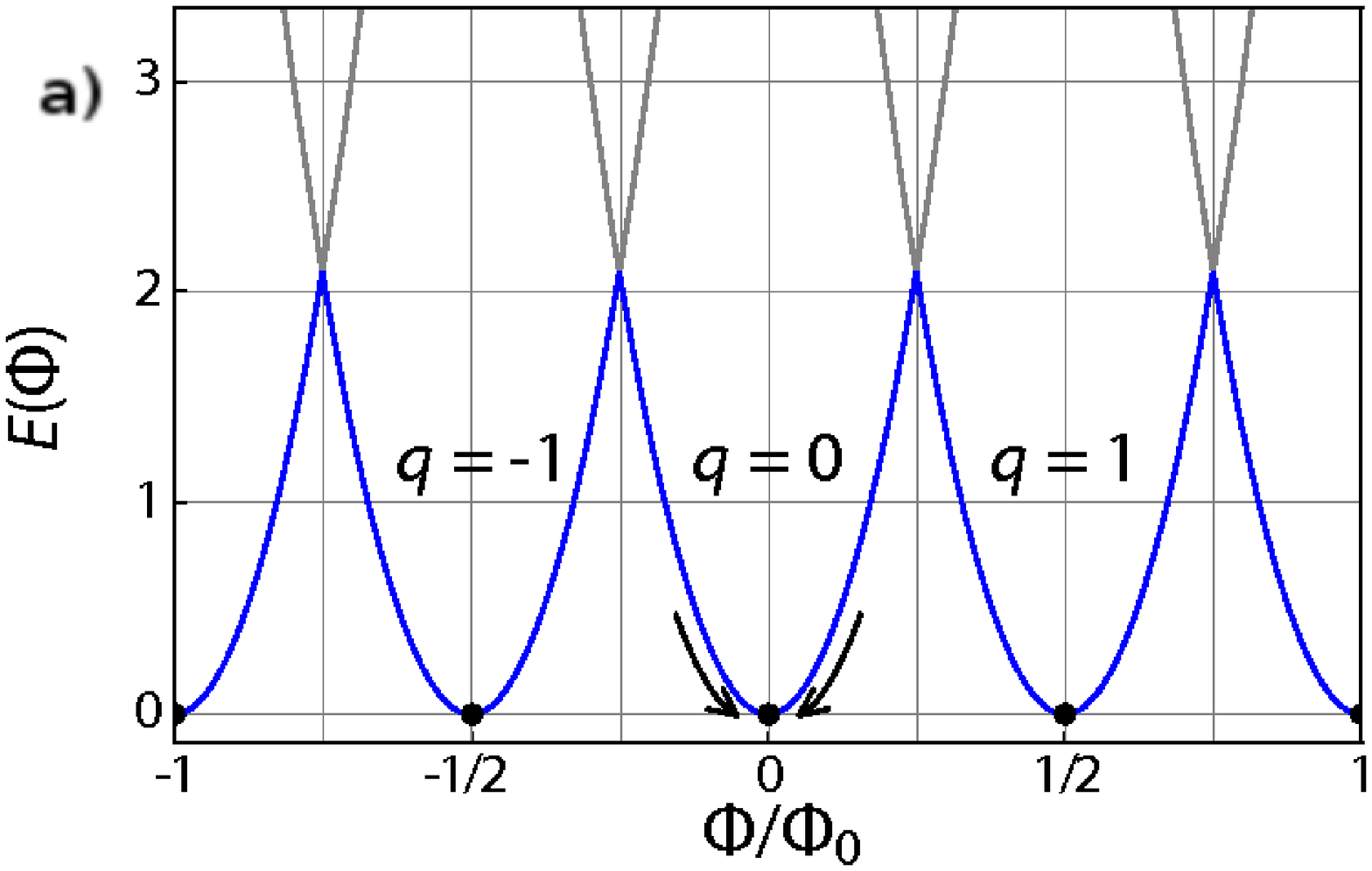}\hspace{.8cm}\includegraphics[width=6.0cm]{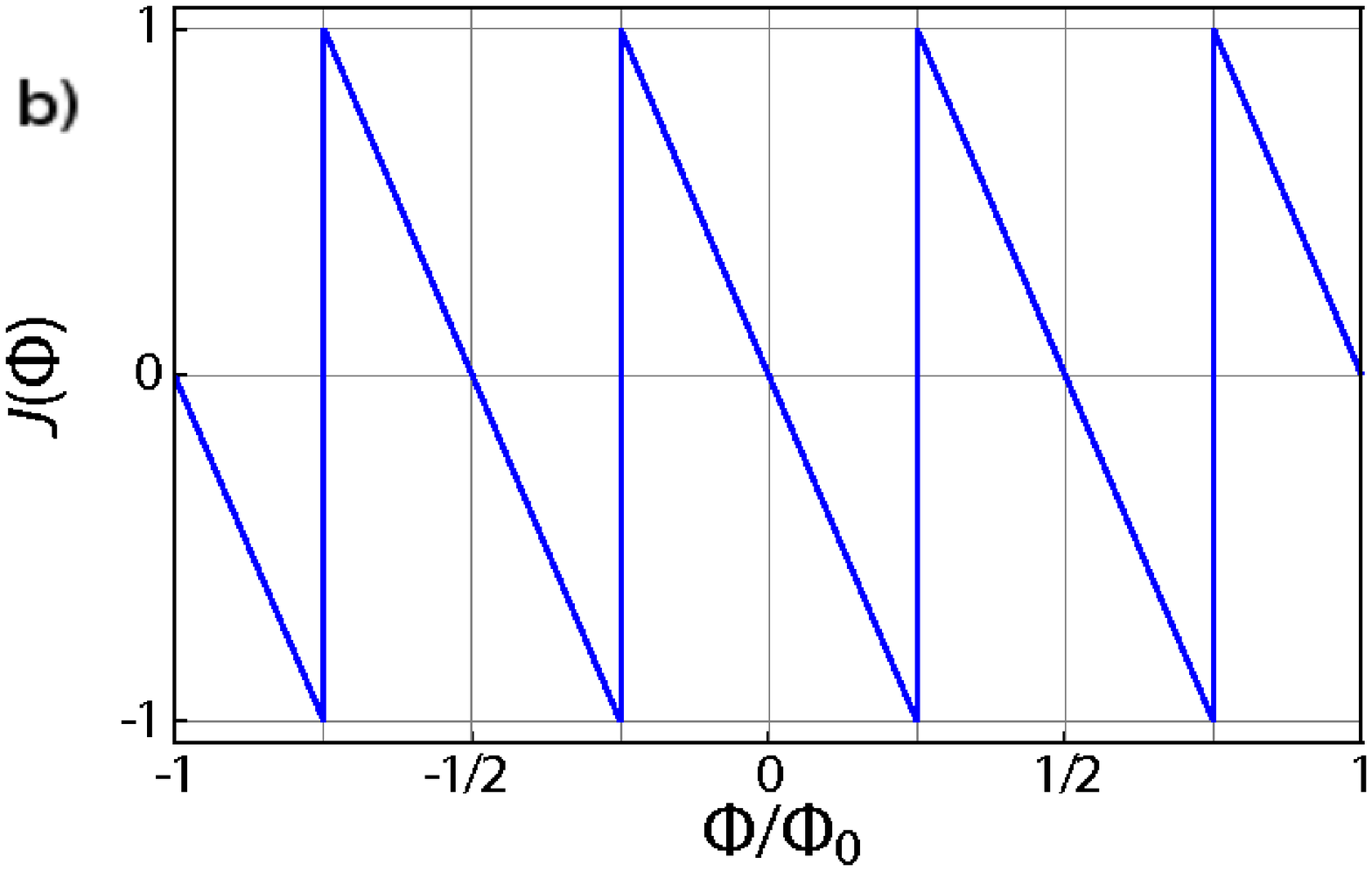}}
\caption{(a) Energy $E(\Phi)$ and (b) supercurrent $J(\Phi)$ as a function of 
flux $\Phi$ for a conventional superconducting loop at $T=0$. The minima in 
$E(\Phi)$ correspond to superconducting states with different pair momenta $q$.
The screening currents in the superconductor drive the system to the closest 
minimum for each flux value (black points), if the walls of the loop are 
thicker than $\lambda$.}
\label{FigI2}
\end{figure}

The two types of pairing states described above are qualitatively different. 
For the $\Phi_0/2$ periodicity, it is further required that the two types of 
states are degenerate. Byers and Yang as well as Brenig showed that this is 
indeed the case in the thermodynamic limit with a continuous density of states.
The energy $E(\Phi)$ is then determined by a series of intersecting parabolae 
with minima at integer multiples of $\Phi_0$ (corresponding to even pair 
momenta $q$) and half integer multiples of $\Phi_0$ (corresponding to odd pair 
momenta $q$) [Fig.~\ref{FigI2}~(a)]. If the loop is thicker than $\lambda$, the
system locks into the minimum closest to the value of the external flux. In 
finite systems however, the degeneracy of the even and odd $q$ minima is 
lifted, but their position is fixed by gauge invariance. The flux periodicity 
in thin loops is thus not necessarily $\Phi_0/2$, but the superconducting flux 
quantum remains $\Phi_0/2$. The circulating supercurrent $J(\Phi)$ is 
proportional to $\partial E(\Phi)/\partial\Phi$ and forms a $\Phi_0/2$ periodic
saw-tooth pattern in the thermodynamic limit as shown in Fig.~\ref{FigI2}~(b).

\section{Flux periodicities in cylinders: An analytic approach}
\label{sec:1}

\begin{figure}[b]
\center{\includegraphics[width=4.5cm]{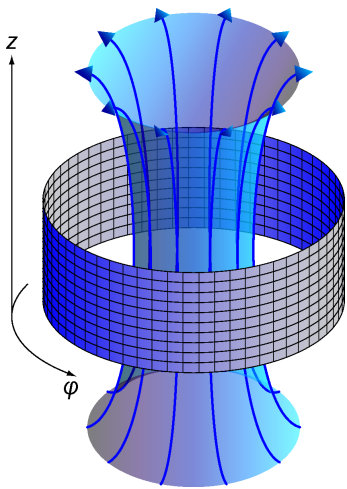}}
\caption{As a model system we use a thin-wall cylinder constructed from a 
two-dimensional discrete lattice. The interior of the cylinder is threaded by a
magnetic flux $\Phi$; we assume that the flux does not penetrate into the 
cylinder wall. $\Phi$ can be chosen arbitrarily, since quantization applies to 
the fluxoid and not the flux itself.}
\label{Figd0}
\end{figure}

For the discussion of the magnetic flux periodicity of $d$-wave superconductors
we choose to bend a discrete two-dimensional $N\times M$ square lattice to a 
cylinder (Fig.~\ref{Figd0}) with circumference $Na$ and height $Ma$. For two 
reasons we expect nodal superconductors to support a $\Phi_0=hc/e$ rather than 
a $\Phi_0/2$ periodicity. The first arises from the discrete nature of the 
eigenenergies in a finite system. For the thin cylinder shown in 
Fig.~\ref{Figd0} the mean level spacing in the vicinity of the Fermi energy 
$E_{\rm F}$ is $\delta_F\propto1/(NM)$; in $s$-wave superconductors with an 
order parameter $\Delta\gg\delta_F$, $\delta_F$ matters little. For 
superconducting states with gap nodes, the situation is different. In $d$-wave 
superconductors with an order parameter $\Delta_{\kv}\propto \kp^2-\kz^2$, the 
nodal states closest to $E_{\rm F}$ have to fulfill the condition $\kz=\kp$, 
thus there are fewer possible eigenstates and $\delta_F\propto 1/N$.

The second reason is that for gapless superconductors with a finite density of 
states close to $E_{\rm F}$, the occupation probabilities of these states 
change with flux. The flux dependence of the occupation enhances the difference
of current matrix elements for integer and half-integer flux values 
\cite{loder:08,loder:08.2,juricic:07,barash:07}. This effect is best understood
in terms of the spatial extent of a Cooper pair. In $s$-wave superconductors, 
the occupation probability remains constant for all $\Phi$, if the diameter of 
the cylinder is larger than the coherence length $\xi_0$. If this condition is 
fulfilled, the constituents of a Cooper pair cannot circulate separately, i.e. 
the pair does not feel the multiply connected geometry of the cylinder. But for
nodal superconducting states, the lengthscale which characterizes their 
coherence, diverges in the nodal directions and there are always Cooper pairs 
which extend around the circumference of the cylinder. Therefore nodal 
superconductors have no characteristic length scale above which the 
superconducting state is unaffected by the geometry of the system. These two 
combined effects are investigated on the basis of an analytic model in 
Sec.~\ref{sec3}.

\subsection{Superconductivity in a flux-threaded cylinder} 
\label{sec2}
\begin{figure}[tb]
\center{\includegraphics[width=9.0cm]{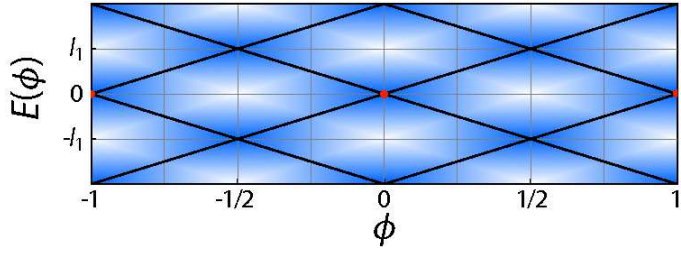}}
\caption{The energy spectrum of a cylinder in the normal state depends on the 
numbers $N$ and $M$, which parametrize the circumference and the height of the 
cylinder~\cite{loder:09}. The black lines represent the energy levels for a 
one-dimensional ring with $M=1$ and $N$ an integer, where level crossings occur
for integer values of $\phi=\Phi/\Phi_0$. $l_1$ is the maximum Doppler shift 
for $\phi=1/2$ (see Sec.~\ref{sec3}). For $M\gg1$, the levels split and form a 
quasi continuous spectrum that depends on the ratio $N/M$ (blue patches).}
\label{Figd0.1}
\end{figure}

The properties of a finite-size multiply connected superconductor depend 
sensitively on the discrete energy spectrum in the normal state. On the 
$N\times M$ square lattice, the flux values where levels cross have a high 
degeneracy for special ratios $N/M$; for $N=M$, the degree of degeneracy is 
$M$. For the latter case, the differences between the spectrum for integer and 
half-integer flux values are most pronounced. For $N=M\pm1$, the spectrum is 
almost $\Phi_0/2$-periodic. Away from these special choices of $N$ and $M$, 
the degeneracies are lifted, indicated by the blue shaded patches in 
Fig.~\ref{Figd0.1}. The size of the normal persistent current circulating 
around the cylinder is controlled by the change of the density of states near 
$E_{\rm F}$ upon increasing $\phi=\Phi/\Phi_0$. Since normal persistent 
currents in clean metallic rings are typically $\Phi_0$ periodic 
\cite{AB,washburn:92}, we will choose $N=M$ and a half-filled system with the
chemical potential $\mu=0$ for our model study, where  the $\Phi_0$ periodicity
of the spectrum is most clearly established. Whenever an energy level crosses 
$E_{\rm F}$ with increasing flux, the current reverses its sign. The current is 
$\Phi_0$-periodic for even $N$ and either paramagnetic or diamagnetic in the 
vicinity of $\phi=0$. For odd $N$, the current is $\Phi_0/2$-periodic. This 
lattice-size dependence persists also in rings with electron-electron 
interactions \cite{fye,Sharov:05,waintal} or in mesoscopic superconducting 
islands \cite{mineev}. 

We choose in the following $N$ and $M$ even, which leads to a normal state 
spectrum of the type shown in Fig.~\ref{Figd0.1}. This is not an obvious 
choice, but we will see in chapter 3 that one obtains this type of spectrum 
also for a square loop to which we will compare the results obtained for the 
cylinder geometry.

The starting point for our analysiss is the BCS theory for a flux threaded 
cylinder with circumference $Na=2\pi Ra$ and height $Ma$, where $R$ is the 
dimensionless radius of the cylinder and $a$ the lattice constant. The pairing 
Hamiltonian is given by
\begin{equation}
{\cal H}=\sum_{\kv,s}\epsilon_\kv(\phi)c_{\kv s}^\dag c_{\kv s}+
\sum_{\kv}\left[\Delta^*(\kv,\qv)c_{\kv\uparrow}c_{-\kv+\qv\downarrow}+
\Delta(\kv,\qv)c^\dag_{-\kv+\qv\downarrow}c^\dag_{\kv\uparrow}\right],
\label{s0}
\end{equation}
where $\kv=(\kp,\kz)$ with $\kp=n/R$ and $n\in\{-N/2+1,\dots,N/2\}$. The open 
boundary conditions in the $z$-direction along the axis of the cylinder allow 
for even-parity solutions with $\kz=(2m_e-1)\pi/M$ and odd-parity solutions 
with $\kz=2\pi m_o/M$, where $m_e,m_o\in\{1,\dots,M/2\}$. The operators 
$c^\dag_{\kv s}$ and $c_{\kv s}$ create and annihilate electrons with angular 
momentum $\hbar\kp/a$ and momentum $\hbar\kz/a$ in $z$ direction. For 
convenience, we choose $\kp$, $\kz\in[0,2\pi]$. The eigenenergies of free 
electrons moving on a discrete lattice on the surface of the flux threaded 
cylinder have the form
\begin{equation}
\epsilon_\kv(\phi)=-2t\left[\cos\left(\kp-\frac{\phi}{R}\right)+\cos\kz\right]
-\mu .
\label{s1}
\end{equation}
For $R\gg 1$, $\epsilon_\kv(\phi)$ is expanded to linear order in $\phi/R$;
\begin{equation}
\epsilon_\kv(\phi)-\epsilon_\kv(0)\approx-2t\frac{\phi}{R}\sin\kp
\label{s1.1}
\end{equation}
is commonly called the Doppler shift.

The superconducting order parameter in the pairing Hamiltonian (\ref{s0}) is 
defined through
\begin{equation}
\Delta(\kv,\qv)\equiv\Delta_q(\phi) g(\kv-\qv/2)=\frac{1}{2}\sum_{\kv'}V(\kv,
\kv',\qv)\langle c_{-\kv'+\qv\downarrow}c_{\kv'\uparrow}-c_{-\kv'+\qv\uparrow}
c_{\kv'\downarrow}\rangle ,
\label{s01.11}
\end{equation}
where $V(\kv,\kv',\qv)$ is the pairing interaction. Here we choose a $d$-wave 
interaction in separable form: $V(\kv,\kv',\qv)=V_1g(\kv-\qv/2)g(\kv'-\qv/2)$ 
with $g({\bf k})=\cos(\kp)-\cos(\kz)$; $V_1$ is the pairing interaction 
strength \cite{Loder2010}. The order parameter $\Delta(\kv,\qv)$ represents 
spin-singlet Cooper pairs with pair momentum $\hbar\qv/a$. On the cylinder, the
coherent motion of the Cooper pairs is possible only in the azimuthal 
direction, therefore $\qv=(q/R,0)$ with $q\in\{-N/2+1,\dots,N/2\}$. The quantum
number $q$ is obtained from minimizing the free energy. The $\phi$-dependence 
of $\Delta_q(\phi)$ enters through the self-consistency condition and has been 
discussed extensively in \cite{czajka:05} and \cite{loder:08.2} for $s$-wave 
pairing, where $g(\kv)\equiv{\rm const}$. As verified numerically, 
$\Delta_q(\phi)$ varies only little with $\phi$, and we start our analytic 
calculation with a $\phi$ and $q$ independent order parameter $\Delta(\kv,\qv)
\equiv\Delta(\kv)$ and $\Delta_q(\phi)\equiv\Delta$. As in our preceding work 
\cite{loder:08.2}, we take $q={\rm floor}(2\phi+1/2)$ in a first step. 
Since the Hamiltonian (\ref{s0}) is 
invariant under the simultaneous transformation $\phi\rightarrow\phi\pm1$ and 
$q\rightarrow q\pm2$, it is sufficient to consider $q=0$ or $1$ and the 
corresponding flux sectors $-1/4\leq\phi<1/4$ and $1/4\leq\phi<3/4$, 
respectively.

The diagonalization of the Hamiltonian (\ref{s0}) leads to the quasiparticle 
dispersion
\begin{equation}
E_\pm(\kv,\qv ,\phi)=\frac{\epsilon_\kv(\phi)-\epsilon_{-\kv+\qv}(\phi)}{2}\pm
\sqrt{\Delta^2(\kv)+\epsilon^2(\kv,\qv,\phi)},
\label{s02}
\end{equation}
with $\epsilon(\kv,\qv,\phi)=[\epsilon_\kv(\phi)+\epsilon_{-\kv+\qv}(\phi)]/2$.
Expanding $E_\pm(\kv,\qv,\phi)$ to linear order in both $\phi/R$ and $q/R$ 
gives
\begin{equation}
E_\pm(\kv,\qv,\phi)\approx-e_q(\kv)\pm
\sqrt{\Delta^2(\kv)+\left[\epsilon_\kv(0)-l_q(\kv)\right]^{2}},
\label{s02.1}
\end{equation}
where
\begin{equation}
e_q(\kv)=\frac{\phi-q/2}{R}2t\sin\kp \hskip1.0cm {\rm and} \hskip1.0cm  
l_q(\kv)=\frac{tq}{R}\sin\kp.
\end{equation}
In the normal state $\Delta=0$, the additive combination of $e_q(\kv)$ and 
$l_q(\kv)$ leads to the $\qv$-independent dispersion~(\ref{s1}). For 
$\Delta>0$, the dispersion (\ref{s02.1}) differs for even and odd $q$, except 
for special ratios of $N$ and $M$, as discussed above. This difference is 
crucial for nodal superconductors: The condition $\kp\approx\kz$ for levels 
close to $E_{\rm F}$ causes a level spacing $\delta_F\approx2l_1(\kv_F)$ for 
small $\Delta$, where $\kv_F$ is the Fermi momentum. For $N$ and $M$ even and 
$q=0$, the degenerate energy level at $E=E_{\rm F}=0$ splits into $M$ levels 
for increasing $\Delta$, which spread between $-\Delta$ and $\Delta$. For 
$q=1$, the degenerate levels closest to $E_{\rm F}$ are located at $E=\pm 
|l_1(\kv_F)|$, thus a gap of $2l_1(\kv_F)$ remains in the superconducting 
spectrum. If $N$ and $M$ are odd, the spectra for even and odd $q$ are 
interchanged, and if either $N$ or $M$ is odd, the spectrum is a superposition.

The gauge invariant circulating supercurrent is given by
\begin{equation}
J(\phi)=\frac{e}{h}\sum_{\kv,s}v_\kv n_{s}(\kv),
\label{s13}
\end{equation}
where $v_\kv={\partial\epsilon_\kv(\phi)}/{\partial(R\kp)}$ is the group 
velocity of the single-particle state with eigenenergy $\epsilon_\kv(\phi)$.
The spin independent occupation probability of this state is
\begin{equation}
n_{s}(\kv)=\langle c_{\kv s}^\dag c_{\kv s}\rangle
=u^2(\kv,\qv,\phi)f(E_+(\kv,\qv,\phi))+v^2(\kv,\qv,\phi)f(E_-(\kv,\qv,\phi))
\label{s14}
\end{equation}
with the Fermi function $f(E)$ and the Bogoliubov amplitudes
\begin{equation}
u^2(\kv,\qv,\phi)=\frac{1}{2}\left[\frac{\epsilon(\kv,\qv,\phi)}{E(\kv,\qv,
\phi)}+1\right]\hskip1.0cm {\rm and}\hskip1.0cm v^2(\kv,\qv,\phi)=\frac{1}{2}
\left[\frac{\epsilon(\kv,\qv,\phi)}{E(\kv,\qv,\phi)}-1\right].
\end{equation}
From Eqs.~(\ref{s13}) and (\ref{s14}), the supercurrent in the cylinder is
obtained by evaluating the sum either numerically or from the approximative 
analytic solution in Sec.~\ref{sec3}, which allows insight into the origin of 
the $\Phi_0$-periodicity in nodal superconductors. First, the analytic 
solution, which was introduced in Ref.~\cite{loder:09}, is reviewed.

\subsection{Analytic solution and qualitative discussion} 
\label{sec3}
An analytic evaluation of the supercurrent is possible in the thermodynamic 
limit where the sum over discrete eigenstates is replaced by an integral. For a
multiply connected geometry, this limit is not properly defined because the 
supercurrent or the Doppler shift vanish in the limit $R\rightarrow\infty$. 
Care is needed to modify the limiting procedure in a suitable way to access the
limit of a large but non-infinite radius of the cylinder~\cite{loder:09}. In 
this limit it is mandatory to consider the supercurrent density $j(\phi)=
J(\phi)/M$ rather than the supercurrent $J(\phi)$. In this scheme, we treat the
 density of states as a continuous function in any energy range where the level
spacing is $\propto 1/NM$, but we keep the finite energy gap of width $2l_q
(\kv_F)\propto1/R\propto1/N$ around $E_{\rm F}$ in the odd-$q$ sectors. For the
tight-binding dispersion in Eq.~(\ref{s1}), the density of states is a complete
elliptic integral of the first kind. For the purpose of an analytic 
calculation, a quadratic dispersion with a constant density of states is 
therefore a more suitable starting point. We use the expanded form of 
Eq.~(\ref{s1}):
\begin{equation}
\epsilon_\kv(\phi)=t\left[\left(\kp-\frac{\phi}{R}\right)^2+\kz^2\right]-\mu',
\label{s21.1}
\end{equation}
where $\mu'=\mu+4t$.

Some algebraic steps are needed to rearrange the sum in Eq.~(\ref{s13}) 
suitably to convert it into an integral. For finite $\phi$, $\epsilon_\kv(\phi)
\neq\epsilon_{-\kv}(\phi)$, and consequently the sum has to be decomposed into 
contributions with $\kp\geq0$ and $\kp<0$. We therefore take $\kp\geq0$ and 
decompose $v_\kv$ as
\begin{equation}
v_{\pm\kv}=\frac{2t}{R}\left(\pm\kp-\frac{\phi}{R}\right)=v_d(\kv)\pm v_p(\kv),
\label{s242}
\end{equation}
into a diamagnetic contribution $v_d(\kv)=-2t\phi/R^2$ and a paramagnetic 
contribution $v_p(\kv)=2t\kp/R$ \cite{scalapino:93}.

In a continuous energy integration, the Doppler shift is noticeable only in the
vicinity of $E_{\rm F}$. On the Fermi surface $\kp$ and $\kz$ are related by
\begin{equation}
k_{\varphi,F}(\kz)=\sqrt{\frac{\mu'}{t}-\kz^2}.
\label{s244}
\end{equation}
We therefore approximate $e_q(\kv)$ and $l_q(\kv)$ by $e_q(\kz)\approx2t(\phi-
q/2)k_{\varphi,F}(\kz)/R$ and $l_q(\kz)\approx tq k_{\varphi,F}(\kz)/R$, 
respectively. The eigenenergies (\ref{s02.1}) near $E_{\rm F}$ are thereby 
rewritten as
\begin{eqnarray}
E_+(\pm\kp,\kz,\qv,\phi)=\mp e_q(\kz)+\sqrt{\Delta_\kv^2+\left(\epsilon_\kv(0)
\mp l_q(\kz)\right)^2} 
\nonumber\\
E_-(\pm\kp,\kz,\qv,\phi)=\mp e_q(\kz)-\sqrt{\Delta_\kv^2+\left(\epsilon_\kv(0)
\mp l_q(\kz)\right)^2}
\label{s02.2}
\end{eqnarray}

The supercurrent $J(\phi)$ in Eq.~(\ref{s13}) is now evaluated by an integral 
over $\kp$ and $\kz$, which is decomposed into an integral over the normal 
state energy $\epsilon$ and an angular variable $\theta$. Within this scheme 
the density of states becomes gapless in the limit $M\rightarrow\infty$ for 
$q=0$, although $N$ is kept finite. For $q=1$ instead, a $\kz$-dependent gap 
$2|l_1(\kz)|$ remains. Thus we replace $\epsilon_\kv(0)\mp |l_q(k_z)|$ by the 
continuous quantity $\epsilon\pm|l_q(E_{\rm F},\theta)|$ where we use the 
parametrization
\begin{equation}
\left(
\begin{array}{l}\kp\\\kz\end{array}\right)=\left(\begin{array}{l}k\cos\theta\\ 
k\sin\theta \end{array}\right)=\sqrt{\frac{\epsilon+\mu'}{t}}
\left(\begin{array}{l}\cos\theta\\\sin\theta\end{array}\right),
\label{s246}
\end{equation}
with $\epsilon=tk^2-\mu'$. The energy integral extends over the whole 
tight-binding band width with $E_{\rm F}=0$ in the center of the band. 
Correspondingly, we integrate from $-\mu'$ to $\mu'$. Furthermore, the Doppler 
shift is parametrized for $\epsilon\approx E_{\rm F}$ as
\begin{equation}
e_q(\theta)=\frac{\phi-q/2}{R}2t\sqrt{\mu'/t}\cos\theta\hskip1.5cm {\rm and} 
\hskip1.5cm l_q(\theta)=\frac{tq}{R}\sqrt{\mu'/t}\cos\theta,
\end{equation}
where the function $l_q(\theta)$ is positive for $|\theta|\leq\pi/2$.
The supercurrent thus becomes
\begin{eqnarray}
j(\phi)&&=\frac{1}{M}\frac{e}{h}\left[\sum_{\kp>0,\kz,s}v_\kv n_{\kv s}(\qv)+
\sum_{\kp<0,\kz,s}v_\kv n_{\kv s}(\qv)\right]
\nonumber\\
&&\approx2{\cal N}\frac{e}{h}\!\int_{-\pi/2}^{\pi/2}\!\!{\rm d}\theta\int_{-
\mu'}^{\mu'}\!{\rm d}\epsilon[n_{q+}(\epsilon,\theta)v_{+}(\epsilon,\theta)+
n_{q-}(\epsilon,\theta)v_{-}(\epsilon,\theta)],
\label{s247}
\end{eqnarray}
where $n_{q\pm}(\epsilon,\theta)= n_{\pm\kv(\epsilon,\theta)}(\qv)$ and $v_\pm
(\epsilon,\theta)=v_{\pm\kv(\epsilon,\theta)}$. The constant density of 
states in the normal state is ${\cal N}=R/4\pi t$. We collect the terms 
proportional to $v_d(\epsilon,\theta)=-2t\phi/R^2$ into a diamagnetic current 
contribution $j_{\rm d}$ and those proportional to $v_p(\epsilon,\theta)=2t
k_{\varphi,F}(\epsilon,\theta)/R$ into a paramagnetic contribution $j_{\rm p}$.
Using $f(-E)=1-f(E)$, $j_{\rm d}$ and $j_{\rm p}$ become
\begin{eqnarray}
j_{\rm d}(q,\phi)&&=4{\cal N}\frac{e}{h}\!\int_{-\pi/2}^{\pi/2}{\rm d}\theta
\int_{l_q(\theta)}^{\mu'}{\rm d}\epsilon\, v_d\left(\epsilon,\theta\right)
\frac{\epsilon}{\sqrt{\Delta^2+\epsilon^2}}\left[f(E+e_q(\theta))-f(-E+e_q(
\theta))\right],
\label{s81.0}
\\[3mm]
j_{\rm p}(q,\phi)&&=4{\cal N}\frac{e}{h}\!\int_{-\pi/2}^{\pi/2}{\rm d}\theta
\int_{l_q(\theta)}^{\mu'}{\rm d}\epsilon\, v_p\left(\epsilon,\theta\right)
\left[f(-E-e_q(\theta))-f(-E+e_q(\theta))\right],
\label{s82.0}
\end{eqnarray}
Here, the integration is over positive $\epsilon$ only, and the lower boundary 
of the energy integration is controlled by $l_q(\theta)$. In Eq.~(\ref{s81.0})
we used the abbreviations $\Delta=\Delta(\theta)$ and $E=E(\epsilon,\theta)=
\sqrt{\Delta^2(\theta)+\epsilon^2}$. The current $j_{\rm d}$ is diamagnetic in 
the even-$q$ flux sectors and paramagnetic in the odd-$q$ sectors. For even 
$q$, it is equivalent to the diamagnetic current obtained from the London 
equations \cite{pethick:79,tinkham}. The current $j_{\rm p}$ has always the 
reverse sign of $j_{\rm d}$ and is related to the quasiparticle current as 
shown below. To analyze the flux dependent properties of the spectra and the 
current in the even-$q$ and odd-$q$ sectors, we explicitly distinguish $s$-wave
pairing and $d$-wave pairing with nodes in the gap function.

\subsubsection{$s$-wave pairing symmetry}
For $s$-wave pairing, $\Delta(\epsilon,\theta)\equiv\Delta$ is constant. 
Therefore, if we assume that $\Delta\geq e_q(\theta)$ for all $\theta$, the 
lower energy integration boundary in Eqs.~(\ref{s81.0}) and (\ref{s82.0}) is 
$\Delta$. Thus $j(\phi)=j_{\rm d}+j_{\rm p}$ is equal in both the even-$q$ and 
the odd-$q$ flux sectors and the flux periodicity is $\Phi_0/2$. However, if 
$\Delta<\max_\theta e_q(\theta)$, different calculational steps have to be 
followed in the evaluation of Eq.~(\ref{s13}), the results of which have been 
presented in \cite{loder:08.2}.

With $\epsilon=\sqrt{E^2-\Delta^2}$, Eqs.~(\ref{s81.0}) and (\ref{s82.0}) 
transform into integrals over $E$ with ${\rm d}\epsilon=D_s(E)\,{\rm d}E$, 
where
\begin{equation}
D_s(E)=\frac{\partial\epsilon}{\partial E}=\left\{\begin{array}{lll}E\,(E^2-
\Delta^2)^{-1/2}\hskip0.5cm {\rm for} \hskip0.5cm E\geq\Delta\\0 \hskip2.9cm 
{\rm for} \hskip0.5cm E<\Delta\end{array}\right.
\label{s6.1}
\end{equation}
is the density of states for $s$-wave pairing. This leads to
\begin{eqnarray}
j_{\rm d}&=&4{\cal N}\frac{e}{h}\!\int_{-\pi/2}^{\pi/2}{\rm d}\theta
\int_\Delta^{\mu'}{\rm d}E v_d\left(\sqrt{E^2-\Delta^2},\theta\right)\left[
f(E+e_q(\theta))-f(-E+e_q(\theta))\right],
\label{s81}
\\[3mm]
j_{\rm p}&=&4{\cal N}\frac{e}{h}\!\int_{-\pi/2}^{\pi/2}{\rm d}\theta
\int_\Delta^{\mu'}{\rm d}E D_s(E)v_p\left(\sqrt{E^2-\Delta^2},\theta\right)
\left[f(-E-e_q(\theta))\!-\!f(-E+e_q(\theta))\right] .
\label{s82}
\end{eqnarray}
At $T=0$, we obtain
\begin{eqnarray}
j_{\rm d}&=&-4{\cal N}\frac{e}{h}\!\int_{-\pi/2}^{\pi/2}{\rm d}\theta
\int_\Delta^{\mu'}{\rm d}E\,2t\frac{\phi-q/2}{R^2}=-2(\mu'-\Delta)\frac{e}{h} 
\frac{\phi-q/2}{R},
\label{s9}\\[3mm]
j_{\rm p}&=&4{\cal N}\frac{e}{h}\!\int_{-\pi/2}^{\pi/2}{\rm d}\theta
\int_\Delta^{e_q(\theta)}{\rm d}E D_s(E)\frac{2t}{R}\sqrt{\frac{\epsilon+\mu'}
{t}}\cos\theta
\nonumber\\
&=&\frac{8t{\cal N}}{R}\frac{e}{h}\sqrt{\frac{\mu'}{t}}\int_{-\pi/2}^{\pi/2}
{\rm d}\theta \cos\theta\int_\Delta^{e_q(\theta)}{\rm d}E D_s(E)+{\cal O}
\left(\frac{\epsilon}{t}\right)^2.
\label{s10.2}
\end{eqnarray}
The current $j_{\rm d}$ becomes independent of the superconducting density of 
states. Its size is proportional to $E_{\rm F}$, as long as $\mu'\gg\Delta$ 
holds. 

If $\Delta>e_q(\theta)$ for all values of $\theta$, then $j_{\rm p}=0$ and the
supercurrent $j(\phi)=j_{\rm d}$ is diamagnetic. For $T>0$, $j_{\rm d}$
decreases slightly. The current $j_{\rm p}$ increases with increasing $T$ and 
reaches its maximum value at $T_{\rm c}$. For finite temperatures $j_{\rm p}$ 
is referred to as the quasiparticle current. The supercurrent is always the sum
of the diamagnetic current $j_{\rm d}$ and the quasiparticle current 
$j_{\rm p}$, and therefore decreases with increasing temperature and vanishes 
at $T_{\rm c}$ \cite{vonoppen:92}. The quasiparticle current has the same flux 
periodicity as the supercurrent, even though it is carried by quasiparticle 
excitations. In the normal state ($\Delta=0$),
\begin{equation}
j_{\rm p}=\frac{8t{\cal N}}{R}\frac{e}{h}\sqrt{\frac{\mu'}{t}}
\int_{-\pi/2}^{\pi/2}{\rm d}\theta\cos\theta\int_0^{e_q(\theta)}{\rm d}E=4\mu'
\frac{e}{h}\frac{\phi-q/2}{R\pi}\int_{-\pi/2}^{\pi/2}{\rm d}\theta\cos^2\theta
=2\mu'\frac{e}{h}\frac{\phi-q/2}{R}
\label{s52}
\end{equation}
which cancels $j_{\rm d}$ exactly in the 
limit\footnote{In this procedure, the normal persistent current vanishes, but 
this is of no concern here because the normal current above $T_{\rm c}$ is 
exponentially small for $T_{\rm c}\gg\delta_F$.} $M\rightarrow\infty$. 

\subsubsection{Unconventional pairing with gap nodes}

\begin{figure}[tb]
\center{\includegraphics[width=7.0cm]{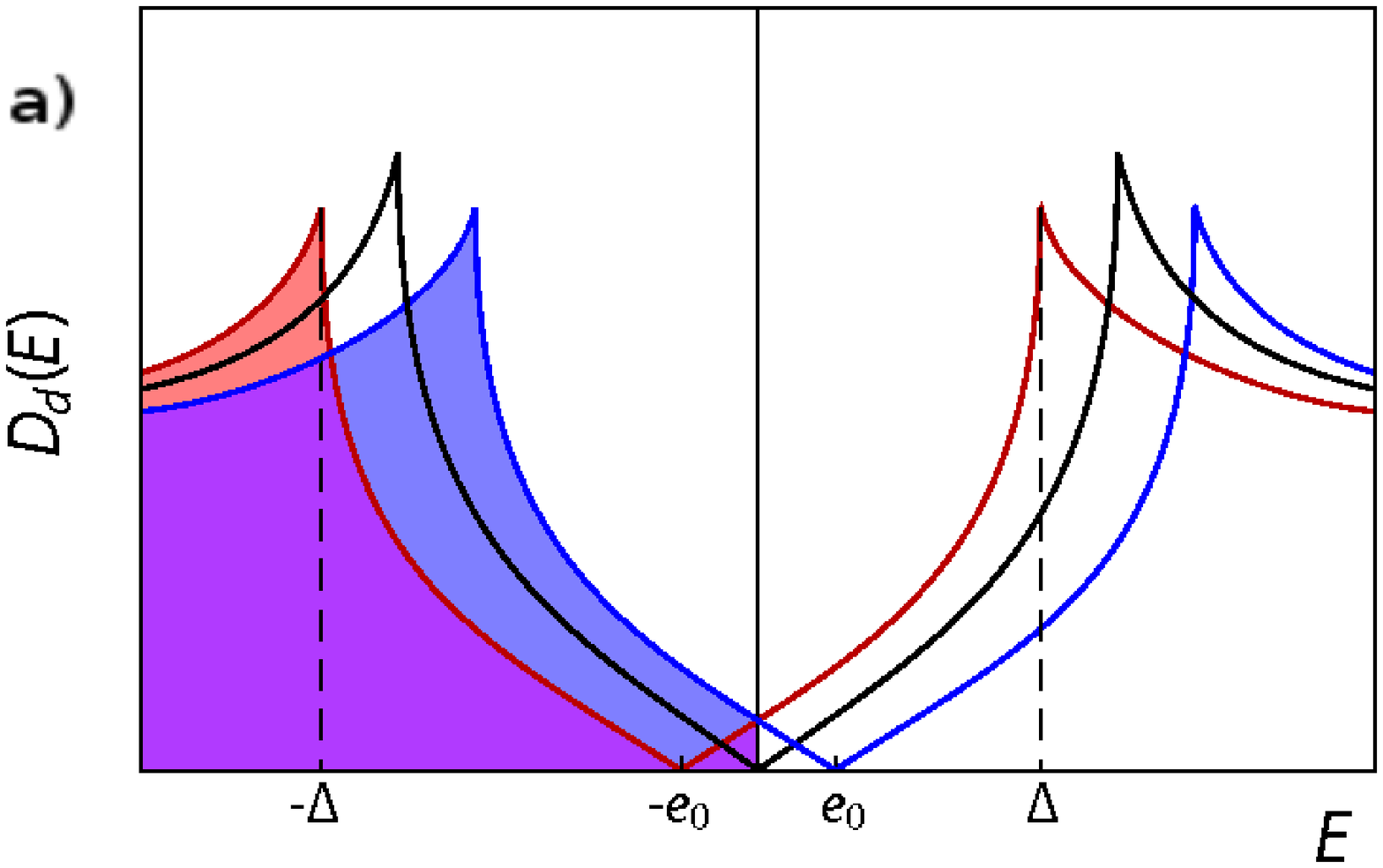}\hspace{.1cm}\includegraphics[width=7.0cm]{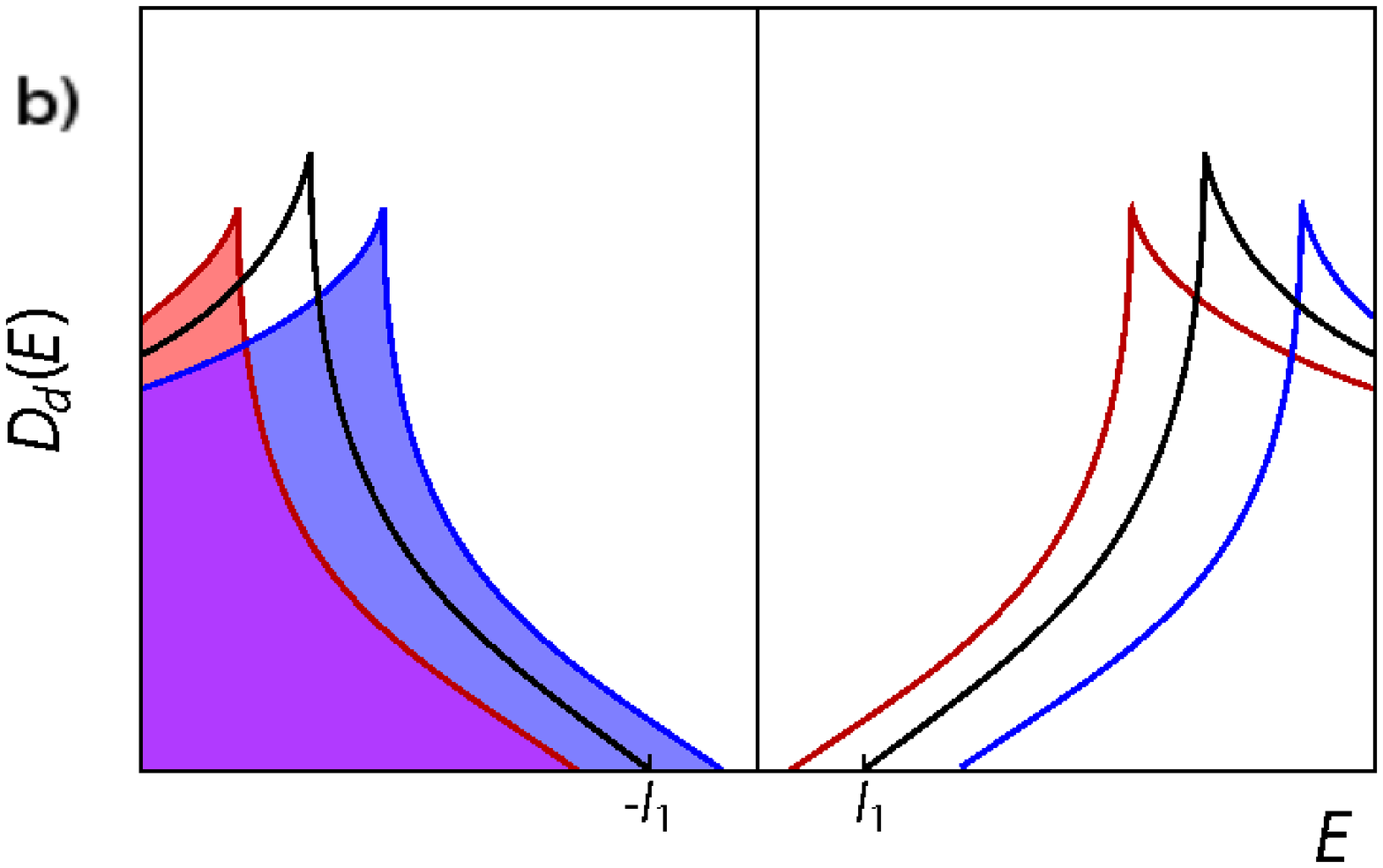}}
\caption{Scheme for the density of states of a $d$-wave superconductor for 
$\phi=1/4$, where $e_q=l_1/2$~\cite{loder:09}. The center-of-mass angular momentum $\hbar 
q/aR$ of the Cooper pairs is (a) $q=0$ and (b) $q=1$. The energies are Doppler 
shifted to higher (red) or lower energies (blue). This results in a double-peak
structure; for $q=0$ the upper and lower band overlap in the region $-e_0<E<e_0$
\cite{khavkine:04} and states in the upper band become partially occupied. For 
$q=1$ there is a gap $l_1$ of the size of the maximum Doppler shift at 
$\phi=1/4$. The black line represents the density of states (a) for $\phi=0$ 
and (b) for $\phi=1/2$.}
\label{Figd1}
\end{figure}

Equation~(\ref{s9}) for $j_{\rm d}$ is valid also for unconventional order 
parameter symmetries. Physically, $j_{\rm d}$ reflects the difference in the 
density of states of quasiparticle states with orbital magnetic moments 
parallel and anti-parallel to the external magnetic field. The former states are
Doppler shifted to lower energies, whereas the latter are Doppler shifted to 
higher energies. This is schematically shown in Fig.~\ref{Figd1} for $d$-wave 
pairing (c.f. \cite{khavkine:04}). In this picture, $j_{\rm d}$ is proportional
to the difference between the area beneath the red and and blue curves 
representing the density of states arising from $E_-(\pm|\kv|,\qv,\phi)<0$. 
Therefore we approximate $j_{\rm d}$ for $\Delta(\theta)\ll\mu'=E_{\rm F}+4t$ 
by
\begin{equation}
j_{\rm d}=-2\mu'\frac{e}{h} \frac{\phi-q/2}{R},
\end{equation}
as given in equation (\ref{s9}) with $\Delta=0$. On the other hand, $j_{\rm p}$
is represented by the occupied quasiparticle states in the overlap region of 
$E_+(\kv,\qv,\phi)$ and $E_-(\kv,\qv,\phi)$ with width $2e_q(\kv_F)$. It 
therefore strongly depends on the density of states in the vicinity of 
$E_{\rm F}$. In Fig.~\ref{Figd1}~(a), which refers to even $q$, the current 
$j_{\rm p}$ is determined by the small triangular patch where the upper and 
lower bands overlap. For odd $q$, the two bands do not overlap, therefore 
$j_{\rm p}=0$.

We will now analyze such a scenario for $d$-wave pairing with an order 
parameter $\Delta_\kv=\Delta(\kp^2-\kz^2)\approx\Delta\cos2\theta$. Again, we
assume $\Delta>e_q(\theta)$ for all $\theta$; then the integral in 
Eq.~(\ref{s82.0}) contains only the nodal states closest to $E_{\rm F}$, for 
which the $d$-wave symmetry demands $\kp\approx\kz$. Jointly with 
Eq.~(\ref{s244}) this condition fixes the Doppler shift at $E_{\rm F}$ to
the $\kv$-independent value $e_q=(\phi-q/2)\sqrt{2t\mu'}/R$ and $l_q=(q/R)
\sqrt{t\mu'/2}$. With the density of states 
\begin{equation}
D_d(E)=\frac{1}{\sqrt{E^2-\Delta^2\cos^22\theta}},
\label{s54}
\end{equation}
Eq.~(\ref{s82.0}) for the paramagnetic current $j_{\rm p}$ at $T=0$ takes the 
form
\begin{equation}
j_{\rm p}=4{\cal N}\frac{e}{h}\!\int_{l_q}^{e_q}{\rm d}E\int_{-\pi/2}^{\pi/2}
{\rm d}\theta D_d(E)\frac{2t}{R}\sqrt{\frac{\epsilon+\mu'}{t}}\sin\theta.
\end{equation}
In the odd-$q$ flux sectors, $l_q\geq e_q$ for all values of $\phi$, therefore
$j_{\rm p}=0$. In the $q=0$ sector, $l_q=0$ and
\begin{eqnarray}
j_{\rm p}\approx\frac{2e}{h\pi}\sqrt{\frac{\mu'}{t}}\int_{0}^{e_q}{\rm d}E
\int_{-\pi/2}^{\pi/2}{\rm d}\theta\sin\theta\frac{1}{\sqrt{E^2-\Delta^2\cos^2
2\theta}}\approx\frac{2e}{\pi h}\sqrt{\frac{\mu'}{t}}\int_{0}^{e_q}{\rm d}E
\frac{E}{\Delta}
\nonumber\\
=\frac{e}{\pi h\Delta}\sqrt{\frac{\mu'}{t}}e_q^2
=\frac{2}{\pi\Delta}\sqrt{t\mu'^3}\frac{e}{h}\left(\frac{\phi-q/2}{R}\right)^2,
\label{s55}
\end{eqnarray}
where the same approximations as in the $s$-wave case are applied. The dominant
contribution to the integral over $\theta$ originates from the nodal parts (see
e.g. \cite{mineev}).

In the even-$q$ sectors, the total current $j(\phi)=j_{\rm d}+j_{\rm p}$ 
becomes
\begin{equation}
j(\phi)=-2\mu'\frac{e}{h}\frac{\phi}{R}\left[1-\frac{\sqrt{t\mu'}}{\pi\Delta}
\frac{\phi}{R}\right],
\label{s34}
\end{equation}
which results in the ratio of the two current components
\begin{equation}
\frac{j_{\rm p}}{j_{\rm d}}=\frac{\sqrt{t\mu'}}{\pi\Delta}\frac{\phi}{R}\equiv 
b\phi .
\label{s45}
\end{equation}

\begin{figure}[tb]
\center{\includegraphics[width=6.0cm]{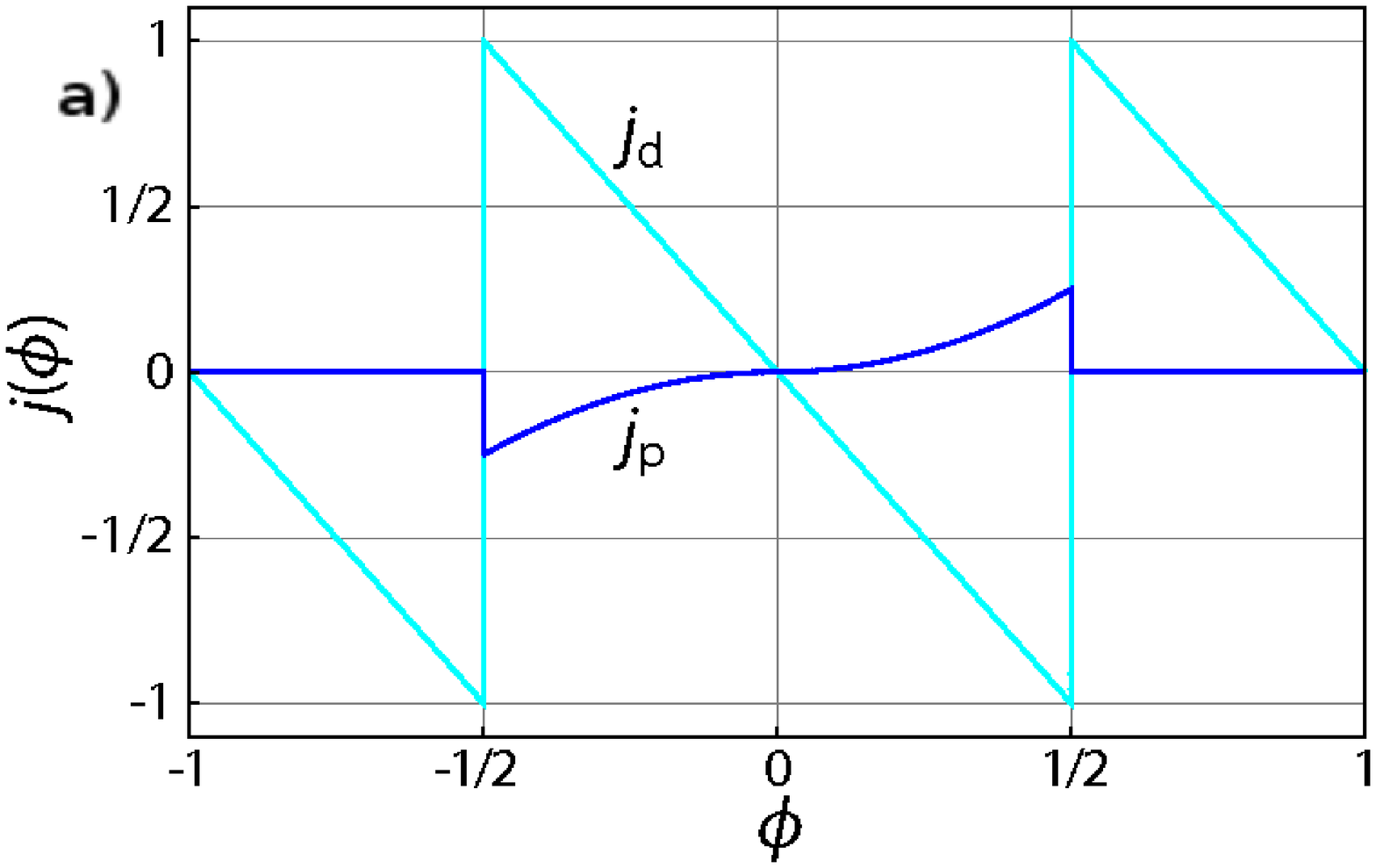}\hspace{.8cm}\includegraphics[width=6.0cm]{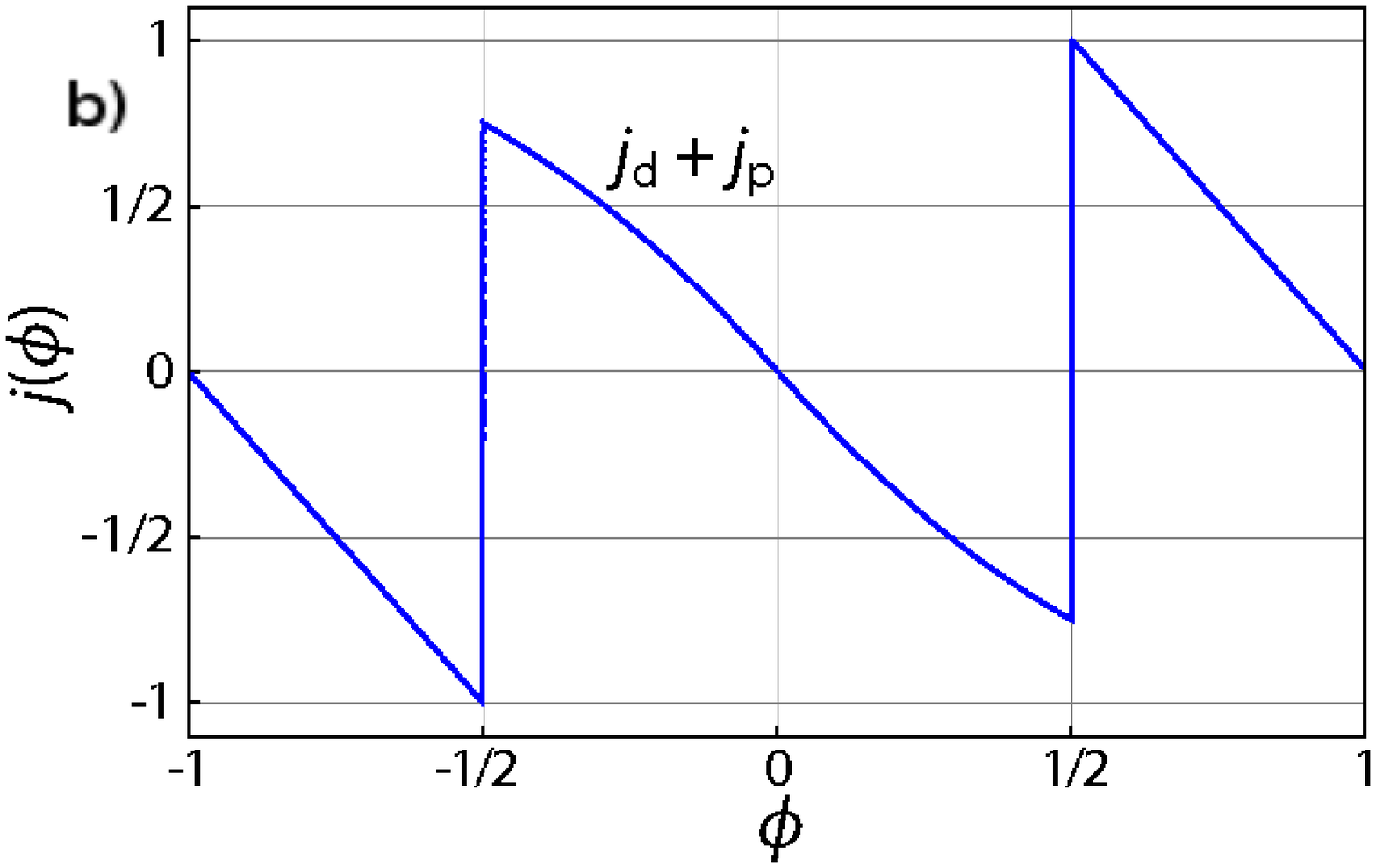}}
\caption{The supercurrent density $j(\phi)=j_{\rm d}+j_{\rm p}$ in a thin 
$d$-wave cylinder as a function of flux $\phi$ (arbitrary 
units)~\cite{loder:09}. Shown is the result of the analytic model calculation 
[Eq.~(\ref{s48})] for the characteristic value $b=0.4$. For $-1/4 < \phi<1/4$, 
where $q=0$, the current is reduced by a contribution proportional to $\phi^2$,
whereas it is linear in $\phi$ otherwise. This leads to an overall flux 
periodicity of $\Phi_0$.}
\label{Figd2}
\end{figure}

In the odd-$q$ flux sectors $j_{\rm p}=0$ and the supercurrent is $j(\phi)=
j_{\rm d}$. $j(\phi)$ is consequently $\Phi_0$ periodic; within one flux period
from $-1/2$ to $1/2$ we represent it as
\begin{equation}
j(\phi)=-2\frac{\mu'}{R}\frac{e}{h}\left\{\begin{array}{llrl}
\phi+1/2&{\rm for}&-1/2\leq&\phi<-1/4,\\
\phi(1-b\phi)&{\rm for}& -1/4\leq&\phi<1/4,\\
\phi-1/2&{\rm for}& 1/4\leq&\phi<1/2,
\end{array}\right.
\label{s48}
\end{equation}
(c.f. Fig.~\ref{Figd2}). The difference of the supercurrent in the even-$q$ 
and odd-$q$ flux sectors is represented best by the Fourier components $j_n=
\int_{-1/2}^{1/2}{\rm d}\phi\;j(\phi)e^{2\pi in\phi}$. For the first ($j_1$) 
and the second Fourier component ($j_2$) we obtain
\begin{equation}
j_1=-2\frac{\mu'}{R}\frac{e}{h}b\frac{8-\pi^2}{16\pi^3} \hskip1.0cm {\rm and} 
\hskip1.0cm j_2=-2\frac{\mu'}{R}\frac{e}{h}\frac{4\pi i-b}{16\pi^2}.
\label{s49}
\end{equation}
To leading order in $1/R$, the ratio of the $\Phi_0$ and the $\Phi_0/2$ Fourier
component is therefore
\begin{equation}
\left|\frac{j_1}{j_2}\right|=\frac{\pi^2-8}{4\pi^2}\frac{\sqrt{2t\mu'}}{\Delta 
R}, \hskip1.5cm \lim_{\mu\rightarrow 0}\left|\frac{j_1}{j_2}\right| 
\approx0.07\frac{2t}{\Delta R},
\label{s51}
\end{equation}
and scales with the inverse ring diameter. This $1/R$-law is the direct 
consequence of the $d$-wave density of states $D_d(E)\propto E$. Using 
Eq.~(\ref{s51}) to estimate this ratio for a mesoscopic cylinder with a 
circumference $Ra=2600a\approx 1\,\mu$m and a ratio $\Delta/t=0.01$, we obtain 
$j_1/j_2\approx0.03$.

\subsection{Further aspects} 
\label{sec6}
We have shown that in rings of unconventional superconductors with gap nodes, 
there is a paramagnetic, quasiparticle-like contribution $j_{\rm p}>0$ to the 
supercurrent at $T=0$. This current is generated by the flux-induced 
reoccupation of nodal quasiparticle states slightly below and above 
$E_{\rm F}$. Formally a coherence length $\hbar v_F/\Delta(\kv,\qv)>2R$ can be 
ascribed to these reoccupied states, which are therefore affected by the 
symmetry of the system. If the normal state energy spectrum has a flux 
periodicity of $\Phi_0$, than the superconducting spectrum is $\Phi_0$ 
periodic, too. The normal state spectrum of a cylinder with a discrete lattice 
strongly depends on the number of lattice sites. This problem is characteristic
for rotationally symmetric systems and is much less pronounced in geometries 
with lower symmetry, such as the square frame discussed in Sec. 3. In the 
latter system impurities do not change the spectrum qualitatively. For 
modelling an experimental arrangement a square loop geometry is therefore 
preferable. 

The $\Phi_0$ periodicity is best visible in the current component $j_{\rm p}$ 
at $T=0$. For $d$-wave-pairing $j_{\rm p}\propto1/R^2$, and the $\Phi_0$ 
periodic Fourier component decays like the inverse radius of the cylinder, 
relative to the $\Phi_0/2$ periodic Fourier component. The lack of a 
characteristic length scale in nodal superconductors, such as the coherence 
length for $s$-wave pairing, generates this algebraic decay. Although 
$j_{\rm p}$ is larger for small $\Delta$, it almost vanishes close to 
$T_{\rm c}$, if $\Delta\gg\delta_F$, and variations of $T_{\rm c}$ with flux, 
as in the Little-Parks experiment \cite{Little,Parks}, do not differ for 
$s$- and $d$-wave superconductors.

\section{Flux periodicity in square frames: Bogoliubov -- de Gennes approach}
\label{sec:squaresbdg}
So far we have presented the principles of the crossover from $\Phi_0$ to 
$\Phi_0/2$ flux periodicity in conventional and unconventional superconductors 
and the mechanisms that leads to the persistence of $\Phi_0$ periodicity in 
large loops of nodal superconductors. Now we present an alternative approach in
real space via the Bogoliubov -- de Gennes equations, which we introduce in 
Sec.~\ref{sec:bdg}. The information we obtain from this technique is 
complementary to Sec.~\ref{sec:1} where we followed the momentum-space 
formulation. The latter proved useful to understand the physical concepts and 
to describe large systems. The price paid was the restriction to highly 
symmetric systems with intriguing energy spectra in the normal state. This 
raises the question whether the $\Phi_0$ periodicity is detectable in realistic
setups, or whether it is rather an artifact of the high degeneracy of energy 
levels in clean and highly symmetric systems? On the other hand, the Bogoliubov
-- de Gennes equations in real space allow to determine the spectrum of 
``natural'' system geometries with reduced symmetry or systems containing 
lattice defects, impurities, magnetic fields or correlations in real space. 
Limitations of computational power, however, restrict the system size, and 
therefore the particular effects introduced by discreteness are unavoidably 
present.

The combination of momentum- and real-space methods can provide answers to the 
questions above. In the following, we first discuss the multi-channel loop for 
a square lattice: a square frame, as shown in Fig.~\ref{figrs0}, with a square 
hole at the center, threaded by a magnetic flux $\Phi$. We use this system in 
Sec.~\ref{sec:square} to study the flux periodicity in clean symmetric square 
frames; a part of this section is contained in~\cite{loder:08}. In 
Sec.~\ref{sec:junction}, we investigate different Josephson junction devices 
that respond periodically to magnetic fields. Junctions are modeled in real 
space by inserting potential barriers. In this context, we investigate also the
effect of impurities and lattice defects on the energy spectrum of the square 
frame.

\subsection{The Bogoliubov -- de Gennes equations}
\label{sec:bdg}
The Hamiltonian which we use in the following section has the form
\begin{equation}
{\cal H}=\!\sum_{\langle ij\rangle, s}t_{ij}c^\dag_{is}c_{js}+\!\sum_{i}\left[
\Delta^{\!*}_{i}c_{i\da}c_{i\ua}+\Delta_{i}c^\dag_{i\ua}c^\dag_{i\da}\right]+
\!\sum_{\langle ij\rangle}\left[\Delta^{\!*}_{ji}c_{j\da}c_{i\ua}+\Delta_{ij}
c^\dag_{i\ua}c^\dag_{j\da}\right]+\!\sum_{i,s}(U_i-\mu) c^\dag_{is}c_{is},
\label{j1}
\end{equation}
where $c^\dag_{is}$, $c_{is}$ are creation and annihilation operators for an 
electron on lattice site $i$ with spin $s$, and $\mu$ is the chemical 
potential. The sum $\sum_i$ runs over all lattice sites and the sum $\sum_{
\langle ij\rangle}$ is restricted to nearest-neighbor sites $i$ and $j$ only, 
and $t_{ij}=te^{\varphi_{ij}}$ with the hopping amplitude $t$ and the Peierls 
phase factor
\begin{equation}
\varphi_{ij}=\frac{e}{\hbar c}\int_i^j{\rm d}{\bf r}\cdot{\bf A}({\bf r}).
\label{j2}
\end{equation}
Additionally, we include an impurity term consisting of potential scatterers 
with repulsive potentials $U_i>0$, which we align to model tunnel junctions. A 
Hamiltonian of the form~(\ref{j1}) has often been used before for the numeric 
investigation of vortices in $d$-wave superconductors and the technique is 
described in detail in a number of 
articles~\cite{soininen:94,wang:95,zhu:00,Zhu:01,ghosal:02,chen:03}.

In the Hamiltonian Eq.~(\ref{j1}) two types of spin-singlet pairing are 
included. The on-site order parameter $\Delta_i$ represents conventional 
$s$-wave pairing originating from an on-site interaction. The order parameter 
$\Delta_{ij}$ originates from a nearest-neighbor interaction between the sites 
$i$ and $j$. They are defined through
\begin{equation}
\Delta_i=V_0\langle c_{i\da}c_{i\ua}\rangle
\quad {\rm and} \quad
\Delta_{ij}=\frac{V_1}{2}\left[\langle c_{j\da}c_{i\ua}\rangle-\langle c_{i\da}
c_{j\ua}\rangle\right].
\label{j3}
\end{equation}
with the interaction strengths $V_0$ and $V_1$. To diagonalize the Hamiltonian 
(\ref{j1}) we use the Bogoliubov transformation
\begin{equation}
c_{i\ua}=\sum_n\left[u_{n_i}a_{n\ua}-v^*_{n_i}a^\dag_{n\da}\right],\hskip1.5cm 
c_{i\da}=\sum_n\left[u_{n_i}a_{n\da}+v^*_{n_i}a^\dag_{n\ua}\right],
\label{j4}
\end{equation}
where the coefficients $u_{n_i}$ and $v_{n_i}$ are obtained from the eigenvalue
equation
\begin{equation}
\left(\begin{array}{c@{\quad}c} {\hat t} & {\hat \Delta} \\ {\hat\Delta^{\!*}} 
& -\hat t^* \end{array}\right)
{{\bf u}_n\choose{\bf v}_n}=E_n {{\bf u}_n\choose{\bf v}_n} .
\label{j5}
\end{equation}
The operators $\hat t$ and $\hat\Delta$ act on the vectors ${\bf u}_n$ and 
${\bf v}_n$ as
\begin{equation}
\hat tu_{n_i}=\sum_{j}t_{ij}u_{nj}+(U_i-\mu)u_{n_i}
\quad {\rm and} \quad
\hat\Delta v_{n_i}=\Delta_iv_{n_i}+\sum_{j}\Delta_{ij}v_{nj},
\label{j6}
\end{equation}
where $j$ labels the nearest-neighbor sites of site $i$. Inserting the 
transformation (\ref{j4}) into Eq.~(\ref{j3}) leads to the self-consistency 
conditions
\begin{equation}
\Delta_{i}=V_0\sum_nu_{n_i}v^*_{n_i}\tanh\left(\frac{E_n}{2T}\right)\; ,
\qquad
\Delta_{ij}=\frac{V_1}{2}\sum_n\left[u_{n_i}v^*_{nj}+u_{nj}v^*_{n_i}\right]
\tanh\left(\frac{E_n}{2T}\right).
\label{j7}
\end{equation}
Equations~(\ref{j7}) together with Eq.~(\ref{j5}) represent
the Bogoliubov -- de Gennes equations. 

The bond order parameters $\Delta_{ij}$ can be projected onto a $d$-wave 
component and an extended $s$-wave component defined as
\begin{equation}
\Delta^s_{i}=\frac{1}{4}\left[\Delta_{i,i+\hat x}e^{i\varphi_{i,i+\hat x}}+
\Delta_{i,i-\hat x}e^{i\varphi_{i,i-\hat x}}+\Delta_{i,i+\hat y}
e^{i\varphi_{i,i+\hat y}}+\Delta_{i,i-\hat y}e^{i\varphi_{i,i-\hat y}}\right],
\label{j8}
\end{equation}
\begin{equation}
\Delta^d_{i}=\frac{1}{4}\left[\Delta_{i,i+\hat x}e^{i\varphi_{i,i+\hat x}}+
\Delta_{i,i-\hat x}e^{i\varphi_{i,i-\hat x}}-\Delta_{i,i+\hat y}
e^{i\varphi_{i,i+\hat y}}-\Delta_{i,i-\hat y}e^{i\varphi_{i,i-\hat y}}\right].
\label{j9a}
\end{equation}
In a uniform system with nearest-neighbor pairing interaction only, the 
self-consistency Eq.~(\ref{j7}) selects a pure $d$-wave superconducting state, 
i.e. $\Delta^s_i=0$. Impurities, potentials or boundaries generate an extended 
$s$-wave contribution $\Delta^s_i>0$ \cite{franz:96}. The expectation value of 
the current $J_{ij}$ (cf. \cite{bagwell:94}) from site $i$ to $j$ is given by
\begin{equation}
J_{ij}=-8t\,\Phi_0\sum_n\mbox{Im}\left(u_{nj}u_{in}^*e^{-i\varphi_{ij}}\right)
f(E_n).
\label{J711}
\end{equation}

\subsection{Flux periodicity in square frames}
\label{sec:square}

\begin{figure}[tb]
\center{\includegraphics[width=7.0cm]{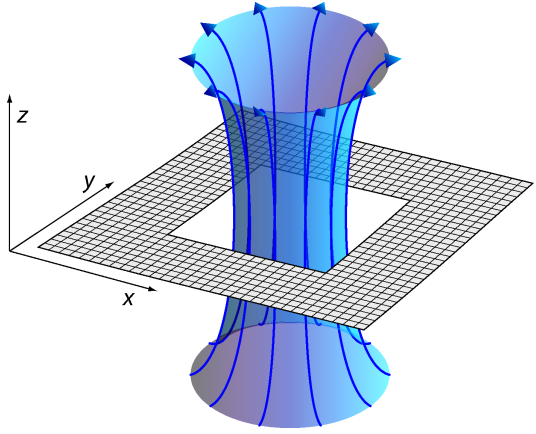}}
\caption{Illustration of a square loop threaded by a magnetic flux.
For the investigation of the flux periodicity of $d$-wave 
superconductors using the Bogoliubov -- de Gennes equations in real space, we 
use a discrete square lattice with open boundary conditions and a square hole 
in the center the frame, which is pierced by the magnetic flux $\Phi$.}
\label{figrs0}
\end{figure}

The Bogoliubov -- de Gennes equations introduced above are now applied to the 
square frame geometry shown in Fig.~\ref{figrs0}, consisting of a discrete 
$N\times N$ lattice with a centered $L\times L$ square hole threaded by a 
magnetic flux $\phi$, where $\phi=\Phi/\Phi_0$. The external magnetic field 
${\bf B}$ threading the hole is supposed not to penetrate into the frame, and 
we restrict it to the center of the hole. ${\bf B}$ is generated by a vector 
potential of the form ${\bf A}({\bf r})=2\pi \phi/|{\bf r}|^2(y,-x,0)$. 

\begin{figure}[b]
\center{\includegraphics[width=6.5cm]{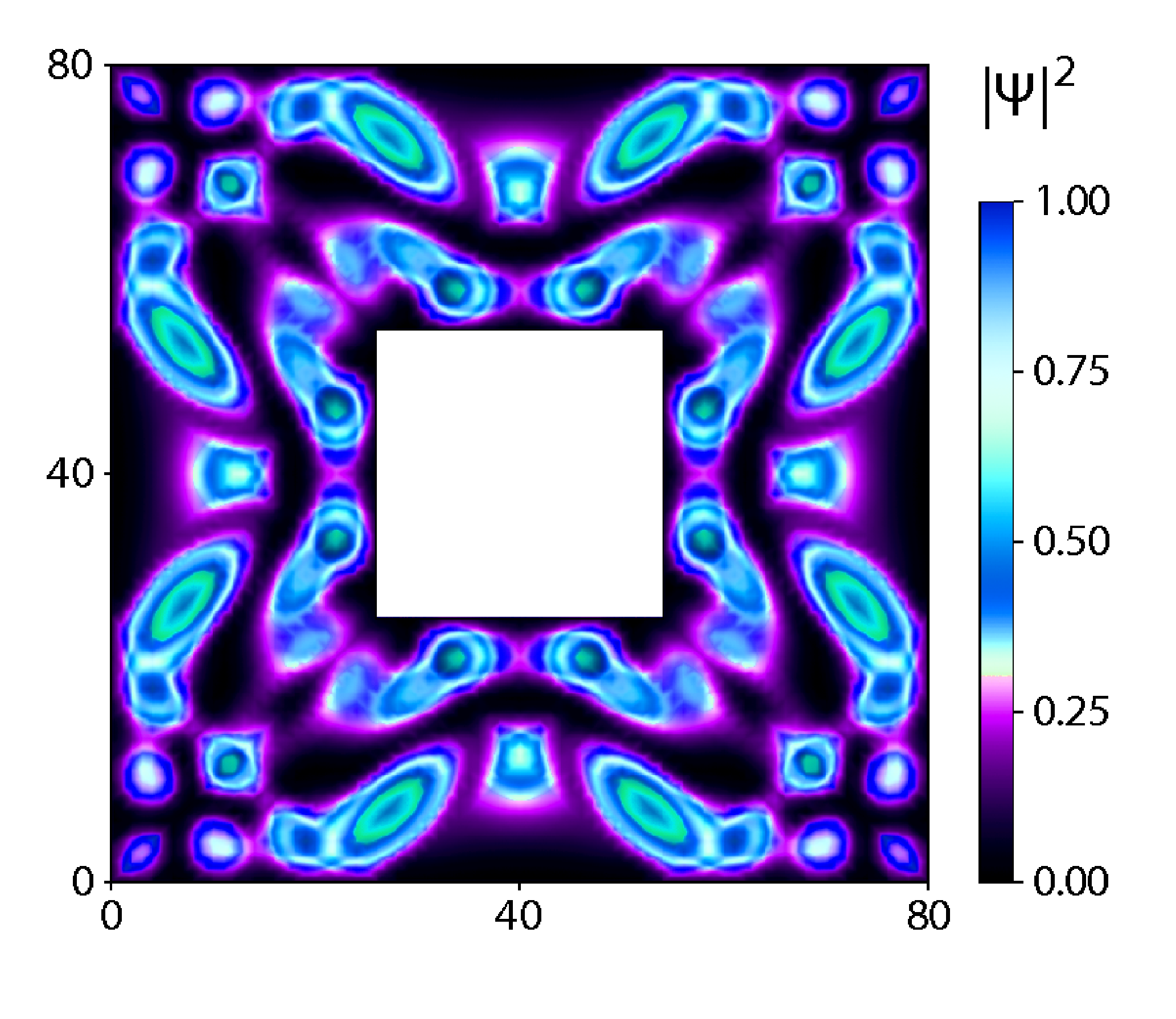}\hspace{0.8cm}\includegraphics[width=6.5cm]{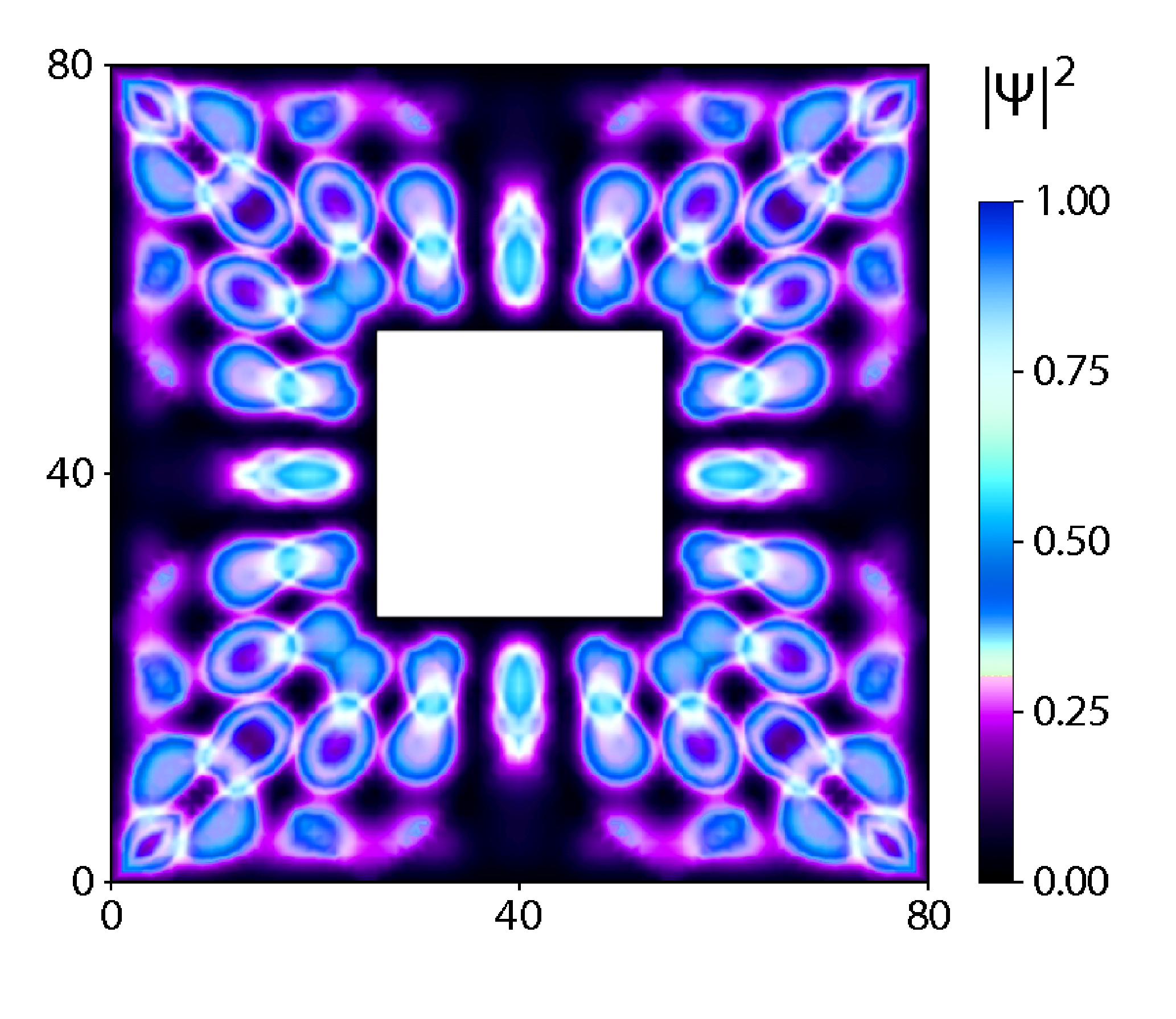}}
\caption{Real-space representations of a square loop with a typical electronic 
probability density $|\Psi|^2$. We show two eigenstates of the $d$-wave pairing
Hamiltonian with slightly different energies in the gap region, calculated for 
a square-loop with 80$\times$80 lattice sites and a pairing interaction
$V_1=0.3t$. The hole in the center has a size of 28$\times$28 unit cells.
To enhance the contrast of the complicated pattern, the special color
code shown on the right is used and the discrete lattice points are smoothly
interpolated.}
\label{figrs1}
\end{figure}

\begin{figure}[tb]
\center{\includegraphics[width=6.0cm]{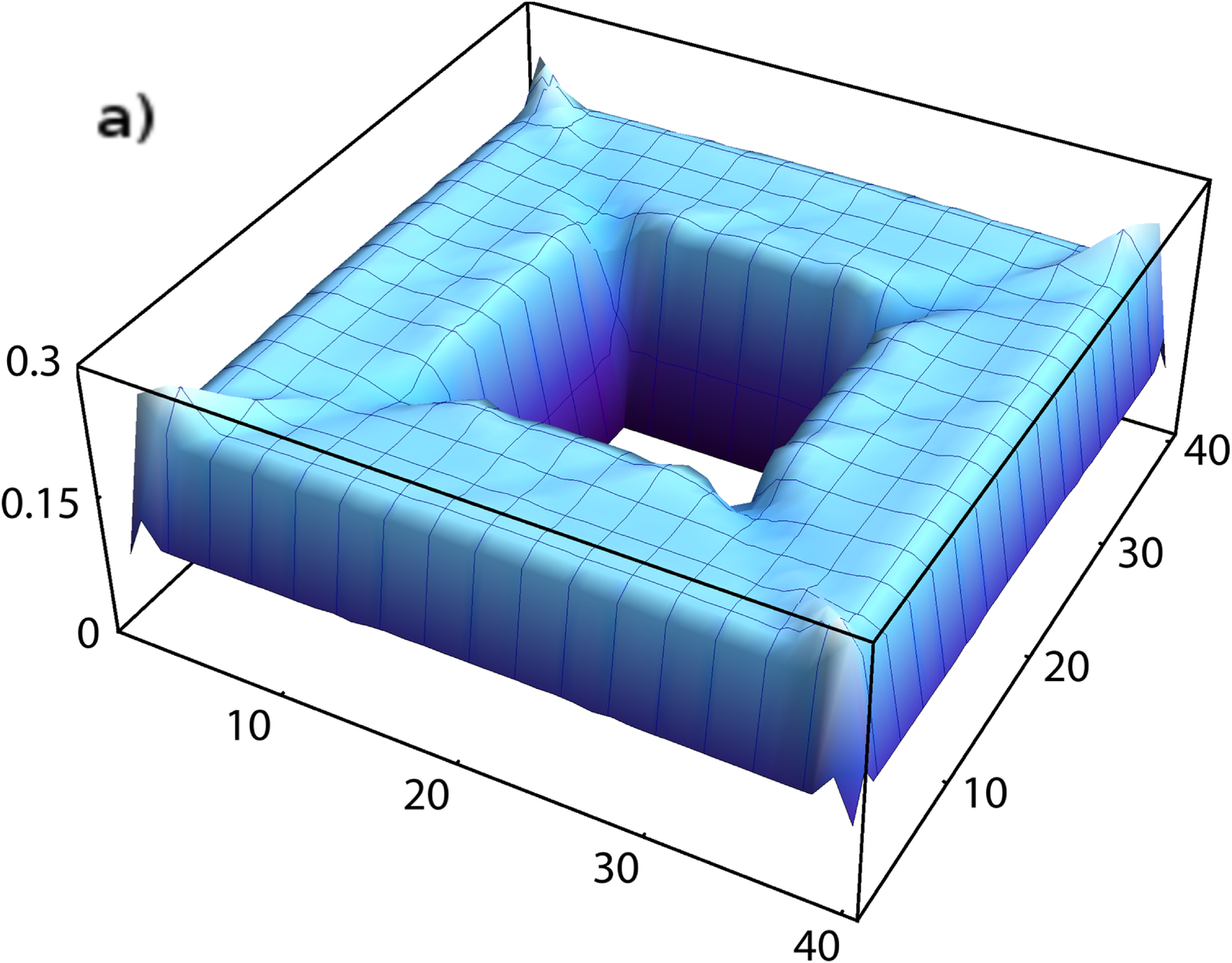}\hspace{.8cm}\includegraphics[width=6.0cm]{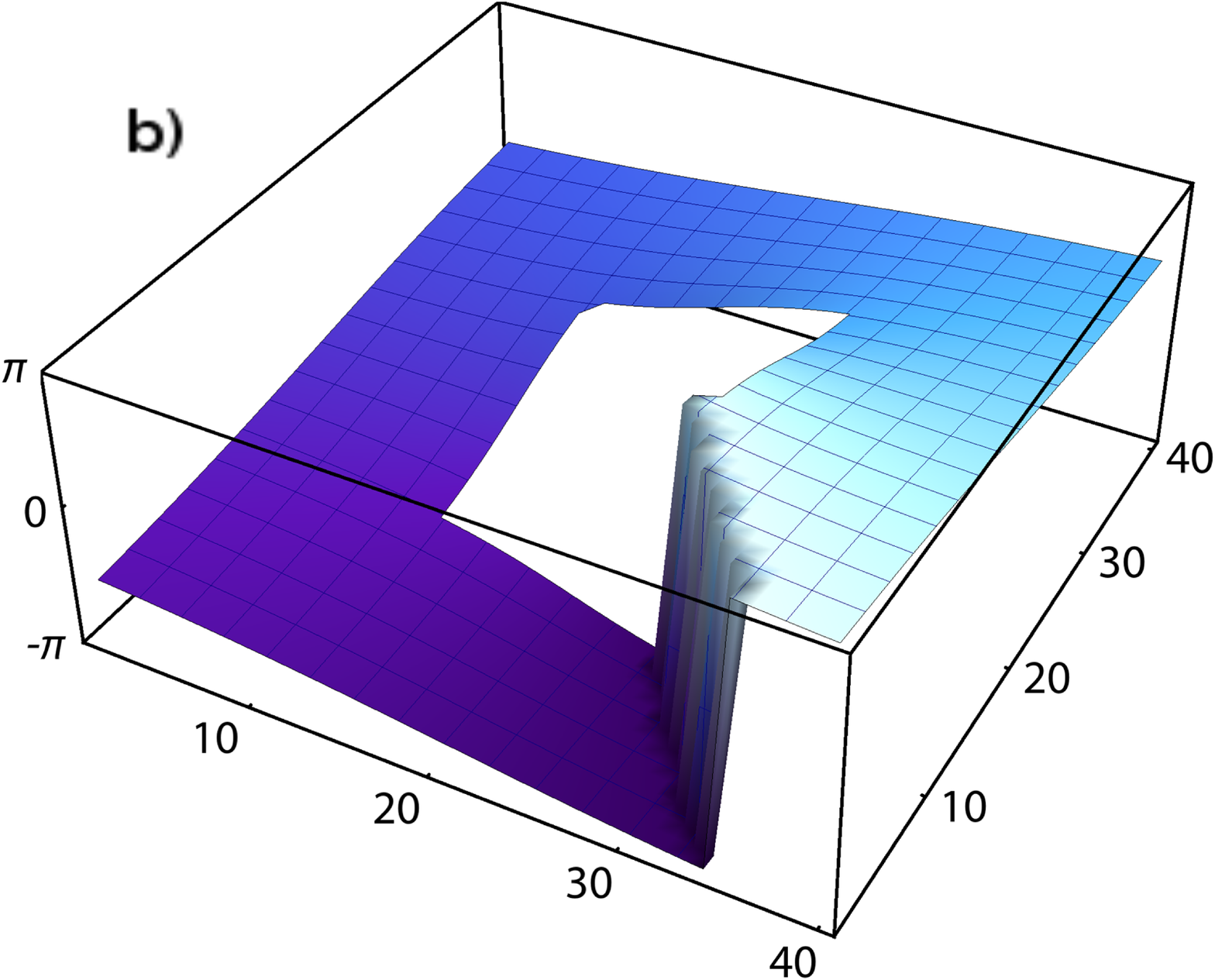}}
\vspace{\FigSpace}
\center{\includegraphics[width=6.0cm]{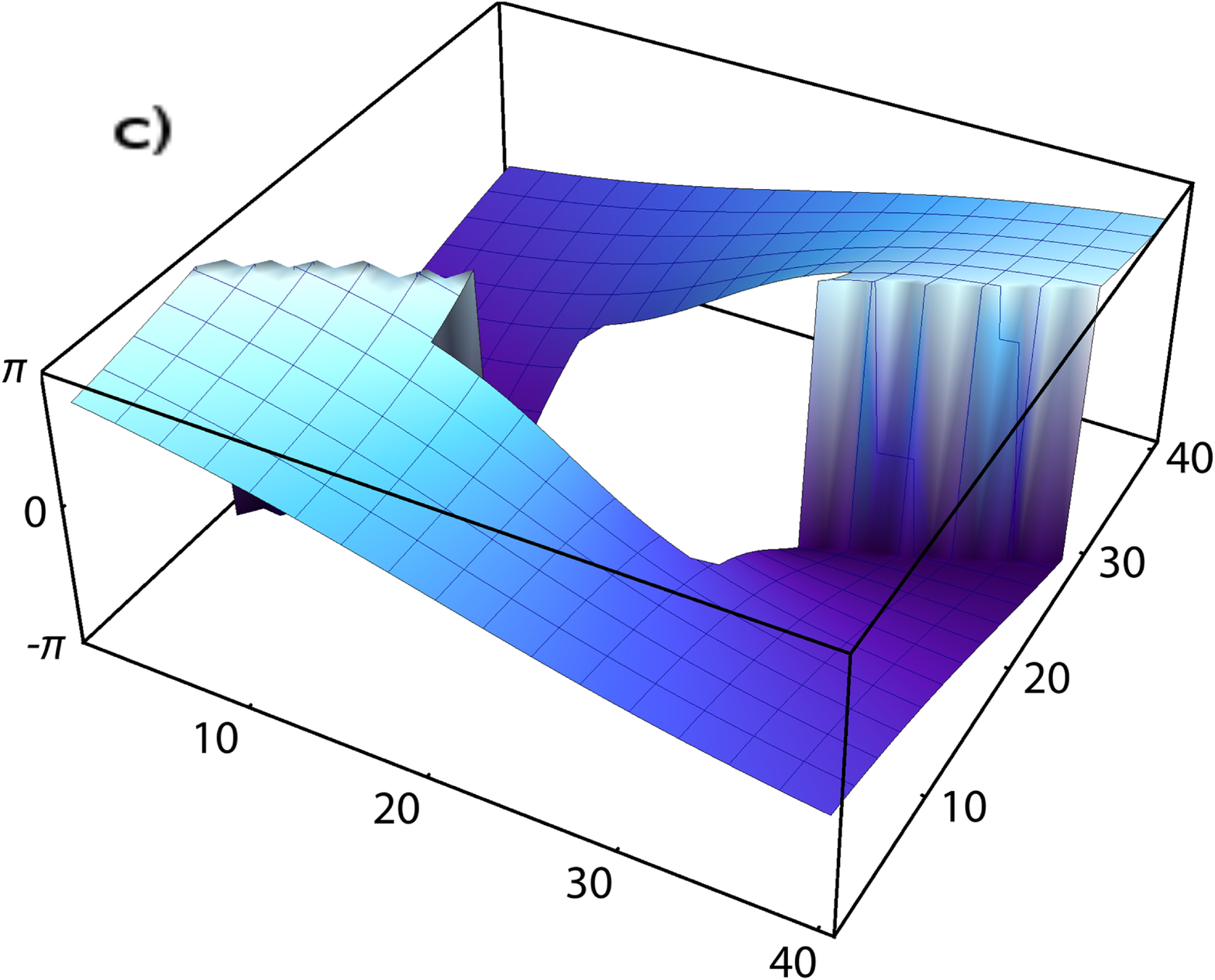}\hspace{.8cm}\includegraphics[width=6.0cm]{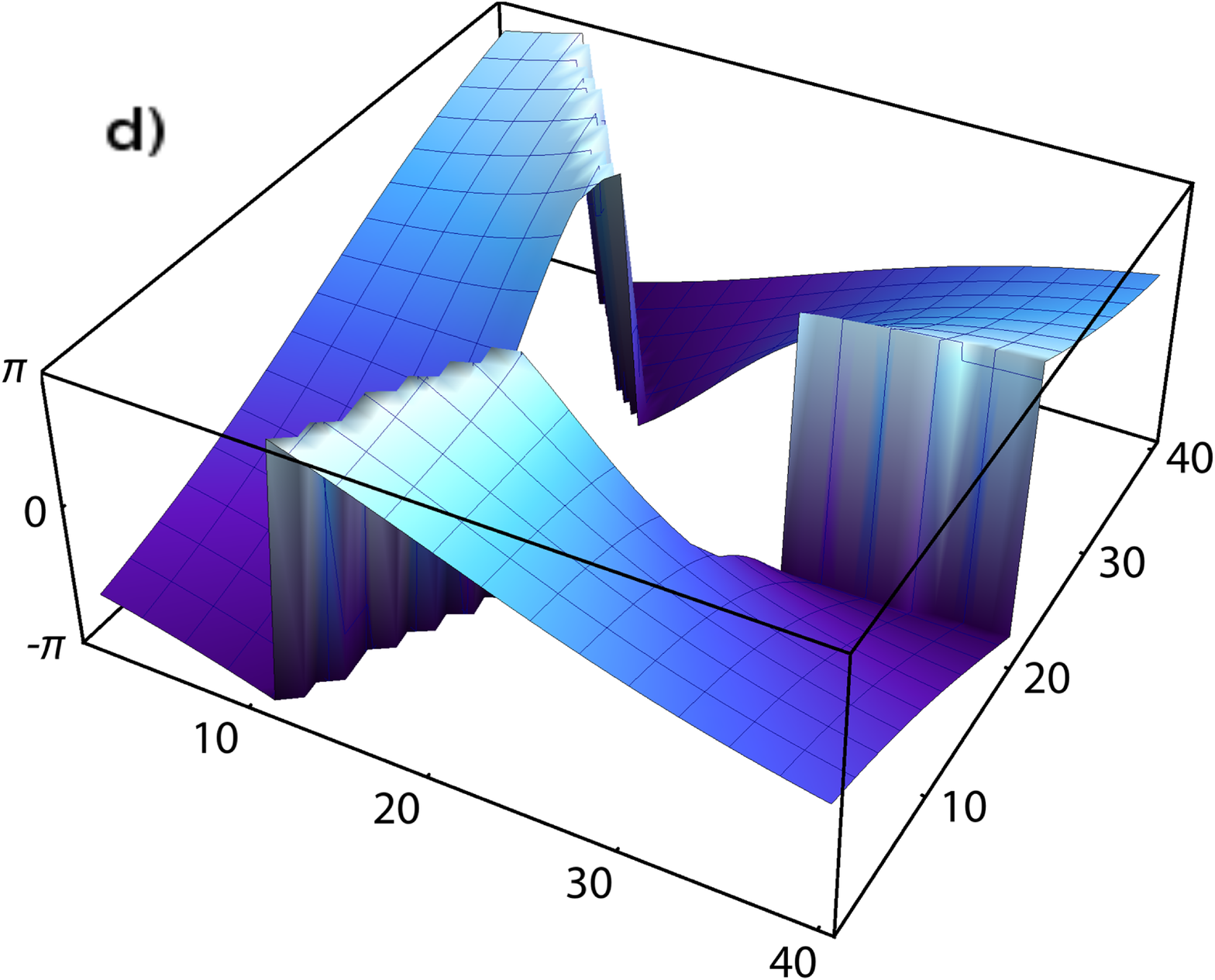}}
\caption{(a) Absolute value of the $d$-wave order parameter $\Delta^d_i$ (in 
units of $t$) in a $40\times40$ square frame with a $14\times14$ hole at the 
center for $q=0$, $\phi=0$, and $V_1=0.3t$. For this interaction strength, the 
suppression of $\Delta^d_i$ for $\phi\neq0$ is small and not visible in this 
plot. The phase of $\Delta^d_i$ is shown for winding numbers $q=1,2,3$ in (b),
(c) and (d), respectively.}
\label{figrs000}
\end{figure}

In the normal state the Bogoliubov -- de Gennes equations reduce to the 
discrete Laplace equation. While the low-energy states do not differ much from 
free plane waves, the higher-energy states near $E_F$ on the square frame 
develop some peculiar, frame-specific features. The wavelength of a state near 
$E_{\rm F}$ is close to two lattice constants, therefore the probability 
density divides into two sublattices. In the square frame, structures on 
different sublattices can overlap, which results in the characteristic 
real-space density profiles which persist in the nodal states of a $d$-wave 
superconductor. Figure~\ref{figrs1} shows two such examples.

The characterization of the superconducting solutions of the Bogoliubov -- de 
Gennes equations in the square frame is analogous to those on the cylinder in
the momentum space analysis. The absolute value of the $d$-wave order parameter
$|\Delta^d_i|$ is shown in Fig.~\ref{figrs000}~(a) for $\phi=0$. The open 
boundary conditions cause a decrease on the boundaries and are responsible for 
Friedel oscillations visible along the diagonal. In multiply connected 
geometries, the Bogoliubov -- de Gennes equations generally allow for solutions
where $\Delta^d_i$ acquires a phase gradient such that the phase difference on 
a closed path around the hole is $2\pi q$ with integer $q$. As in 
Secs.~\ref{secI2} and \ref{sec:1}, this phase winding number $q$ represents the
center-of-mass motion of a Cooper pair, although it cannot be identified with 
the angular momentum in the square geometry. The different numerical solutions 
are obtained by choosing appropriate initial values for the phase of 
$\Delta^d_i$, and the phases of the self-consistent results are shown in 
Figs.~\ref{figrs000}~(b), (c) and~(d) for $q=1$, $2$ and $3$ and flux values 
$\phi=1/2$, $1$ and $3/2$, respectively. 

\begin{figure}[tb]
\center{\includegraphics[width=8.0cm]{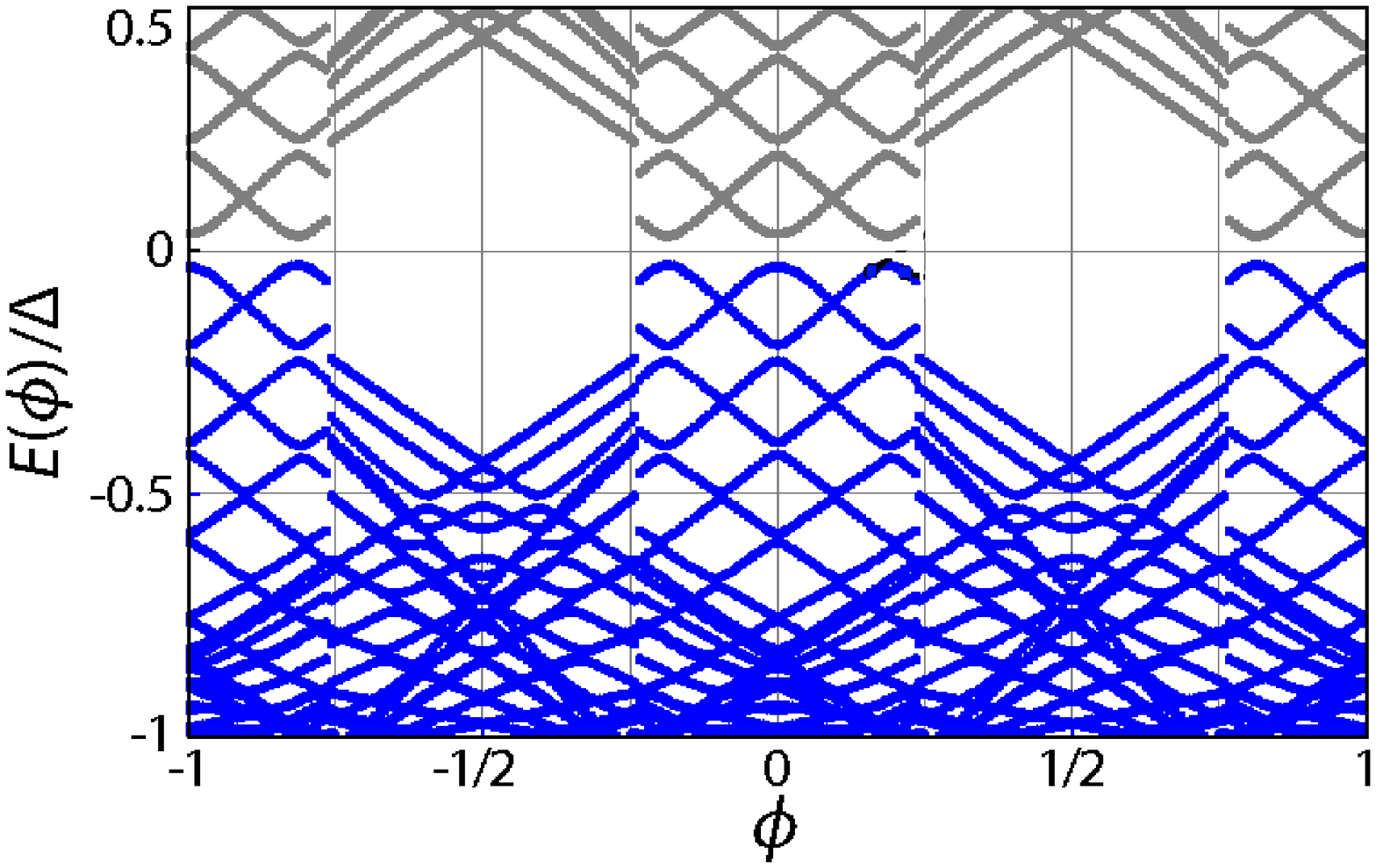}}
\caption{Energy spectrum for a $d$-wave superconductor on a square frame. The 
eigenenergies in the gap region are shown for a square 40$\times$40 loop with a
14$\times$14 hole and pair interaction $V_1=0.3t$ as a function of flux $\phi$
(in units of $\Phi_0$). The energies are given in units of the superconducting 
order parameter $\Delta$ at $\phi=0$ ($\Delta\approx0.22t$). The 
superconducting condensate consists of the states below $E_{\rm F}=0$. 
Reconstruction of the condensate takes place near $\phi=\pm(n+1)/2$, where the 
eigenenergies jump abruptly (after Ref.~\cite{loder:08}.}
\label{figrs2}
\end{figure}

To assess the $E(\phi)$ and the current $J(\phi)$, the evolution of the 
eigenenergies with magnetic flux has to be calculated first. The eigenstates 
with energies below $E_{\rm F}$ form the ground-state condensate 
(Fig.~\ref{figrs2}). Here we discuss only flux values $\phi$ between 0 and 
$1/2$, because all quantities are either symmetric or antisymmetric with 
respect to flux reversal $\phi\rightarrow -\phi$. The spectrum for a square 
frame with $N=40$ and $L=14$ is shown in Fig.~\ref{figrs2} for half filling, 
i.e., $\mu=0$. Because the number of lattice sites on straight paths around the
hole is a multiple of four in a square frame, the spectrum is almost identical 
to the one for a cylinder with an even number of lattice sites and with the 
same number $N-L=26$ of transverse channels (compare to Fig.~6 in 
Ref.~\cite{loder:09}). For the square frame, the energy levels do not actually 
cross $E_{\rm F}$, because the lack of rotational symmetry leads to
hybridization of the levels and level repulsion. Nevertheless, the same clearly
distinct flux regimes are found: the flux intervals between 0 and $1/4$ and
from $1/4$ to $1/2$ (in units of $\Phi_0$).

Up to $\phi\simeq 1/4$ the current $J(\phi)$ generates a magnetic field which 
tends to reduce the applied field by a continuous shift of the eigenenergies in
the condensate. At $\phi=0$, pairs of states with opposite circulation 
compensate their respective currents, thus $J=0$. The well separated states at 
$\phi=0$ in Fig.~\ref{figrs2} are the states in the vicinity of the nodes of 
the $d$-wave superconductor. Away from $E_{\rm F}$, the density of states 
increases towards the states near the maximum energy gap $\Delta$ that provide 
most of the condensation energy. For $\phi>0$, the energy of the states with 
orbital magnetic moment anti-parallel (parallel) to the magnetic field is 
increased (decreased). Correspondingly the supercurrent, which is carried by 
these states, depends on the details of level crossings and avoidings. The main
contribution to the supercurrent arises from the occupied levels closest to 
$E_{\rm F}$, because the contributions from the lower-lying states tend to 
cancel in adjacent pairs.

\begin{figure}[t]
\center{\includegraphics[width=6.0cm]{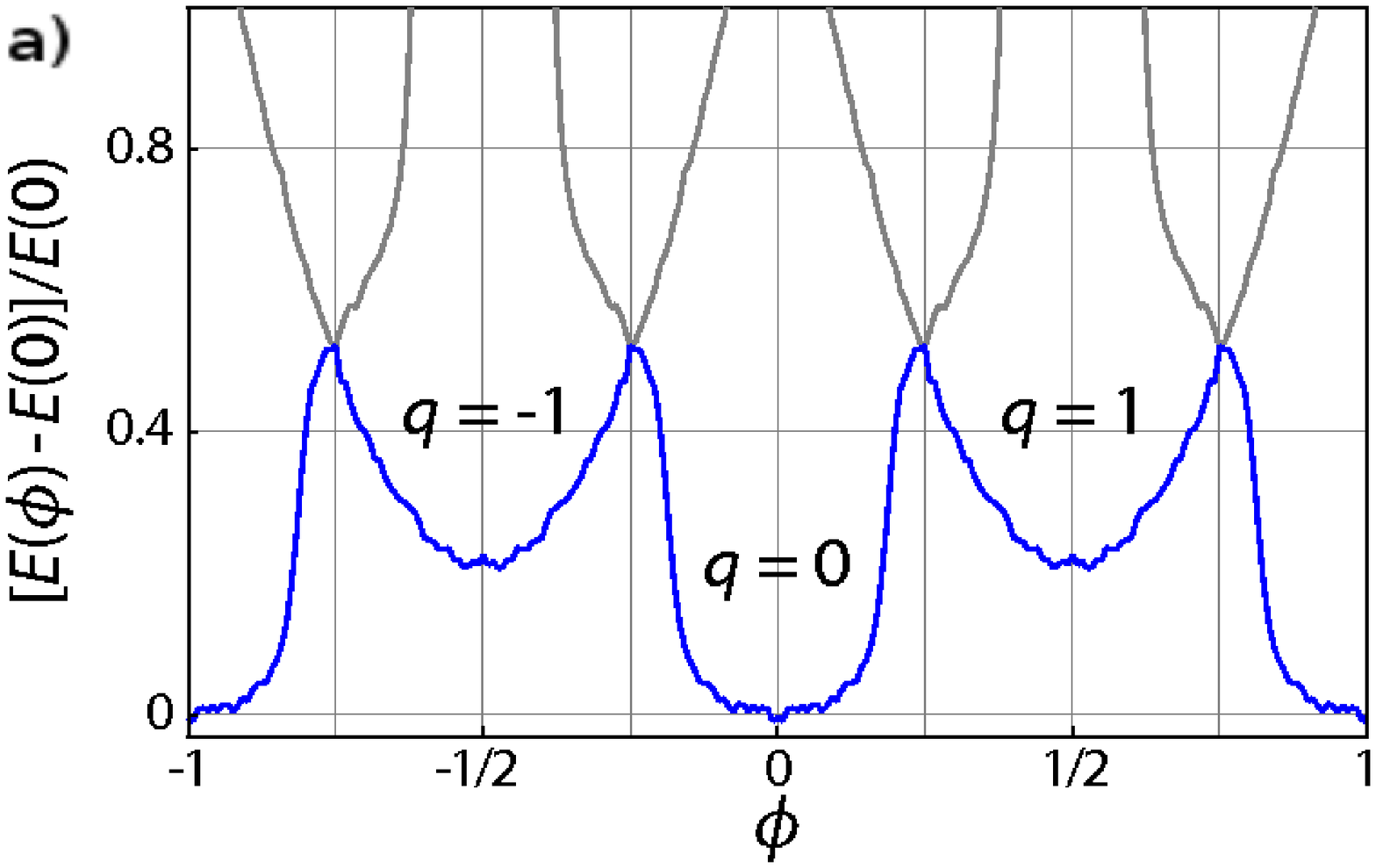}\hspace{.8cm}\includegraphics[width=6.0cm]{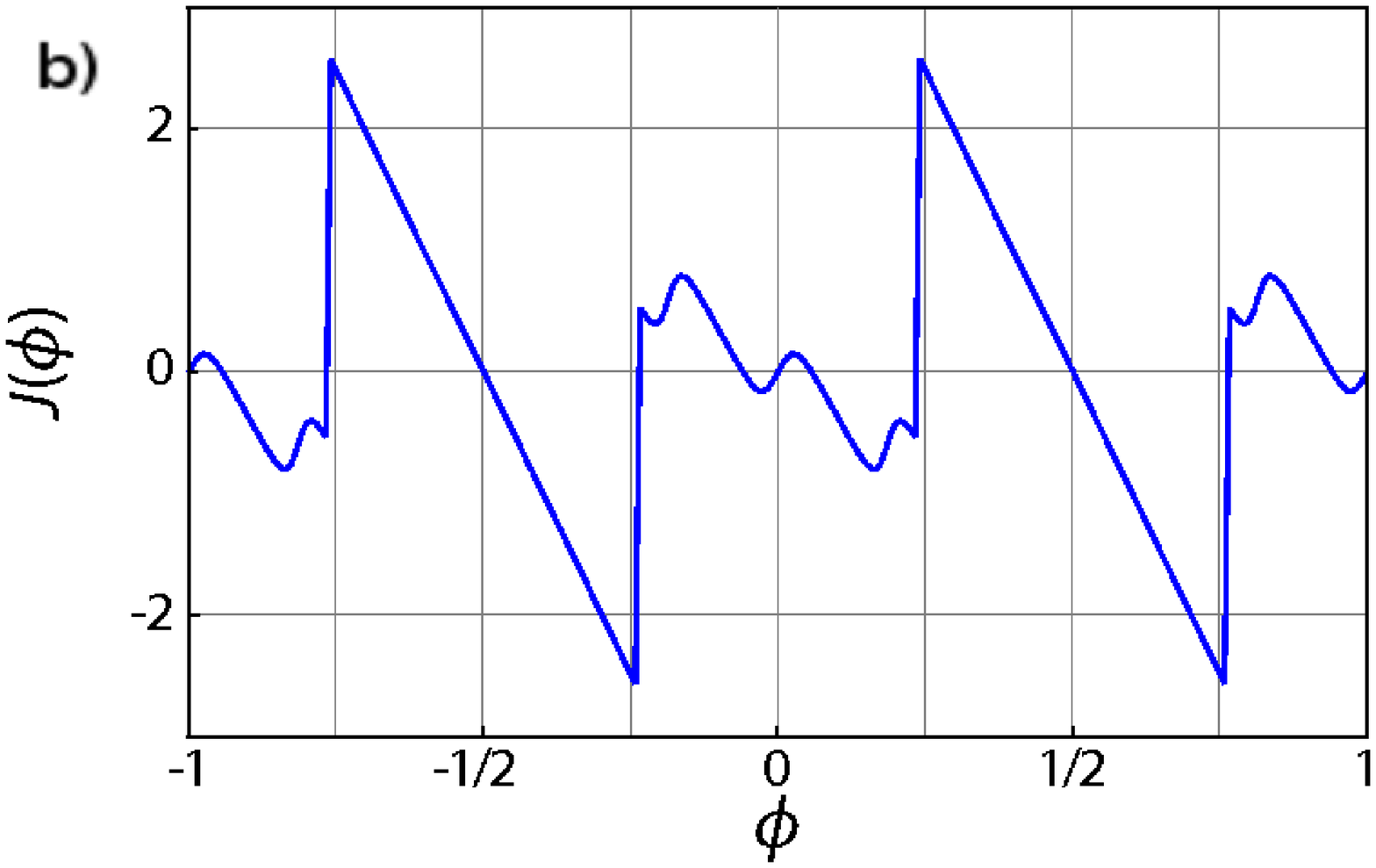}}
\caption{Flux dependence of energy and current for the square frame. Energy 
$[E(\phi)-E(0)]/E(0)$ (a) and circulating current $J(\phi)$ (b) for a square 
40$\times$40 loop with a 14$\times$14 hole and pair interaction $V_1=0.3t$. 
$J(\phi)$ is given in units of $t/\Phi_0=6\times10^{-5}{\rm A}$ for the choice 
of $t=250$~meV. The condensate states with even and odd winding number $q$ are 
clearly distinct, which is reflected, e.g., in the deformation of the 
$q=0$-parabola. The overall $\phi$ periodicity for $E(\phi)$ and $J(\phi)$ is 
$\Phi_0$ (from Ref.~\cite{loder:08}).}
\label{figrs3}
\end{figure}

As the highest occupied state shifts with increasing flux to lower energies,
the current in the square loop first increases for small $\phi$ 
(Fig.~\ref{figrs3}), then decreases when the highest occupied level with an 
orbital moment opposite to the applied magnetic field starts to dominate. With 
increasing flux this state approaches $E_{\rm F}$. A current-carrying state in 
the vicinity of the nodes is replaced upon a slight increase of $\phi$ by a 
state of opposite current direction. The states of the condensate are thereby 
continuously changing near the extrapolated crossing points. As a consequence,
the energy \textquotedblleft parabola'' centered at zero flux is different from
the ground-state energy parabola centered at $\phi=1/2$ 
[Fig.~\ref{figrs3}~(a)]. The deviation from a parabolic shape near zero flux is
due to the evolution of the near-nodal states; the vertical offset of the 
energy minima at $\phi=n$ results mostly from the flux dependence of the states
near the maximum value of the anisotropic gap.

For flux values near $\phi=1/4$ the condensate reconstructs. The 
superconducting state beyond $1/4$ belongs to the class of wave functions 
introduced by Byers and Yang~\cite{Byers} in which, for a circular geometry, 
each pair acquires a center-of-mass angular momentum $\hbar$~\cite{schrieffer}.
Remarkably, in the flux interval from near $1/4$ to $1/2$, a full energy gap 
exists also for $d$-wave superconductors (Fig.~\ref{figrs2}). Here the 
circulating current enhances the magnetic field; the paramagnetic orbital 
moment of the current is parallel to the field. The resulting energy gain is 
responsible for the field-induced energy gap. This reconstruction of the 
condensate is the origin of the $\Phi_0$ periodicity in energy and current.

These calculations show that a $d$-wave superconducting loop in a square 
geometry has almost identical properties to a flux threaded cylinder. This is 
remarkable, because on a closed path in the square frame, the phase of the 
$d$-wave order parameter $\Delta^d_i$ rotates by $2\pi$, whereas in the 
cylinder, the order parameter rotates with the lattice. Therefore, while 
changes in the geometry and the number of transverse channels modify the 
spectrum and the $J(\phi)$ characteristics in detail, they do not eliminate the
$\Phi_0$ periodic component. The reduction of the symmetry, here to the 
four-fold rotational symmetry of the square frame, stabilizes the spectrum 
compared to the cylinder geometry.

\section{Flux periodicity of Josephson junctions}
\label{sec:junction}
All energy levels are Doppler shifted in current carrying systems, not only in 
flux threaded loops but also in wires or at the surface of bulk 
superconductors. In the latter systems, the phase gradient of the 
superconducting order parameter typically does not reach the value necessary to
drive the superconductor into a finite-momentum pairing state with $q\neq0$, 
which is why the influence of finite momentum pairing on the flux periodicity 
has not been discussed in the literature until recently. An exception are 
systems with strong inhomogeneities of the order parameter, which act as 
Josephon junctions. The phase gradient accumulates at the junctions and they 
behave periodically with the phase gradient, as described by the Josephson 
relation. From what has been discussed for the flux periodicity in multiply 
connected geometries, it appears natural that the Doppler shift of nodal states
might also influence the periodicity of Josephson junctions.

A Josephson junction is intrinsically a more complicated system than a  
superconducting loop. Several parameters are needed to characterize the 
junction as well as the superconducting states on each side of the junction. 
Most junctions can be classified either as transparent or as tunnel junctions, 
regardless whether they consist of a geometrical constriction, a potential 
barrier, or a normal metal bridge. This classification is closely related to 
the Doppler shift of single energy levels in the system, as will be explained 
below. In the following we will therefore discuss the Josephson relations in 
both the tunneling and the transparent regimes.

\subsection{Current-phase relation}
\label{secJ1}
The current-phase relation, which expresses the supercurrent $J$ over a 
Josephson junction as a function of the phase difference $\delta\varphi$ of the
order parameters on both sides of the junction is:
\begin{equation}
J=J_{\rm c}\sin(\delta\varphi) .
\label{j15I}
\end{equation}
$J_{\rm c}$ is the critical current over the junction, above which the zero 
voltage state breaks down. This relation was predicted by Josephson in 
1962~\cite{josephson} and can be directly derived from a Ginzburg-Landau 
description~\cite{tinkham}. For transparent junctions, $\sin(\delta\varphi)$ in
Eq.~(\ref{j15I}) distorts into a saw-tooth pattern similar to the current-flux 
relation in superconducting loops~\cite{golubov:04}. It is crucial to realize 
that the phase gradient of the order parameter is twice that of the 
superconducting wave function. If the phase difference of the order parameter 
on both sides of the junction is $\delta\varphi$, then the phase difference of 
the wave function is $\delta\varphi/2$. Because the wave function of the system
has to be $2\pi$-periodic, the periodicity of the energy spectrum and the order
parameter of a finite system is $4\pi$. The current contributions from all 
energy levels add up to a $2\pi$ periodic supercurrent only in the 
thermodynamic limit. In this section we analyze whether the Doppler shift of 
the energy levels leads to the same doubling of the periodicity in 
$\delta\varphi$ of a junction as it does for the flux periodicity of loops. 
While for the tunneling regime we rely on a simple linear-junction model, we 
will analyze transparent junctions by inserting a Josephson junction into a 
square frame. This has the advantage of a remarkable stability of the energy 
spectrum against the insertion of impurities and lattice defects, as will be 
seen in Sec.~\ref{secJ1.2}.

\begin{figure}[tb]
\center{\includegraphics[width=5.0cm]{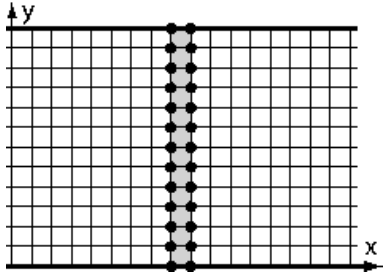}}
\caption{To model a Josephson junction we choose a discrete square lattice with
$N$ sites in $x$-direction, and $M$ sites in $y$-direction. The junction itself
is modeled by one or two lines of potential scatterers (black points) each with
a repulsive potential $U>0$.}
\label{Figj1}
\end{figure}

\subsubsection{Tunnel junctions}
\label{secJ1.1}

A simple model of a tunnel junction is a square lattice with $N$ sites in 
$x$-direction and $M$ sites and periodic boundary conditions in $y$-direction. 
The junction is modeled in the tunneling regime by one or two lines of 
potential scatterers with a repulsive potential $U>4t$ (Fig.~\ref{Figj1}). In 
the absence of a magnetic field, this system is homogeneous in $y$-direction, 
and the Fourier transformation with respect to the $y$-coordinate will allow 
the diagonalization of larger systems~\cite{andersen}.

\begin{figure}[b]
\center{\includegraphics[width=6.0cm]{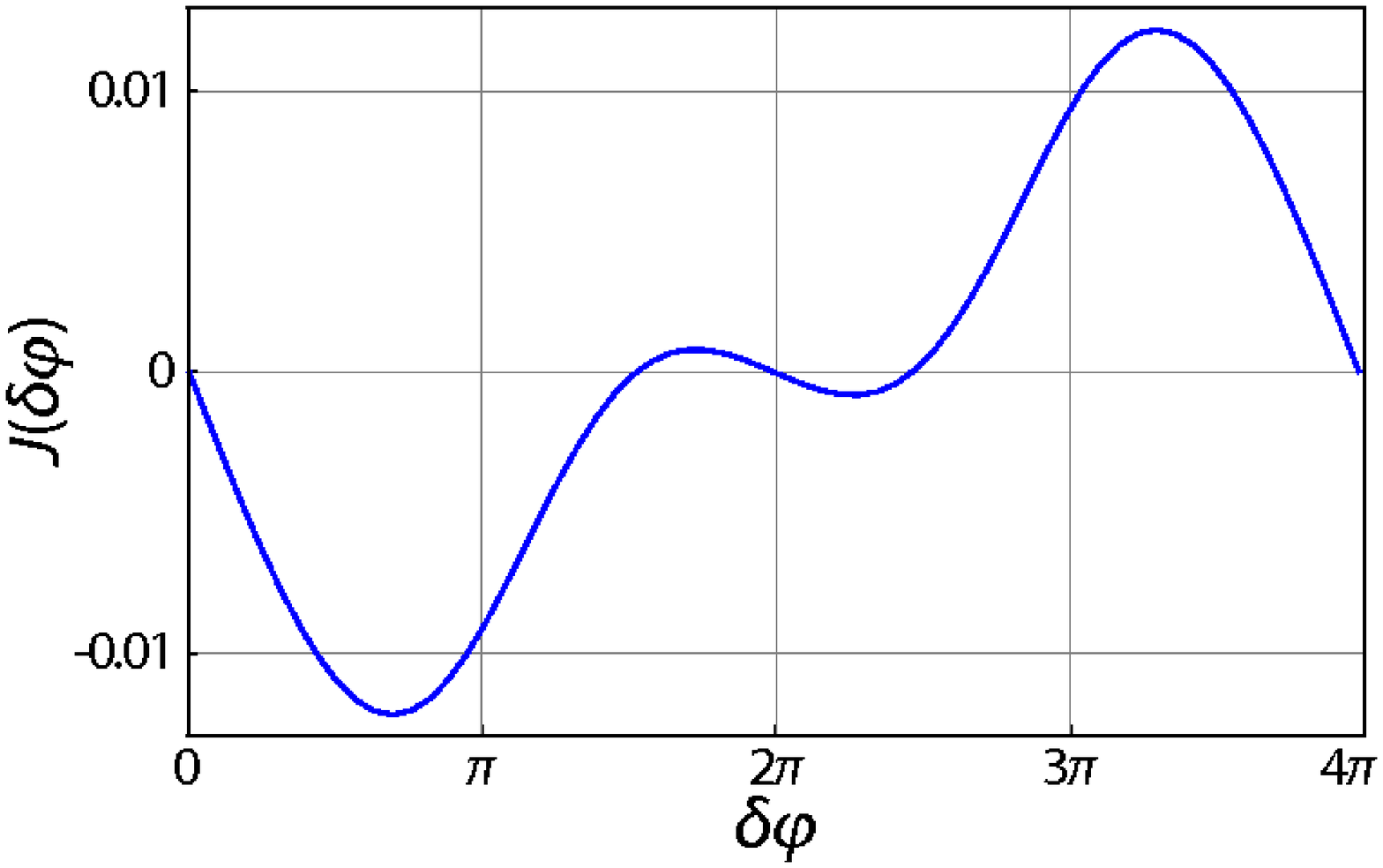}\hspace{.8cm}\includegraphics[width=6.0cm]{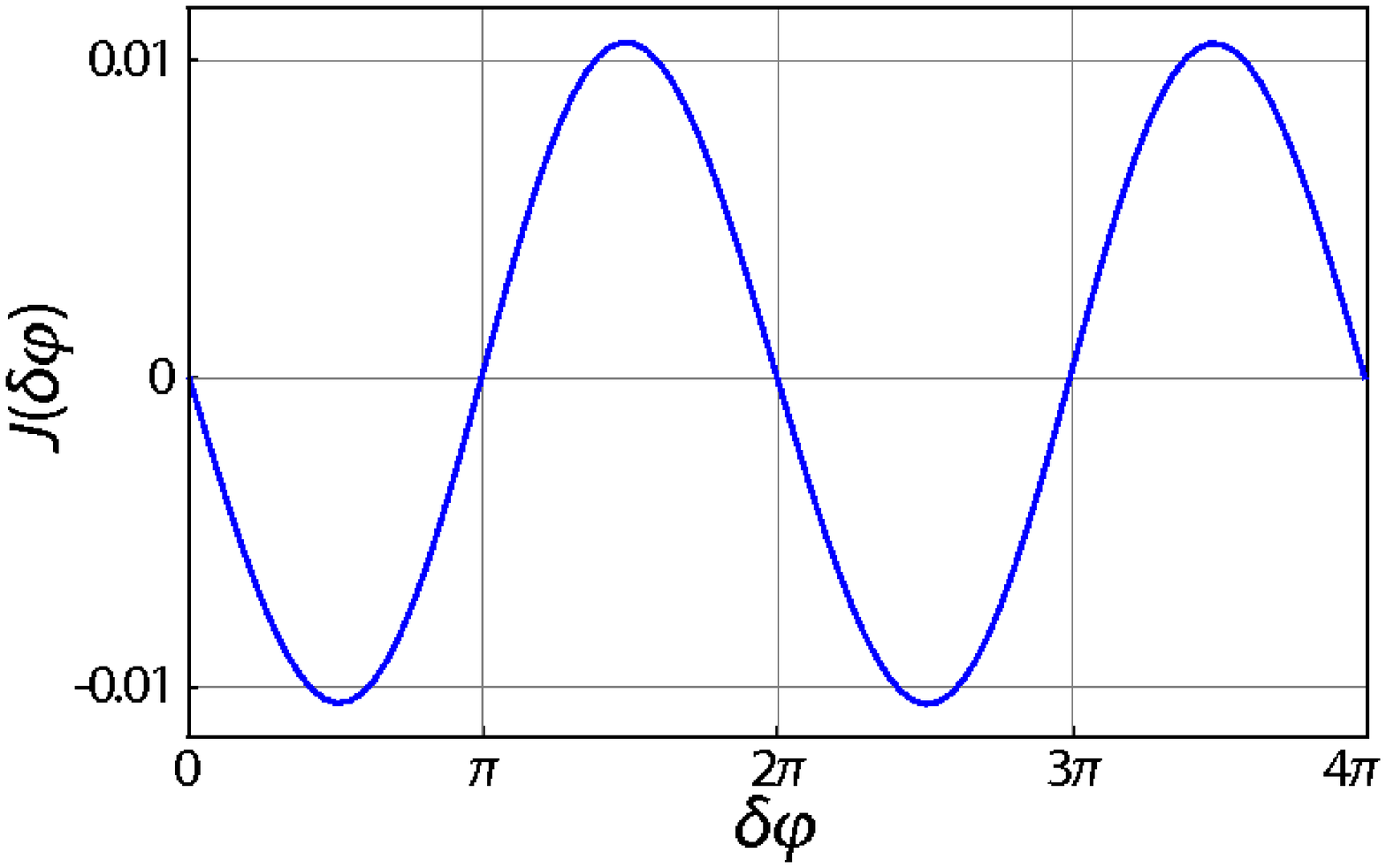}}
\caption{Current-phase relation calculated for Josephson junctions in the 
tunneling regime. Left panel: $N=18$, $M=12$, $V_1=0.3t$, and $U=4.5t$. The 
$\sin(\delta\varphi)$ relation is considerably deformed, which is typical for 
narrow junctions with very few channels. Right panel: $N=20$, $M=200$, $V_1=t$,
and $U=5t$. This junction has a sufficiently many channels to exhibit the known
 current-phase relation. The overall sign depends on the choice of the sign in 
the phase factor of the hopping matrix elements.}
\label{figj3}
\end{figure}

The Bogoliubov -- de Gennes equations are slightly modified in this case: the 
eigenvalue equation~(\ref{j5}) for nearest-neighbor interaction is now defined 
through the relations
\begin{equation}
\hat tu_{n_{i_xk_y}}=\sum_{j_x}t_{ij}u_{n_{j_xk_y}}+(\epsilon_{k_y}+U_{i_x})
u_{n_{i_xk_y}}\hskip0.3cm , \hskip0.3cm
\label{j9b}\\
\hat\Delta v_{n_{i_xk_y}}=\sum_{j_x}\Delta_{ij}v_{n_{j_x,k_y}}+\Delta_{k_y}
v_{n_{i_xk_y}},
\label{j10}
\end{equation}
where $\epsilon_{k_y}=-2t\cos(k_y)-\mu$ and $\Delta_{k_y}=\Delta_y\cos(k_y)$. 
The indices of the eigenvectors ${\bf u}_n$ and ${\bf v}_n$ are the 
$x$-coordinate of the site and the wave number $k_y$ in $y$-direction. The 
corresponding self-consistency equations are
\begin{equation}
\Delta_{ij}=\frac{V_1}{2}\sum_{n,k_y}\left[u_{n_{i_xk_y}}v^*_{n_{j_xk_y}}+
u_{n_{j_xk_y}}v^*_{n_{i_xk_y}}\right]\tanh\left(\frac{E_n(k_y)}{2k_{\rm B}T}
\right) ,
\label{j13}
\end{equation}
if $j=i\pm\hat x$, and if the bonds are along the $y$ direction
\begin{equation}
\Delta_y=V_1\sum_{n,k_y}u_{n_{i_xk_y}}v^*_{n_{i_xk_y}}\cos(k_y)\tanh
\left(\frac{E_n(k_y)}{2k_{\rm B}T}\right).
\label{j14I}
\end{equation}
The self-consistency equation for the $s$-wave order parameter $\Delta_i$ with 
on-site interaction is analogous to~(\ref{j14I}), but without the factor 
$\cos(k_y)$.

To induce a finite phase gradient of the order parameter and a supercurrent, we
introduce a ``phase jump'' $\delta\varphi$ in the matrix elements $t_{ij}$ for 
hopping from $i_x=N-1$ back to $i_x=0$, and a jump $-\delta\varphi$ for the 
corresponding hopping in the opposite direction. An alternative, but physically
 equivalent choice for the phase of $t_{ij}$ is a constant phase factor 
$e^{i\varphi_{ij}}$ with $\varphi_{ij}=\delta\varphi/N$ for all hopping 
processes along the $x$-direction, which is mathematically identical to a 
cylinder threaded by a flux $\Phi=(hc/e)\,\phi$ with $2\pi\phi=\delta
\varphi/2$. In the fully transparent case with $U=0$, this leads to a 
homogeneous phase gradient of $\Delta^d_i=(\Delta_{i,i+\hat x}+\Delta_{i,i-\hat
 x})/2+\Delta_y$ (or $\Delta_i$, respectively), whereas far in the tunneling 
regime for $U>4t$, the phase of the order parameter drops only across the 
junction. The current across the junction is calculated as in Eq.~(\ref{J711}).
The results for two typical situations are shown in Fig.~\ref{figj3}. The left 
panel displays the current-phase relation of a narrow Josephson junction with a
width of $M=12$ sites. The usual current-phase relation is considerably 
deformed in this case, as is typical for junctions with very few 
channels~\cite{golubov:04}. The exact form of the current-phase relation is 
characteristic for each junction; it depends on the structure of the energy 
spectrum, which changes strongly upon increasing or decreasing the system size 
or adding impurities. For increasing $M$, the current-phase relation 
approaches~(\ref{j15I}), as the level spacing becomes negligible. This is the 
regime of wide junctions, shown in Fig.~\ref{figj3}~(right panel), which is 
well described by the Ginzburg-Landau approach.

\begin{figure}[tb]
\center{\includegraphics[width=7.0cm]{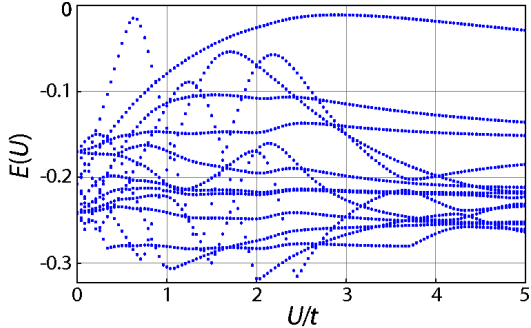}}
\caption{The top 14 energy levels below $E_F$ calculated for a linear Josephson
juntion with $N=18$, $M=12$ and $V_1=0.6t$ as a function of the repulsive 
potential $U$ on the junction.}
\label{figj3.1}
\end{figure}

Our numerical analysis shows that the Josephson relation~(\ref{j15I}) describes
wide junctions in the tunneling regime very well; a doubling of the period is 
not observed, even for $d$-wave superconductors with small antinodal energy 
gap. The reason for this is twofold: (1) Along with the suppression of the 
critical current $J_{\rm c}$ across the junction, the Doppler shift decreases 
strongly with increasing repulsive potential $U$. In the tunneling regime 
($U>4t$), $J_{\rm c}$ decreases by a factor $> 10^3$. Consequently no energy 
levels (or negligibly few in very large systems) approach $E_{\rm F}$ as a 
function of $\delta\varphi$ and the effects related to a reversal of single 
particle currents are absent. (2) For tunnel junctions, the thermodynamic 
Ginzburg-Landau limit is reached also for $d$-wave superconductors, if the 
density of states close to $E_{\rm F}$ becomes quasi-continuous, in contrast to
the flux threaded loop. The deformation of the current-phase relation in narrow
tunnel junctions is generically not due to levels reaching $E_{\rm F}$. The 
deformation is induced, if the total current is carried by very few states, 
each with a period of $4\pi$. The $2\pi$-asymmetric terms do not cancel and the
critical current is $4\pi$-periodic.

\subsubsection{Transparent junctions}
\label{secJ1.2}
Transparent junctions are more involved than tunnel junctions. One reason for 
their complexity is the strong coupling of the superconducting states on both 
sides of the junction,which does not allow to choose the phases of the 
corresponding order parameters independently. Consequently the phase difference
$\delta\varphi$ is not an adequate variable for describing the current across 
the junction. Another reason is that the energy spectrum in a linear junction 
of the type shown in Fig.~\ref{Figj1} changes strongly upon changing 
microscopic details of the system, such as the strength of the repulsive 
potential $U$ on the impurity sites in this case. This is illustrated vividly 
by Fig.~\ref{figj3.1} showing the evolution of the highest occupied energy 
levels with increasing $U$.

\begin{figure}[b]
\center{\includegraphics[width=5.0cm]{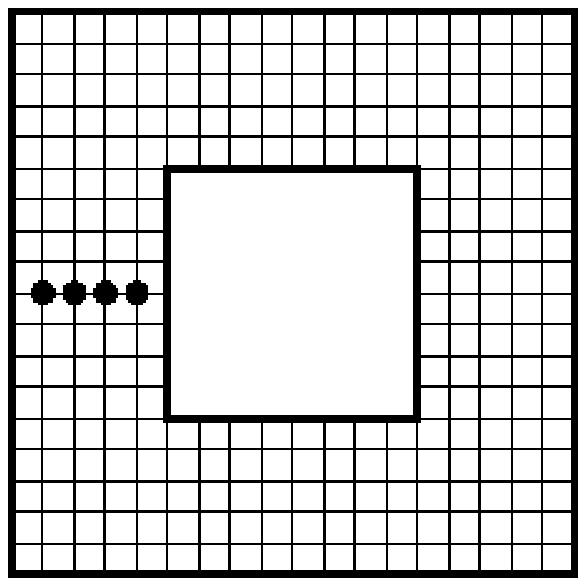}}
\caption{For the description of transparent junctions, we choose the 
square-frame geometry and model the junction with potential scatterers arranged
on a line crossing one side on the frame (black points). The current is driven 
by a magnetic flux threading the frame.}
\label{figj3.2}
\end{figure}

These problems can be resolved by using a square-frame geometry as in 
Sec.~\ref{sec:square}. Here the Josephson junction is modeled by adding 
potential scatterers on a line as shown in Fig.~\ref{figj3.2}, and the current 
is driven by a magnetic flux $\phi$ threading the frame. For a tunnel junction,
this would induce a phase jump of $4\pi\phi$ in the order parameter aross the 
junction and thus a $\sin(4\pi\phi)$ current-flux relation. In transparent 
junctions, the jump is smaller and vanishes in a clean frame. For this topology
 the magnetic flux $\Phi=\phi\cdot hc/e$ is related to the phase variation of 
the order parameter $\delta\varphi$ by $2\pi\phi=\delta\varphi/2$.

\begin{figure}[tb]
\center{\includegraphics[width=7.0cm]{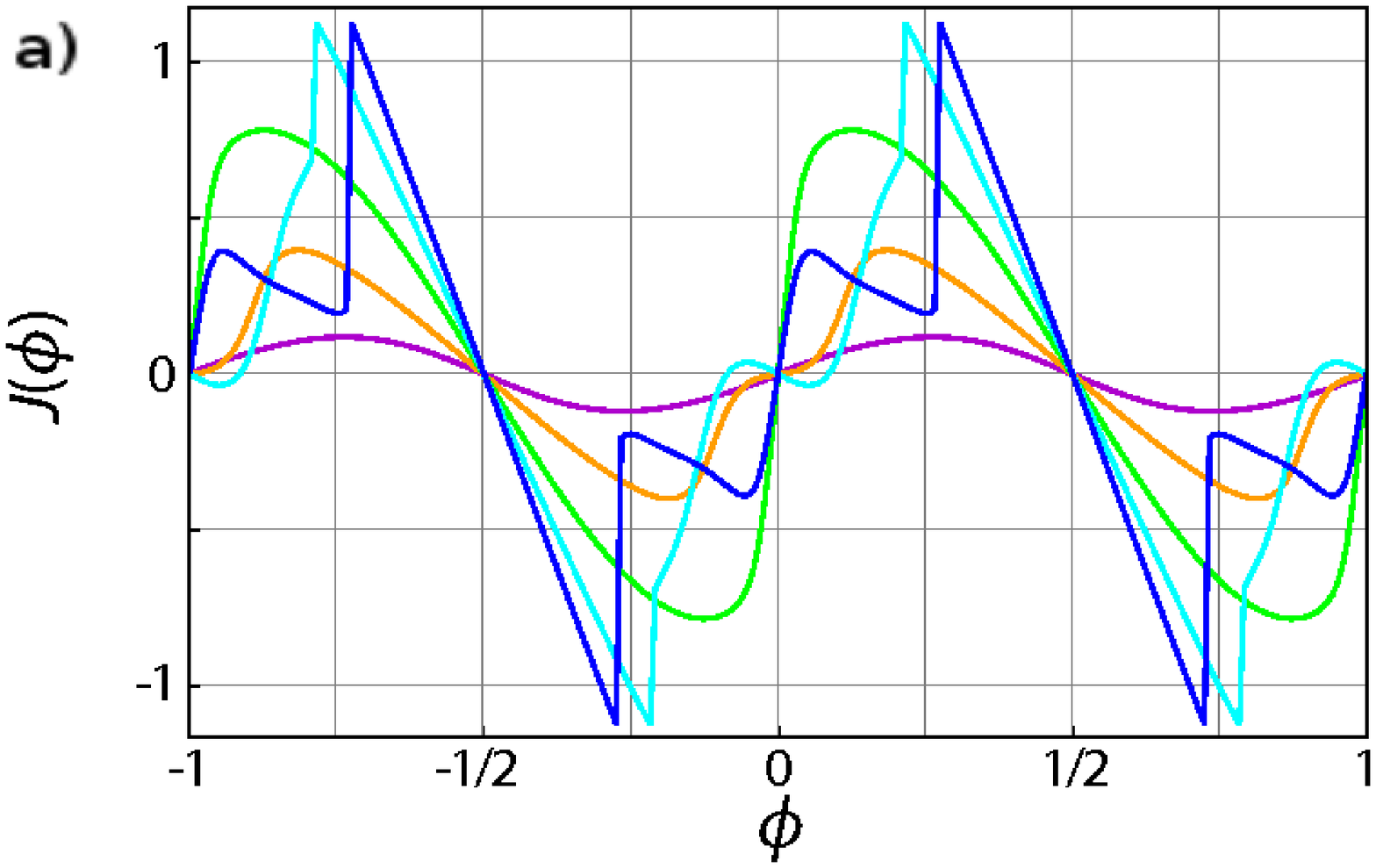}}
\vspace{\FigSpace}
\center{\includegraphics[width=6.0cm]{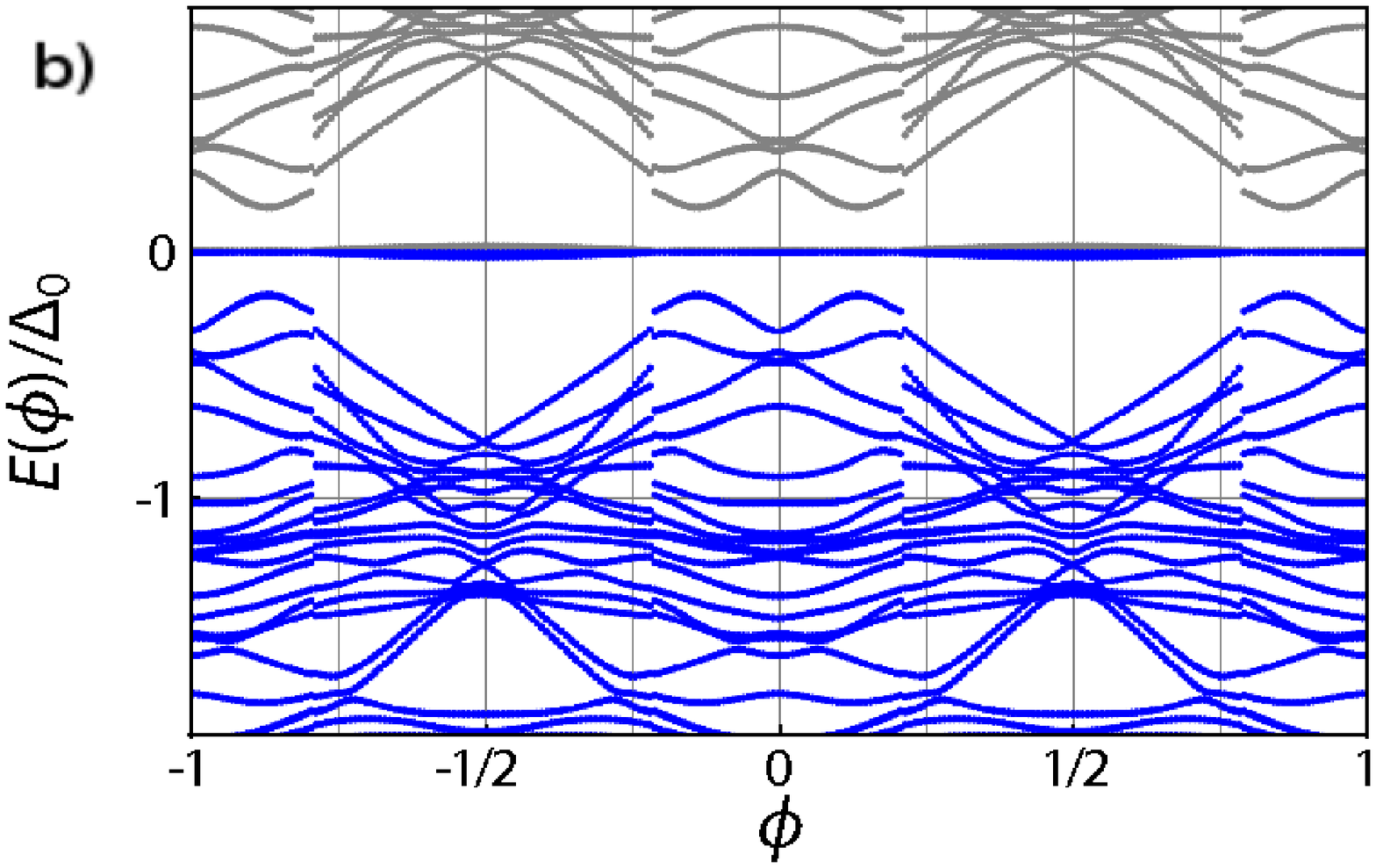}\hspace{.8cm}\includegraphics[width=6.0cm]{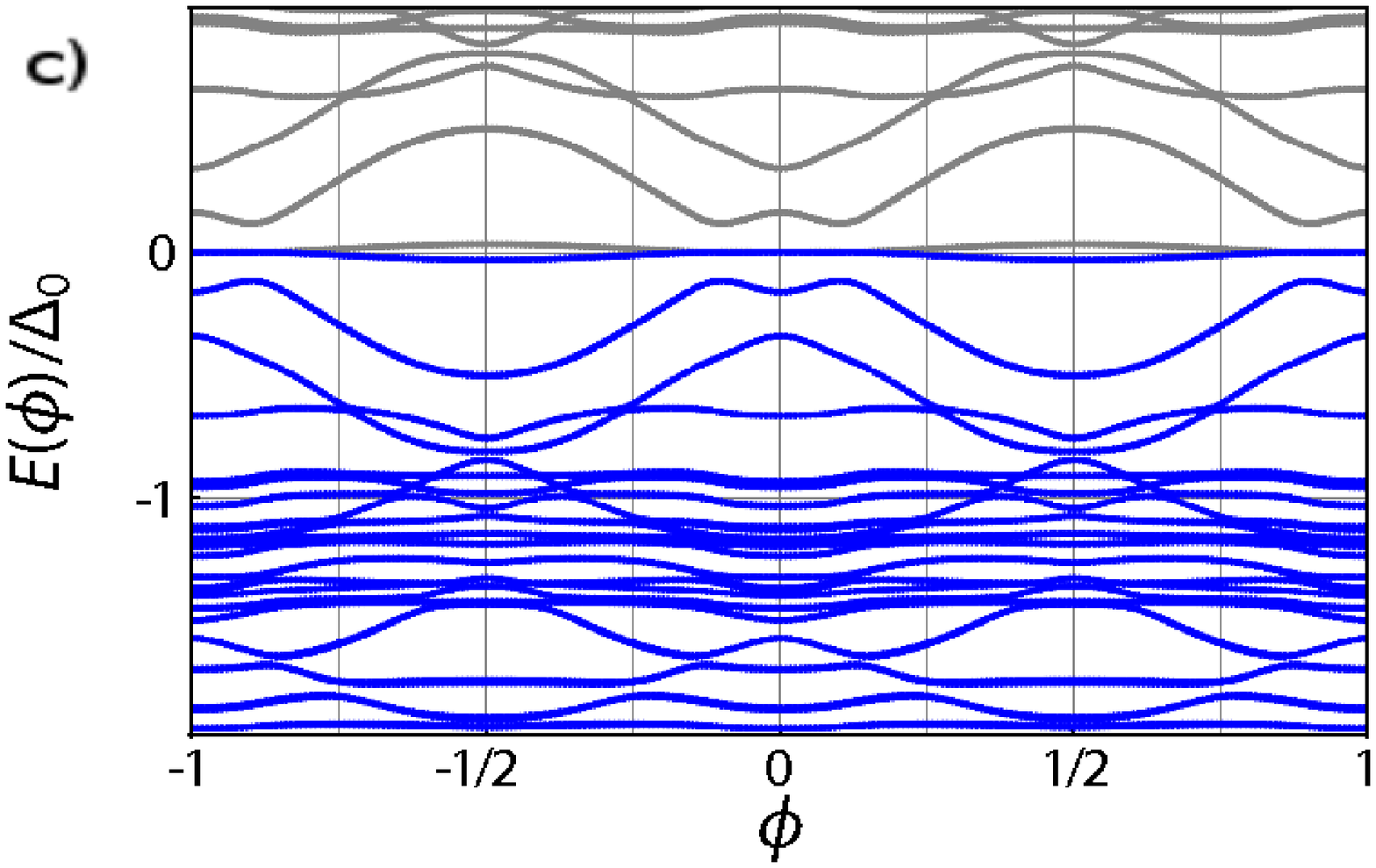}}
\caption{Supercurrent $J(\phi)$ and energy spectrum $E(\phi)$ of a $20\times20$
square frame with an $8\times8$ hole containing a Josephson junction. The width
of the arms for the frame is six sites. The impurity potential is $U=100t$. (a)
$J(\phi)$ for one (blue), two (turquoise), three (green), four (orange), and 
five (purple) impurity sites. Energy spectrum for (b) two and (c) four impurity
sites.}
\label{figj2.3}
\end{figure}

For sufficiently large $U$, say $U=100t$, these impurities act as a geometrical
constriction. Figure~\ref{figj2.3} shows explicitly that the spectrum of a 
square frame remains qualitatively invariant upon inserting a small number of 
impurities, even sufficiently strong to block the current over the impurity 
site completely. Figures~\ref{figj2.3} (b) and (c) show the spectra versus 
$\phi$ for two and four impurities for a $20\times20$ square frame with a 
$8\times8$ square hole. In the presence of impurities, bound states arise at 
$E_{\rm F}$ in a $d$-wave superconductor~\cite{zhu:00,franz:96}, which are
nearly flux independent. These bound states are easily identified in 
Figs.~\ref{figj2.3}~(a) and (b) near the Fermi energy. Otherwise, the spectrum 
in Fig.~\ref{figj2.3}~(b) is very similar to that of the clean frame discussed 
in Sec.~\ref{sec:square} (Fig.~\ref{figrs2}). Clearly visible is the 
discontinuity of the spectrum where the condensate reconstructs to a 
superconducting state with different winding number $q$. The relevance of $q$ 
is a characteristic property of transparency and directly connected to a 
discontinuity of the supercurrent (see Fig.~\ref{figj2.3}~(a) for one and two 
impurities).

For three to five impurities, the supercurrent is continuous, as is the 
spectrum shown in Fig.~\ref{figj2.3}~(c) for four impurities. Nevertheless, the
typical features of the square-frame spectrum are still present, in particular 
the gap in the odd flux regimes and one energy level approaching $E_{\rm F}$ in
the even $q$ regime. This level causes the wiggle in the supercurrent around 
$\phi=0$; its slope and that of a few others remain almost as steep as in a 
clean frame, which indicates the existence of channels with free current flow. 
The Doppler shift of nodal states is therefore not negligible in the 
calculation of the supercurrent across transparent Josephson junctions, and it 
may cause appreciable deviations  from the $\sin(4\pi\phi)$ current-flux 
relation even in the case of wide junctions. The $\sin(4\pi\phi)$ is expected 
for the thermodynamic Ginzburg-Landau limit.

Finally we note that for five impurities, only one channel through the junction
remains, which is almost blocked by the bound state. Thus the spectrum becomes 
nearly flux independent, leading to a junction in the tunneling regime. 
However, the supercurrent does not follow the expected $\sin(4\pi\phi)$ but 
rather a $\sin(2\pi\phi)$ current-phase relation. This is due to the 
point-contact like character of the junction and the extreme limit of the 
deformation of the current-flux relation as shown in Fig.~\ref{figj3} -- 
similar to the left panel, however with $J(\delta\varphi)\sim-\sin(\delta
\varphi/2)$.

\subsection{Field-threaded junctions}
\label{secJ2}

A magnetic field threading a Josephson junction modifies the phase difference 
of the order parameters of the superconductors on both sides and thus alters 
the supercurrent. This behavior is well understood on the basis of the
Ginzburg-Landau approach. The current-flux relation of a linear junction that 
is homogeneous in $y$-direction has the shape of a Fraunhofer diffraction 
pattern~\cite{tinkham}, although it deviates from the Fraunhofer form for all 
other junction geometries. Despite these deviations it preserves the 
characteristic flux periodicity of $\Phi_0/2$ for conventional Josephson 
junctions. The magnetic field dependent critical current of Josephson junctions
is therefore another key property where the Doppler shift might cause a 
doubling of the flux period.

Here we use again the linear junction model of Sec.~\ref{secJ1.1} and fix the 
phase difference to $\delta\varphi=\pi/2$, for which the absolute value of the 
current across the junction in the tunneling regime is largest. In order to 
introduce a magnetic field threading the junction, we construct the junction 
from single plaquettes with potential scatterers on each of its sites. All 
plaquette $l$ which belong to the junction are threaded by a magnetic flux 
$\phi^l$, generating Peierls phase factors $\varphi^l_{ij}$. We restrict our 
discussion to a homogeneous field distribution inside the junction, $\phi^l=
\phi$ for all $l$, and the repulsive potential on the respective sites is $U$.
In the presence of a magnetic field, the system is not homogeneous in 
$y$-direction, and we have to diagonalize it in real space. This restricts the 
maximum system size for our analysis.

\subsubsection{Current-flux relation of tunnel junctions}
\label{secJ2.1}

\begin{figure}[tb]
\center{\includegraphics[width=6.0cm]{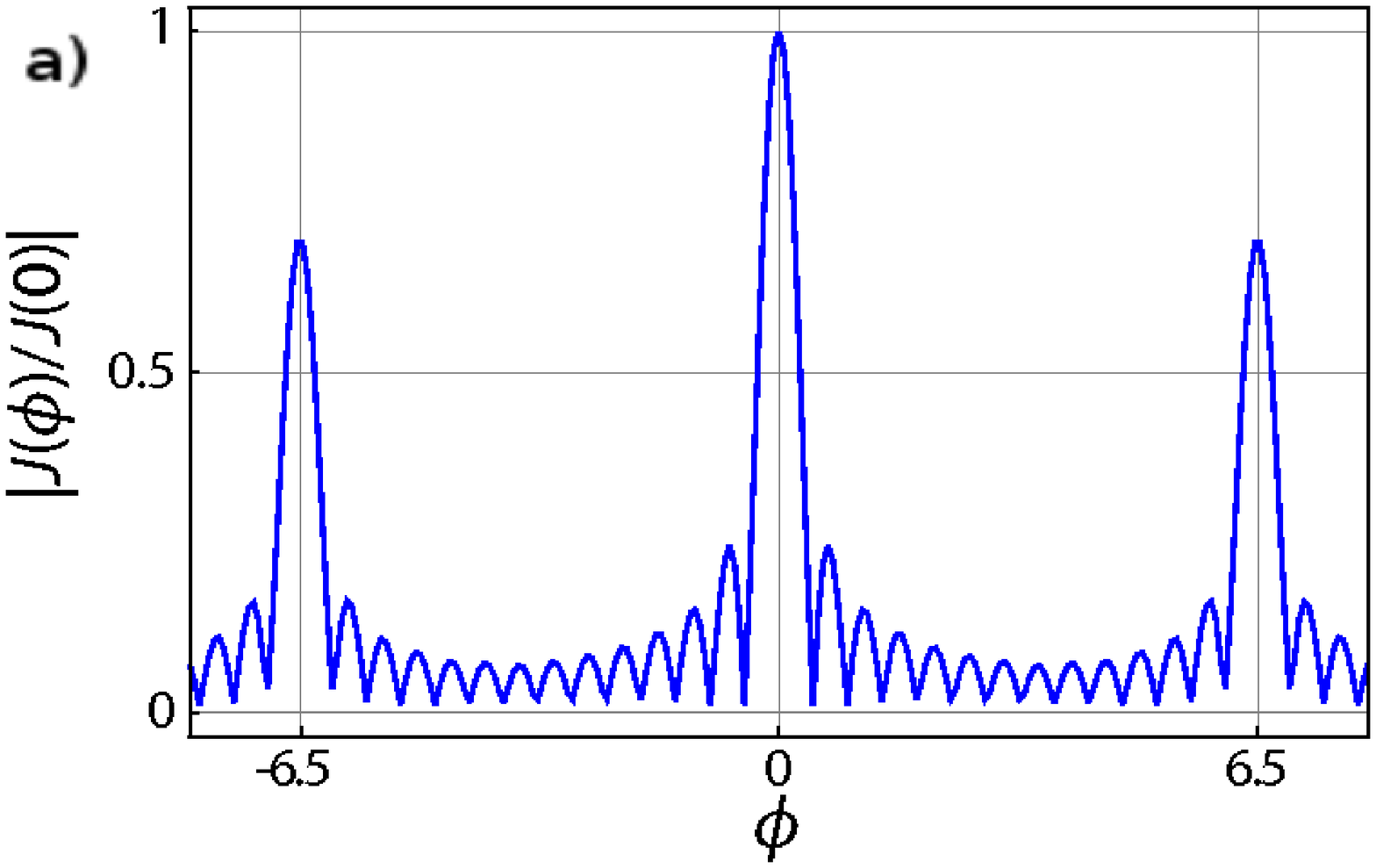}\hspace{.8cm}\includegraphics[width=6.0cm]{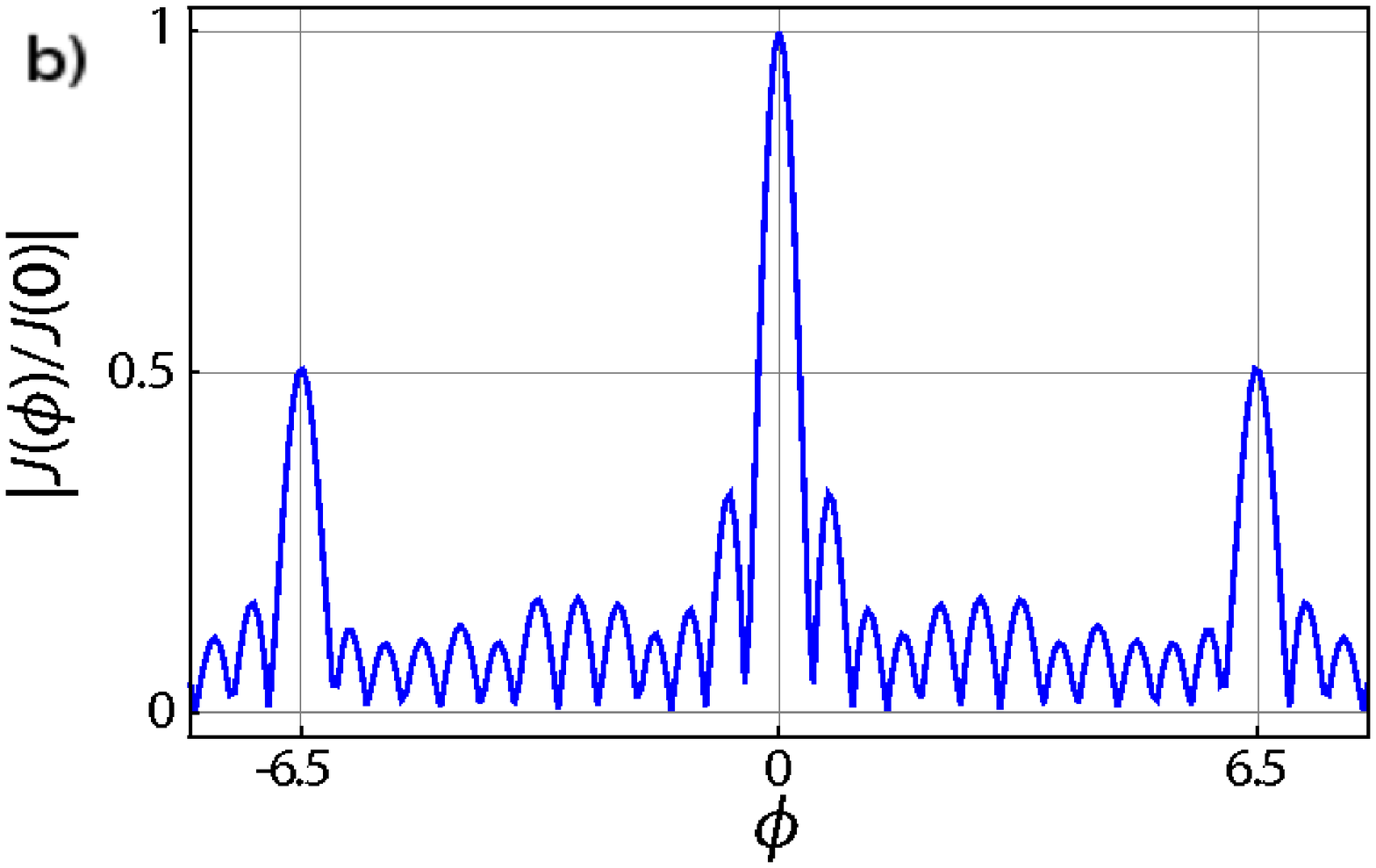}}
\caption{Absolute value of the maximum current flowing across a tunnel junction
versus the total applied magnetic flux $\phi$ (in units of $hc/e$) obtained 
from the Bogoliubov -- de Gennes equations solved on a lattice with $N=16$, 
$M=14$, $U=5t$, and $\delta\varphi=\pi/2$ in units of $J(0)$. (a) $s$-wave 
pairing with $V_0=t$. (b) $d$-wave pairing with $V_1=0.7t$. Here $\phi$ is the 
flux within the junction.}
\label{figj3.4}
\end{figure}

The simplest model of a field-threaded Josephson junction consists of two lines
of impurity sites as used in Sec.~\ref{secJ1.1} (Fig.~\ref{Figj1}). The 
current-flux relation of such a junction as obtained from the Bogoliubov -- de 
Gennes equations is shown in Fig.~\ref{figj3.4} for $s$- and $d$-wave junctions
 with a length of 14 sites and thus 13 plaquettes. Upon first glance, the 
current-flux relation of the $s$-wave junction [Fig.~\ref{figj3.4}~(a)] appears
to be similar to the Fraunhofer pattern known from the Ginzburg-Landau approach
for linear Josephson junctions~\cite{tinkham}, as does the current-flux 
relation for the $d$-wave junction [Fig.~\ref{figj3.4}~(b)]. The 
characteristics are a central peak around $\phi=0$ with width $\Phi_0$ and side
peaks of decreasing height with width $\Phi_0/2$. They display the expected 
global periodicity of $13 \Phi_0$, enforced by gauge invariance, if each 
plaquette is threaded by an integer multiple of $\Phi_0$. On closer inspection 
of Fig.~\ref{figj3.4}~(a) one finds that the $s$-wave junction has one maximum 
surplus in one period of $13 \Phi_0$, whereas the $d$-wave junction has not. 
The width of the peaks in Fig.~\ref{figj3.4}~(a) is therefore slightly smaller 
than the expected value $\Phi_0$. In the following, we explain this effect 
jointly with an investigation of the current-flux relation of inhomogeneous 
junctions by analyzing the Ginzburg-Landau approach for a lattice model.

We consider a two-dimensional superconductor which is divided by a thin, quasi 
one-di\-men\-sio\-nal Josephson  junction of width $d$ oriented along the 
$y$-direction with $d\ll\lambda$, such that screening currents are negligible; 
$\lambda$ is the London penetration depth. If the junction is threaded by a 
constant magnetic field $B_z(x,y)=B_z$, the supercurrent across the junction 
derived from the Ginzburg-Landau equations is
\begin{equation}
J=\int{\rm d}y\,j_{\rm c}(y)\sin(ky),
\label{J1}
\end{equation}
where $k=\pi B_zd/\Phi_0$. The critical current density $j_{\rm c}(y)$ is 
controlled by the microscopic structure of the junction. If $j_{\rm c}(y)$ is 
constant, one obtains the well known Fraunhofer pattern
\begin{equation}
\left|\frac{J(\Phi)}{J(0)}\right|=\left|\frac{\sin(\pi\Phi/\Phi_0)}
{\pi\Phi/\Phi_0}\right|
\label{J2}
\end{equation}
for the current across the junction; $\Phi$ is the total magnetic flux through 
the area of the junction.

\begin{figure}[tb]
\center{\includegraphics[width=6.0cm]{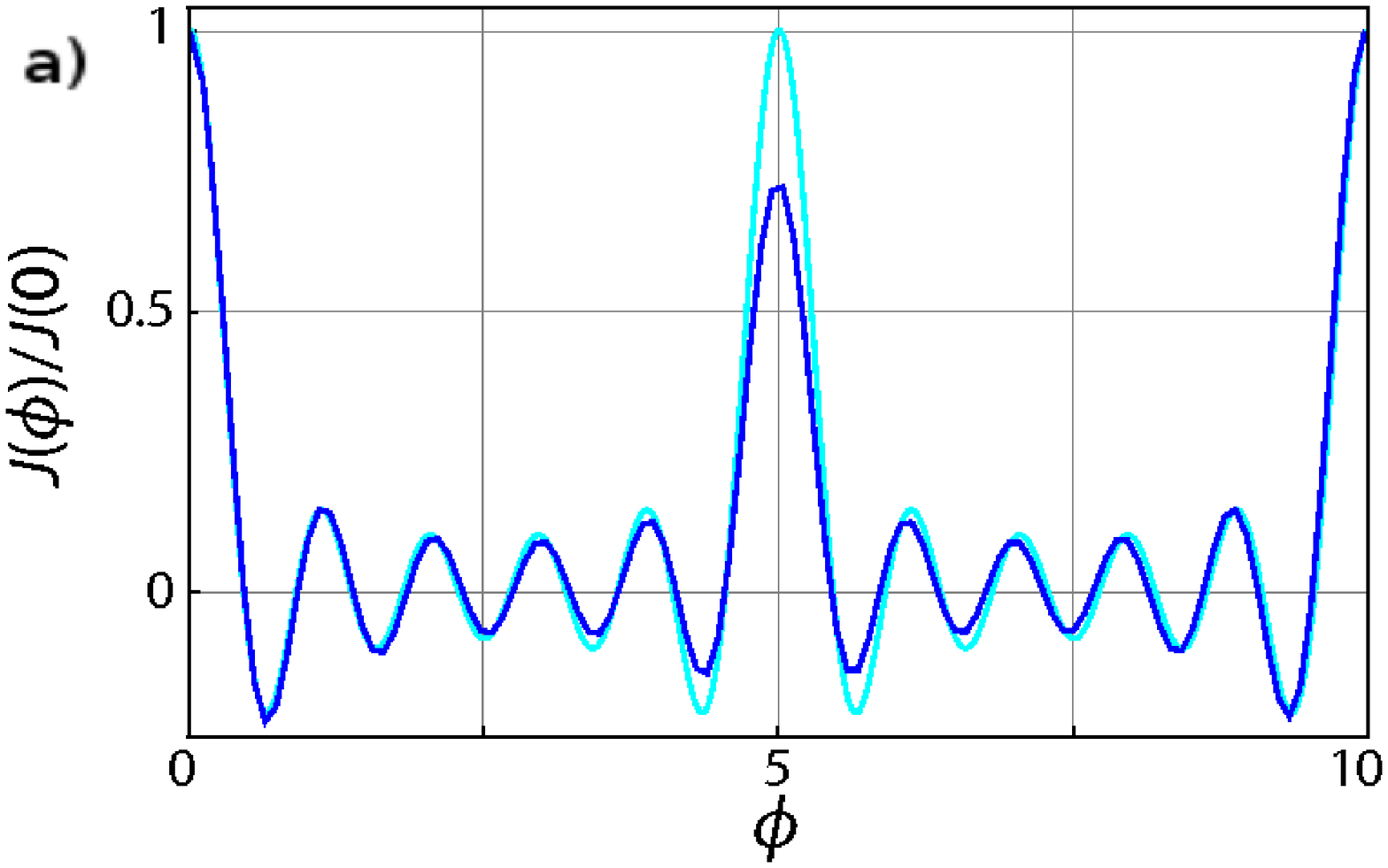}\hspace{.8cm}\includegraphics[width=6.0cm]{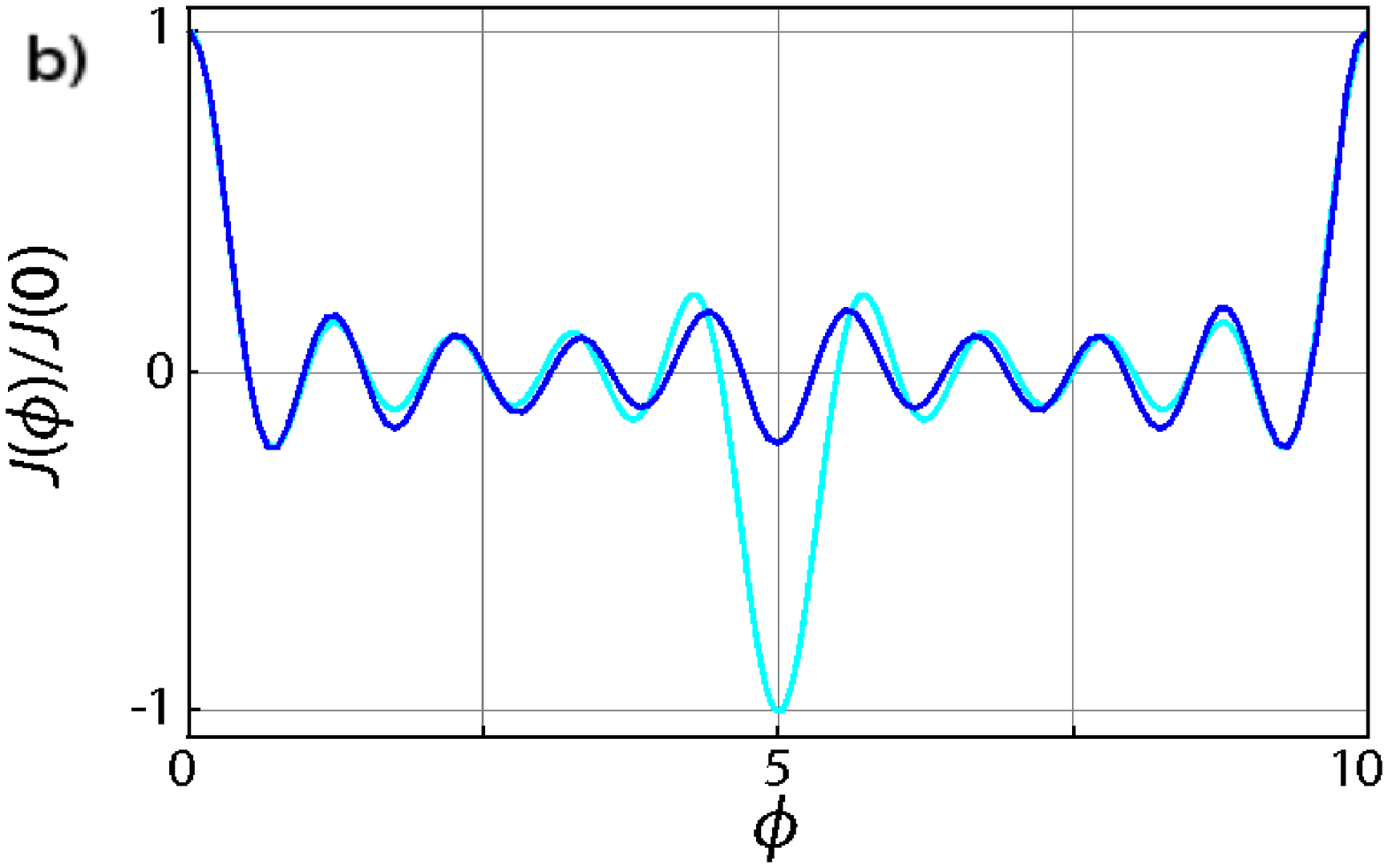}}
\caption{Current-flux relation of a Josephson junction as in Fig.~\ref{figj3.4}
but with $M=11$ (10 plaquettes) as obtained from the Bogoliubov -- de Gennes 
equations (blue) and from the discrete Ginzburg-Landau approach (turquoise). 
(a) $s$-wave pairing: The peaks at $\phi=(M-1)/2$ and $\phi=0$ have the same 
sign. (b) $d$-wave pairing: The peaks at $\phi=( M-1)/2$ and $\phi=0$ have the 
opposite sign. }
\label{figj1.5}
\end{figure}

On a discrete square lattice with $M$ lattice sites in $y$-direction and an 
order parameter defined on the lattice sites ($s$-wave), Eq.~(\ref{J1}) becomes
\begin{equation}
J=\sum_{i=1}^Mj_{{\rm c},i}\sin(ky_i).
\label{J3}
\end{equation}
If $j_{{\rm c},i}$ is equal for all $i$, one obtains a flux dependence similar 
to the Fraunhofer pattern:
\begin{equation}
\frac{J(\phi)}{J(0)}=\sum_{i=1}^M\sin(ky_i)/(M+1)=\frac{\sin\left(k(M+1)/M
\right)}{(M+1)\sin\left(k/M\right)}.
\label{J4}
\end{equation}
This formula reproduces the flux dependence of the supercurrent as obtained 
from the Bogoliubov -- de Gennes equations (shown in Fig.~\ref{figj1.5}), apart
from slight deviations in the amplitude around the central peak at $\phi=
(M-1)/2$. It explains naturally the deviation from the $\Phi_0/2$ periodicity: 
it is  an effect of discreteness, caused by the fact that the number of lattice
sites in $y$-direction exceeds the number of plaquettes by one.

>From what has been explained for an $s$-wave junction, we construct a simple 
Ginzburg-Landau analogon for a $d$-wave junction. In a $d$-wave superconductor,
the order parameter is defined on the bonds between two neighboring lattice 
sites, and we therefore define the corresponding supercurrent as
\begin{equation}
J=\sum_{i=1}^{M-1}j_{{\rm c},i}\sin(k(y_i+1/2)).
\label{J5}
\end{equation}
For a constant $j_{{\rm c},i}$ we obtain
\begin{equation}
\frac{J(\phi)}{J(0)}=\sum_{i=1}^{M-1}\sin(k(y_i+1/2))/M=\frac{\sin\left(
k\right)}{M\sin\left(k/M\right)},
\label{J6}
\end{equation}
which indeed reproduces the $\Phi_0/2$ periodic Fraunhofer pattern obtained 
from the Bogoliubov -- de Gennes equations with nearest-neighbor pairing. The 
deviations in the amplitude are larger than for the $s$-wave junction telling
that the $d$-wave junctions fulfill the Ginzburg-Landau conditions not as well 
as the $s$-wave junction.

\begin{figure}[tb]
\center{\includegraphics[width=5.0cm]{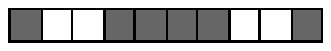}}
\vspace{\FigSpace}
\center{\includegraphics[width=6.0cm]{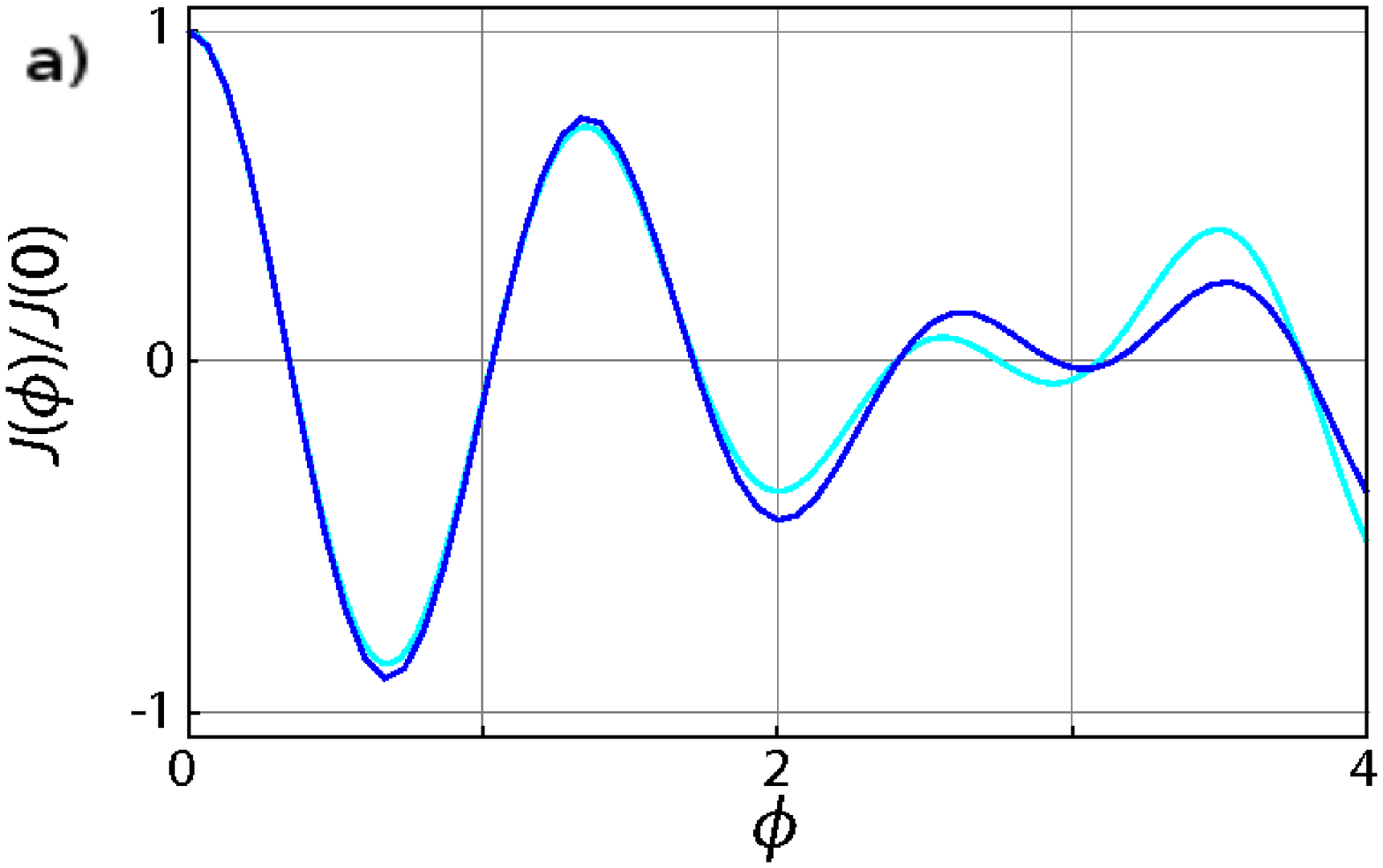}\hspace{.8cm}\includegraphics[width=6.0cm]{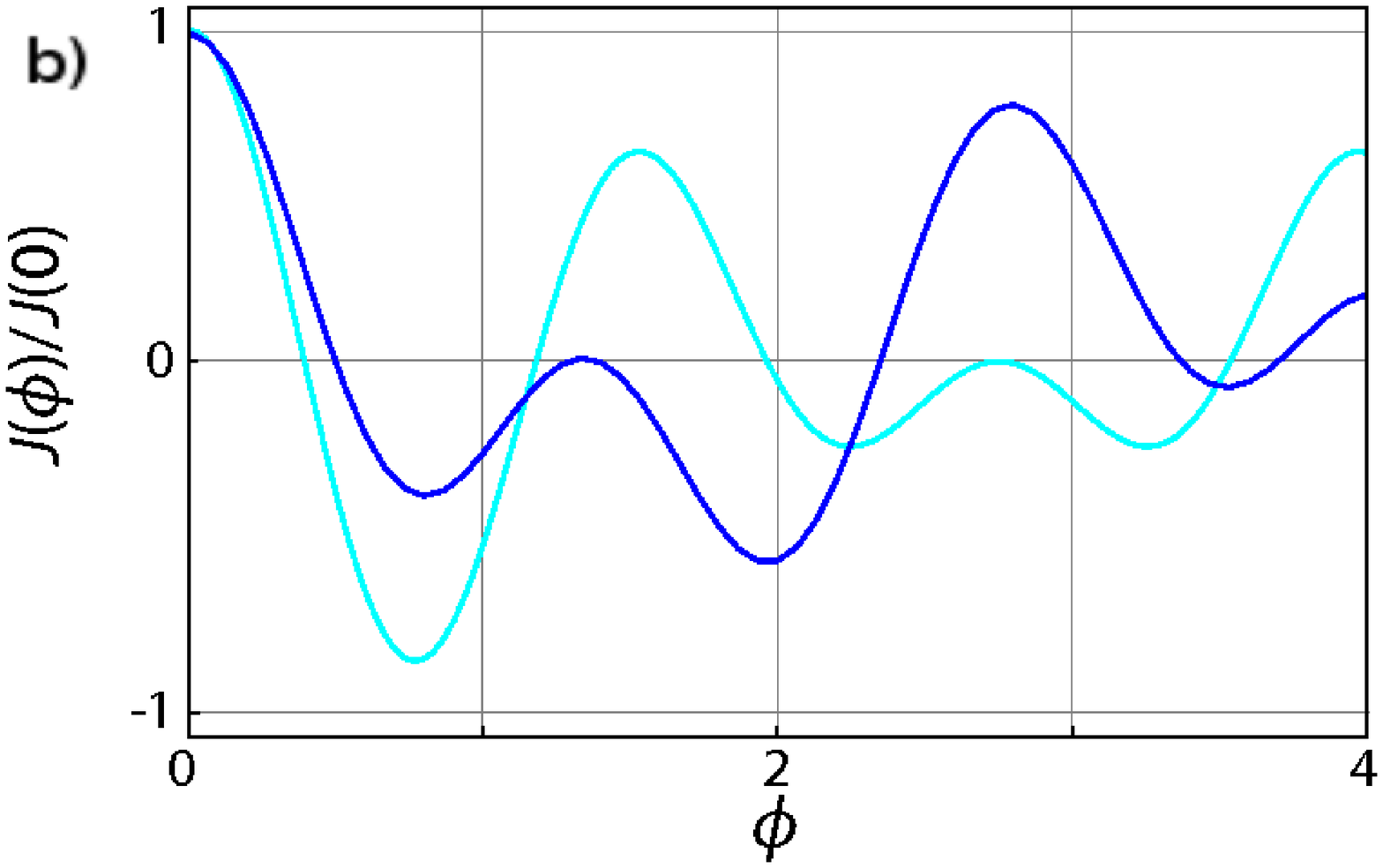}}
\caption{Current-flux relation calculated for a Josephson junction with $M=11$ 
and an inhomogeneous impurity distribution. Top panel: The gray plaquettes in 
the profile of the junction have a scattering potential $U=100t$, while the 
white plaquettes have $U=2t$, thus leaving two transparent channels through 
which almost the entire current flows. Blue: Bogoliubov -- de Gennes equations 
and turquoise: Ginzburg-Landau model for (a) $s$-wave pairing and (b) $d$-wave 
pairing.}
\label{figj2.6}
\end{figure}

The Ginzburg-Landau formulae (\ref{J3}) and (\ref{J5}) are suitable also to 
calculate the supercurrent flowing across junctions with an inhomogeneous 
impurity distribution. It is instructive to compare also in this case the 
supercurrent to results obtained from the Bogoliubov -- de Gennes equations.
Figure~\ref{figj2.6} shows such a comparison for a junction with $M=11$ and 
current flowing only through the two ``gaps'' between the white plaquettes in 
the top panel of Fig.~\ref{figj2.6}. In the microscopic model, this is achieved
by setting strong repulsive potentials $U=100t$ on the sites of the gray 
plaquettes and a small potential $U=2t$ on the white plaquettes. In the 
Ginzburg-Landau approach, we set $j_{{\rm c},i}=0$ except for the two 
transparent channels. This system appears to be quite far from respecting the 
conditions for the validity of the Ginzburg-Landau equations. Nevertheless, for
the $s$-wave junction, the results obtained from the Bogoliubov -- de Gennes 
and Ginzburg-Landau equations are remarkably close. Even for the $d$-wave 
junction, the simple implementation of the Ginzburg-Landau equations reproduces
the same features as the Bogoliubov -- de Gennes equations, in particular it 
has maxima for similar values, but the amplitudes of the oscillations deviate 
strongly.

These considerations jointly lead to the conclusion that, even for small 
junctions where discreteness is pronounced, we do not find any indications that
the Doppler shift has an effect on the current-flux relation of Josephson 
junctions in the tunneling regime. The essential characteristics of the 
current-flux relation, especially the position of the current maxima, agree 
quite well with the Ginzburg-Landau approach, where these effects are not 
included.

\subsubsection{Current-flux relation of transparent junctions}
\label{secJ2.2}
\begin{figure}[t]
\center{\includegraphics[width=6.0cm]{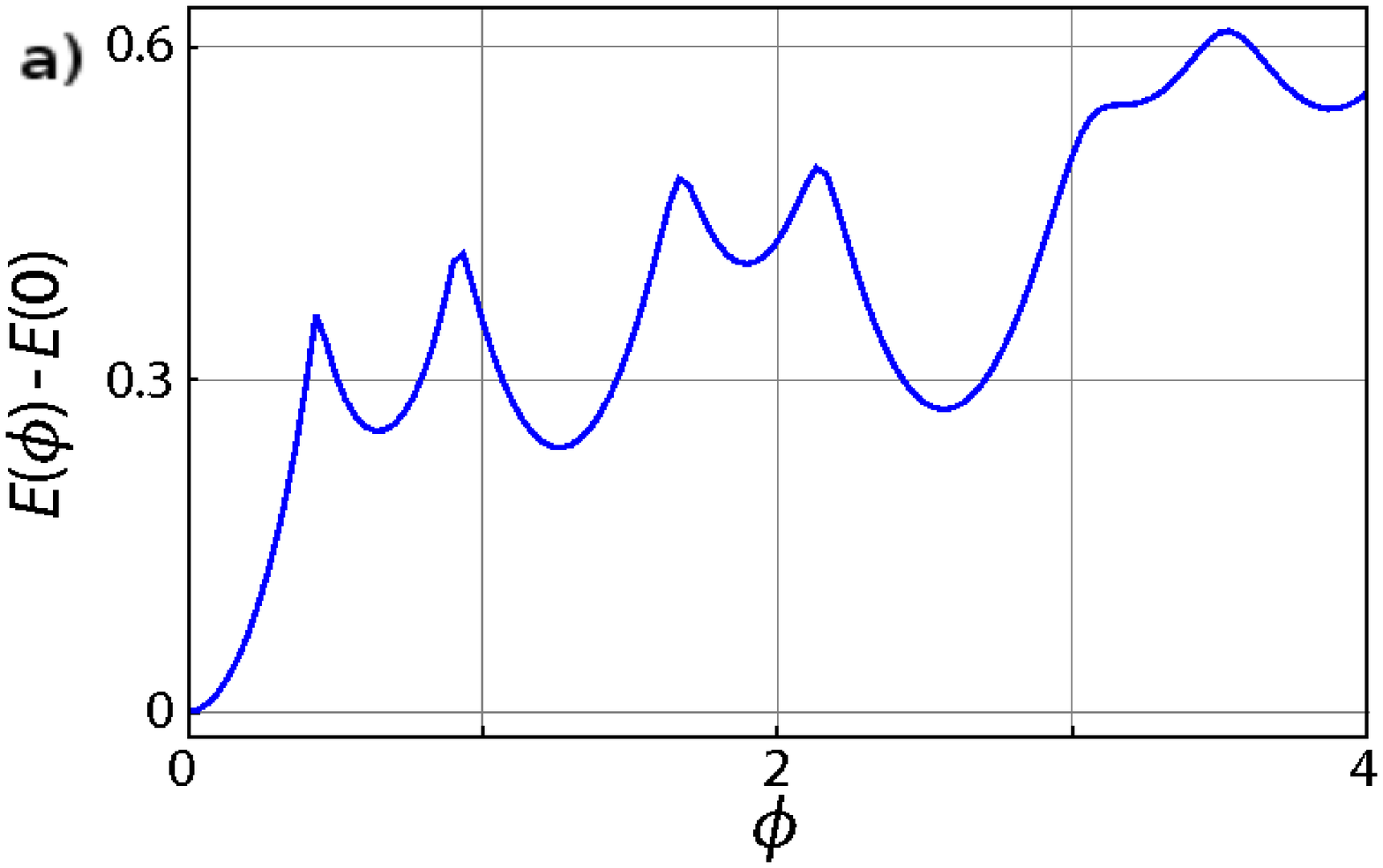}\hspace{.8cm}\includegraphics[width=6.0cm]{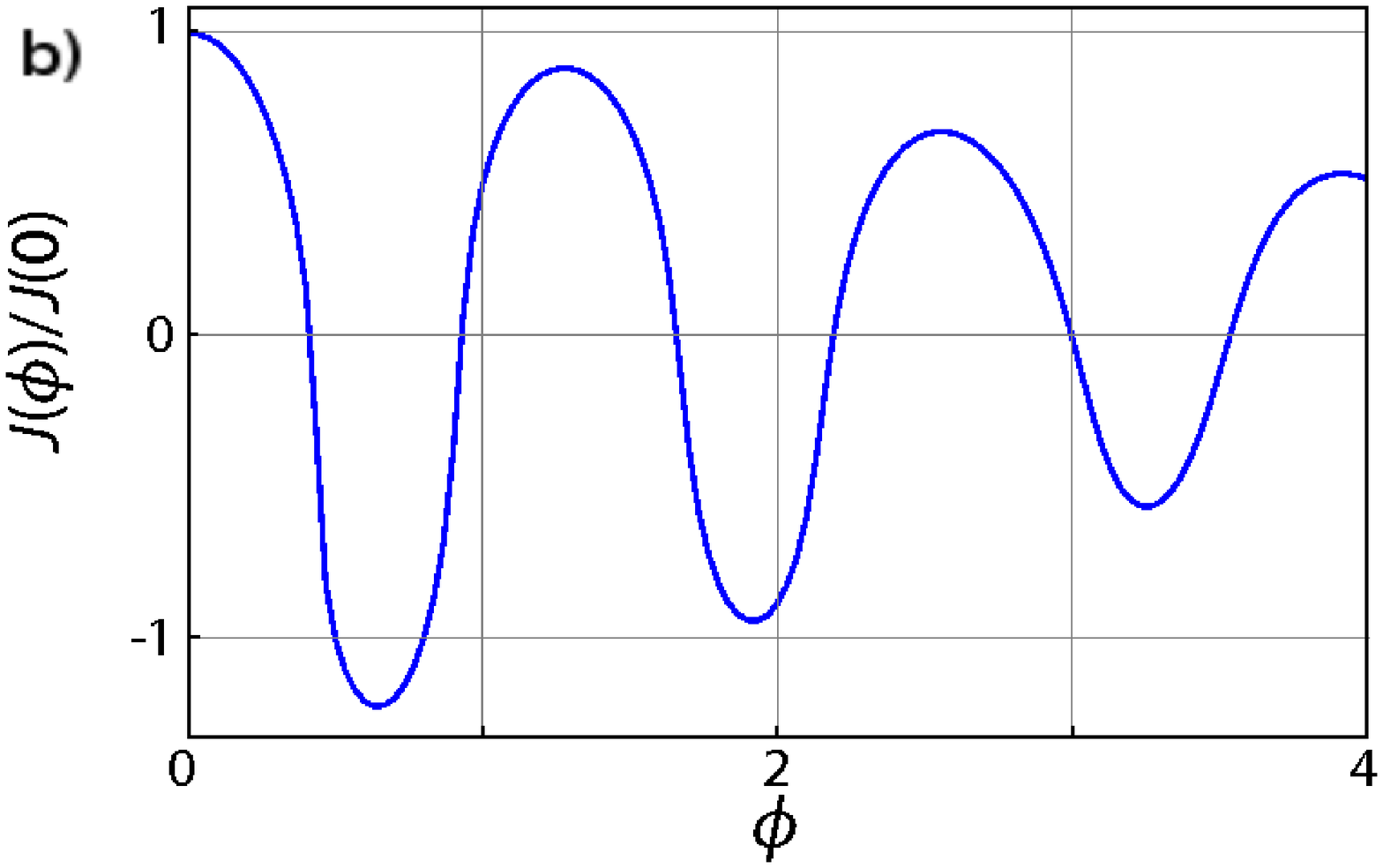}}
\vspace{\FigSpace}
\center{\includegraphics[width=6.0cm]{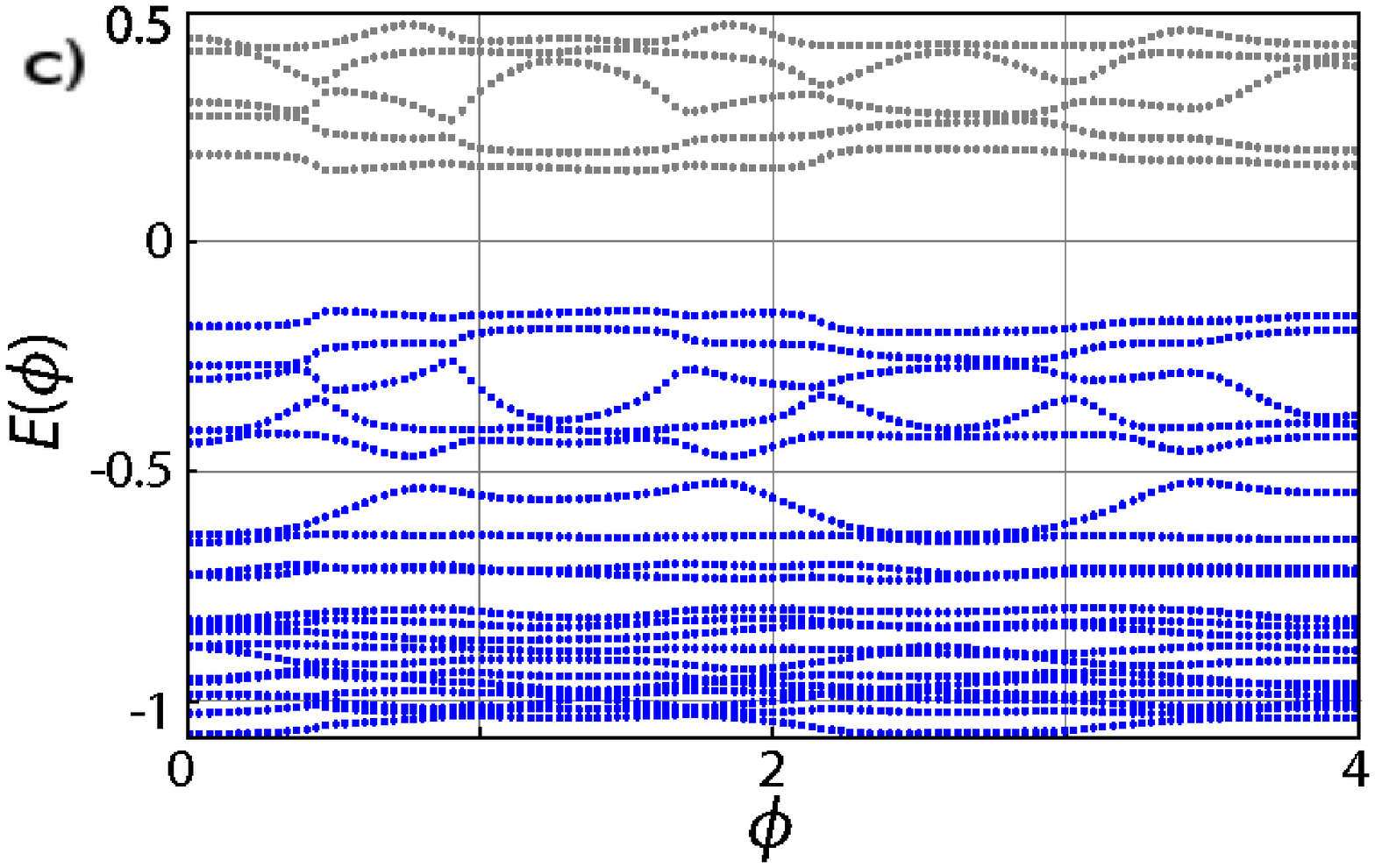}}
\caption{Characteristics of a transparent junction as obtained from the
Bogoliubov -- de Gennes equations in a system with $N=14$ and $M=12$ and a 
homogeneous impurity distribution with repulsive potential $U=2t$. (a) The 
total energy, (b) the Josephson current, and (c) the energy spectrum of the 
system versus the external flux through the junction.}
\label{figj4}
\end{figure}

A magnetic field threading a Josephson junction generates a supercurrent 
circulating around the junction, similar to a vortex in a type II 
superconductor, but with the complete flux confined to the junction. If the 
junction is sufficiently transparent, the order parameter reacts to the current
loop with a phase winding as in a flux-threaded ring, with a winding number $q$
that minimizes the total energy. The superconducting state in a transparent 
junction is therefore characterized similarly as a loop by the quantum number 
$q$ related to a center-of-mass motion of the Cooper pairs and the supercurrent
across the junction changes sign when the condensate reconstructs to another 
$q$. Remarkably, if the transparency is reduced, the discontinuities vanish 
smoothly, the current-flux relation of the superconducting state with fixed $q$
becomes periodic in $\phi$, and in the tunneling regime, all states with 
different $q$ become equivalent. This behavior is illustrated in 
Fig.~\ref{figj4}, which shows $E(\phi)$, $J(\phi)$, and the spectrum for a 
uniform junction with nearest-neighbor pairing and $U=2t$. The total energy 
consists of a series of parabolae, which correspond to different phase winding 
numbers. The kinks in $E(\phi)$ and in the flux dependence of the spectrum are 
sharp for small values of $\phi$, but the finite repulsion on the junction 
smoothens the discontinuities in the supercurrent. Although the Doppler shift 
of the energy levels is not strongly pronounced in Fig.~\ref{figj4}, the 
physical phenomena typical for multiply connected geometries govern the field 
dependence of the supercurrent across a Josephson junction, if its transparency
is sufficiently high.

\section{Conclusions}
For unconventional nodal superconductors we established within a momentum-space
formulation for superconducting loops that $hc/e$ oscillations are present in 
the flux dependence of the ground state. The calculations in momentum space 
were restricted to rotationally symmetric systems like a cylinder, the energy 
spectrum of which depends sensitively on microscopic details. In 
Sec.~\ref{sec:square} we have provided an analysis of the flux periodicity in a
square frame with $d$-wave pairing symmetry analogous to the cylinder geometry 
of Sec.~\ref{sec:1} with remarkably similar results. Nevertheless, the 
real-space calculations contributed to the understanding of the flux 
periodicity. We verified that the characteristic flux dependence of the 
$d$-wave energy spectrum does not depend on the geometry or the absence of 
impurities.  

Within the real-space formulation, we constructed and analyzed more complex 
systems, in particular we investigated the periodicity of Josephson junctions.
The idea that the Doppler shift drives energy levels through the Fermi energy 
in junctions between $d$-wave superconductors, and thereby doubles the 
periodicity of the current-phase relation, seemed natural, but the physics 
turned out to be more subtle. Narrow junctions with only a few channels always 
display a period in the phase difference of $4\pi$, even for $s$-wave 
superconductors, and the Doppler shift in tunnel junctions is too small to 
influence the current-phase relation. Only for transparent junctions does the 
Doppler shift become important; in this regime the supercurrent across a 
Josephson junction behaves similar to the persistent supercurrent in a loop. 
These observations are also valid for the current-flux relation of 
field-threaded junctions. The microscopic theory excellently reproduced the 
results from the Ginzburg-Landau description of Josephson junctions in the 
tunneling regime, even for nanoscopically small systems with $d$-wave pairing.

\medskip
{\small We thank R.~Fr\'esard, S.~Graser, C.~Schneider, and D.~Vollhardt for 
insightful discussions and J. Mannhart for valuable contributions. This work 
was supported by the Deutsche Forschungsgemeinschaft through SFB 484 and TRR 
80.}


\begin{thebibliography}{}

\bibitem{ehrenberg}
W.~Ehrenberg, R.~E.~Siday,
Proc. Phys. Soc. B \textbf{62}, 8 (1949)

\bibitem{AB}
Y.~Aharonov, D.~Bohm,
Phys. Rev. \textbf{115}, 485 (1959)

\bibitem{London}
F.~London, \emph{Superfluids}, John Wiley \& Sons, New York (1950)

\bibitem{bcs}
J.~Bardeen, L.~N.~Cooper, J.~R. Schrieffer,
Phys. Rev. \textbf{108}, 1175 (1957)

\bibitem{Doll}
R.~Doll, M.~N\"abauer,
Phys. Rev. Lett. \textbf{7}, 51 (1961)

\bibitem{Deaver}
B.~S.~Deaver, W.~M.~Fairbank,
Phys. Rev. Lett. \textbf{7}, 43 (1961)

\bibitem{Abrikosov}
A.~A.~Abrikosov, Soviet Physics -- JETP \textbf{5}, 1174 (1957)

\bibitem{Essmann}
U.~Essmann, H.~Tr\"auble,.
Phys. Lett. A \textbf{24}, 526 (1967)

\bibitem{onsager:61}
L.~Onsager, Phys. Rev. Lett. \textbf{7}, 50 (1961)

\bibitem{Byers}
N.~Byers, C.~N.~Yang,
Phys. Rev. Lett. \textbf{7}, 46 (1961)

\bibitem{brenig:61}
W.~Brenig, Phys. Rev. Lett. \textbf{7}, 337 (1961)

\bibitem{schrieffer}
J.~R.~Schrieffer, \emph{Theory of Superconductivity},
Addison Wesley (1964), chapter~8

\bibitem{loder:08}
F.~Loder, A.~P. Kampf, T.~Kopp, J.~Mannhart, C.~Schneider, Yu.~S.~Barash,
Nature Phys. \textbf{4}, 112 (2008)

\bibitem{loder:08.2}
F.~Loder, A.~P.~Kampf, T.~Kopp,
Phys. Rev. B \textbf{78}, 174526 (2008)

\bibitem{juricic:07}
V.~Juri\v{c}i\'{c}, I.~F.~Herbut, Z.~Te\v{s}anovi\'{c},
Phys. Rev. Lett. \textbf{100}, 187006 (2008)

\bibitem{barash:07}
Yu.~S.~Barash,
Phys. Rev. Lett. \textbf{100}, 177003 (2008)

\bibitem{washburn:92}
S.~Washburn, R.~A. Webb,
Rep. Prog. Phys. \textbf{55}, 1311 (1992)

\bibitem{fye}
R.~M.~Fye, M.~J.~Martins, D.~J.~Scalapino, J.~Wagner, W.~Hanke,
Phys. Rev. B \textbf{44}, 6909 (1991); Phys. Rev. B \textbf{45}, 7311 (1992)

\bibitem{Sharov:05}
S.~V.~Sharov, A.~D.~Zaikin,
Phys. Rev. B \textbf{71}, 014518 (2005)

\bibitem{waintal}
X.~Waintal, G.~Fleury, K.~Kazymyrenko, M.~Houzet, P.~Schmitteckert, 
D.~Weinmann,
Phys. Rev. Lett. \textbf{101}, 106804 (2008)

\bibitem{mineev}
V.~P.~Mineev, K.~V.~Samokhin,
\emph{Introduction to Unconventional Superconductivity}, 
Gordon and Breach Science Publishers (1999), chapters~5, 8, and 17

\bibitem{Loder2010}
F. Loder, A. P. Kampf, T. Kopp,
Phys. Rev. B {\bf 81},020511(R) (2010)

\bibitem{czajka:05}
K.~Czajka, M.~M. Ma\'{s}ka, M.~Mierzejewski, Z.~\'{S}led\'{z},
Phys. Rev. B \textbf{72}, 035320 (2005)

\bibitem{scalapino:93}
D.~J. Scalapino, S.~R. White, S.~Zhang,
Phys. Rev. B \textbf{47}, 7995 (1993)

\bibitem{pethick:79}
C.~J.~Pethick, H.~Smith,
Annals of Physics \textbf{119}, 133 (1979)

\bibitem{tinkham}
M.~Tinkham, \emph{Superconductivity},
McGraw-Hill Internation Editions (1996), chapters~3 and 6

\bibitem{vonoppen:92}
F.~von Oppen, E.~K.~Riedel,
Phys. Rev. B \textbf{46}, 3203 (1992)

\bibitem{khavkine:04}
I.~Khavkine, H.-Y. Kee, K.~Maki
Phys. Rev. B. \textbf{70}, 184521 (2004)

\bibitem{Little}
W.~A.~Little, R.~D.~Parks,
Phys. Rev. Lett. \textbf{9}, 9 (1962)

\bibitem{Parks}
R.~D.~Parks, W.~A.~Little,
Phys. Rev. \textbf{133}, A97 (1964)

\bibitem{soininen:94}
P.~I. Soininen, C.~Kallin, A.~J. Berlinsky,
Phys. Rev. B \textbf{50}, 13883 (1994)

\bibitem{wang:95}
Y.~Wang, A.~H.~MacDonald,
Phys. Rev. B \textbf{55}, R3876 (1995)

\bibitem{zhu:00}
J.-X. Zhu, T.~K. Lee, C.~S.~Ting, C.-R.~Hu,
Phys. Rev. B \textbf{61}, 8667 (2000)

\bibitem{Zhu:01}
J.-X. Zhu, C.~S.~Ting,
Phys. Rev. Lett. \textbf{87}, 147002 (2001)

\bibitem{ghosal:02}
A.~Ghosal, C.~Kallin, A.~J.Berlinsky,
Phys. Rev. B \textbf{66}, 214502 (2002)

\bibitem{chen:03}
Y.~Chen, Z.~D.~Wang, C.~S.~Ting,
Phys. Rev. B \textbf{67}, 220501 (2003)

\bibitem{franz:96}
M.~Franz, C.~Kallin, A.~J. Berlinsky,
Phys. Rev. B \textbf{54}, R6897 (1996)

\bibitem{bagwell:94}
P.~F.~Bagwell, Phys. Rev. B \textbf{49}, 6841 (1994)

\bibitem{loder:09}
F.~Loder, A.~P. Kampf, T.~Kopp, J.~Mannhart,
New J. Phys. \textbf{11}, 075005 (2009)

\bibitem{josephson}
D.~B.~Josephson, 
Phys. Lett. \textbf{1}, 251 (1962)

\bibitem{golubov:04}
A.~A.~Golubov, M.~Y.~Kupriyanov, E.~Il'ichev,
Rev. Mod Phys. \textbf{76}, 411 (2004)

\bibitem{degennes}
P.~G.~de~Gennes,
\emph{Superconductivity of Metals and Alloys}, 
Addison Wesley (1966), chapter~5

\bibitem{andersen}
B.~M. Andersen, I.~V. Bobkova, P.~J. Hirschfeld, Yu.~S.~Barash,
Phys. Rev. B \textbf{72}, 184510 (2005); Phys. Rev. Lett \textbf{96}, 117005 
(2006)

\bibitem{cayssol:03}
J.~Cayssol, T.~Kontos, G.~Montambaux,
Phys. Rev. B \textbf{67}, 184508 (2003)

\end{thebibliography}
\end{document}